\documentclass[structabstract]{aa}  

\usepackage{url}
\usepackage{cancel}
\usepackage{graphics}
\usepackage{latexsym}
\usepackage{graphicx}
\usepackage{amssymb}
\usepackage{lscape}
\usepackage{xcolor}
\usepackage{txfonts}
\usepackage{natbib}
\usepackage{chngcntr}
\usepackage{soul}

\usepackage[textwidth=3.5cm]{todonotes}
\usepackage{blindtext}

\renewcommand{\d}[0]{{\rm d}}
\newcommand{\e}[0]{{\rm e}}
\renewcommand{\i}[0]{{\rm i}}
\newcommand{\ave}[1]{\langle #1 \rangle}

\renewcommand{\Ref}[1]{(\ref{#1})}

\newcommand{\mat}[1]{\tens{#1}}

\newcommand{\msol}[0]{{\rm M}_\odot}

\newcommand{\erf}[1]{\mathrm{erf}{\left(#1\right)}}
\newcommand{\cfhtlens}[0]{\mbox{CFHTLenS}}

\begin{document}

\title{{CFHTLenS: Galaxy bias as function of scale, stellar mass, and
  colour}\thanks{MCMCs of biasing functions and estimated galaxy
    redshift distributions are only available in electronic form at
    the CDS via anonymous ftp to cdsarc.u-strasbg.fr (130.79.128.5) or
    via \mbox{\tt http://cdsweb.u-strasbg.fr/cgi-bin/qcat?J/A+A/}}}
\subtitle{{Conflicts with predictions by semi-analytic models}}

\author{Patrick Simon$^1$\thanks{\email{psimon@astro.uni-bonn.de}} and
  Stefan Hilbert$^{2,3}$}

\institute{$^1$Argelander-Institut f\"ur Astronomie, Universit\"at
  Bonn, Auf dem H\"ugel 71, 53121 Bonn, Germany\\
  $^2$ Faculty of Physics, Ludwig-Maximilians University, Scheinerstr. 1, 81679 M\"unchen, Germany\\
  $^3$ Excellence Cluster Universe, Boltzmannstr. 2, 85748 Garching, Germany}

\date{Received \today}

\authorrunning{Simon and Hilbert}
\titlerunning{Galaxy bias in \cfhtlens}

\abstract{Galaxy models predict a tight relation between the
  clustering of galaxies and dark matter on cosmological scales, but
  predictions differ notably in the details. We used this opportunity
  and tested two semi-analytic models by the Munich and Durham groups
  with data from the Canada-France-Hawaii Telescope Lensing Survey
  (\cfhtlens). For the test we measured the scale-dependent galaxy
  bias factor $b(k)$ and correlation factor $r(k)$ from linear to
  non-linear scales of \mbox{$k\approx10\,h\,\rm Mpc^{-1}$} at two
  redshifts $\bar{z}=0.35,0.51$ for galaxies with stellar mass between
  $5\times10^9$ and $3\times10^{11}\,h_{\rm 70}^{-2}\,\msol$. Our
  improved gravitational lensing technique accounts for the intrinsic
  alignment of sources and the magnification of lens galaxies for
  better constraints for the galaxy-matter correlation $r(k)$. Galaxy
  bias in \cfhtlens~increases with $k$ and stellar mass; it is
  colour-dependent, revealing the individual footprints of galaxy
  types.  Despite a reasonable model agreement for the relative change
  with both scale and galaxy properties, there is a clear conflict for
  $b(k)$ with no model preference: the model galaxies are too weakly
  clustered. This may flag a model problem at \mbox{$z\gtrsim0.3$} for
  all stellar masses. As in the models, however, there is a high
  correlation $r(k)$ between matter and galaxy density on all scales,
  and galaxy bias is typically consistent with a deterministic bias on
  linear scales. Only our blue and low-mass galaxies of about
  \mbox{$7\times10^9\,h_{\rm 70}^{-2}\,\msol$} at
  \mbox{$\bar{z}=0.51$} show, contrary to the models, a weak tendency
  towards a stochastic bias on linear scales where
  \mbox{$r_{\rm
      ls}=0.75\pm0.14\,{\rm(stat.)}\pm0.06\,{\rm(sys.)}$}. This result
  is of interest for cosmological probes, such as $E_{\rm G}$, that
  rely on a deterministic galaxy bias. We provide Monte Carlo
  realisations of posterior constraints for $b(k)$ and $r(k)$ in
  \cfhtlens~for every galaxy sample in this paper as online material.}


\maketitle

\section{Introduction}

The broad diversity of galaxy types and their large-scale clustering
requires complex and numerically challenging physics that in models is
approximated or empirically calibrated. This leaves us today with a
variety of modelling techniques that produce a range of theoretical,
not always consistent predictions for galaxy statistics
\citep[e.g.][]{2020NatRP...2...42V,2018MNRAS.475..676S,10.1093/mnras/stw1888,2017MNRAS.469.2626H,2018MNRAS.477..359C,2016arXiv160808629S}. Of
major interest for model tests, due to the key role of dark matter
halos for galaxy physics, is the predicted relative clustering of
matter and galaxies: the galaxy bias.

Galaxy bias varies with galaxy type \citep[and references
therein]{2010gfe..book.....M}, rendering it a type-dependent footprint
for the galaxy-matter connection. Motivated by this idea, as a test we
compare measurements of these footprints in the Canada-France-Hawaii
Lensing Survey (\cfhtlens) data \citep{2013MNRAS.433.2545E} to
predictions by two semi-analytic models (SAMs) for various galaxy
types, redshifts, and spatial scales from the linear down to the
highly non-linear regime at \mbox{$k\approx10\,h\,\rm Mpc^{-1}$}. They
are the Munich model `H15' by \cite{2015MNRAS.451.2663H} and the
Durham model `L12' by \cite{2012MNRAS.426.2142L}.  These SAMs are
different implementations of a technique that populates dark matter
halos in cosmological simulations, here the Millennium Simulation
\citep{2005Natur.435..629S}, with model galaxies by emulating known
galaxy physics \citep[][for a general
overview]{2000MNRAS.319..168C}. Their predictions for the excess
surface matter-density around galaxy pairs were already studied in
\cite{2016arXiv160808629S} and \cite{2019A&A...622A.104S}. Here we
focus on galaxy biasing in these SAMs.

Among the many ways to define galaxy bias (see e.g.
\citealt{2016arXiv161109787D}), our model tests employ the definition
by \cite{1999ApJ...518L..69T} in terms of galaxy and matter power
spectra for the (comoving) Fourier modes $k$,
\begin{equation}
  \label{eq:biasdef}
  b(k):=\sqrt{\frac{P_{\rm g}(k)}{P_{\rm m}(k)}}~;~
  r(k):=\frac{P_{\rm gm}(k)}{\sqrt{P_{\rm m}(k)\,P_{\rm g}(k)}}\;.
\end{equation}
These two biasing functions are measured for a specific galaxy
population and redshift. They incorporate the power spectra of
fluctuations in the matter density field, $P_{\rm m}(k)$; in the
galaxy density field, $P_{\rm g}(k)$; and their cross power
$P_{\rm gm}(k)$. In the definition of $P_{\rm g}(k)$, a Poisson
sampling shot-noise is subtracted, which may produce \mbox{$r(k)>1$}
\citep{peebles80,2001MNRAS.321..439G}. The Eqs. \Ref{eq:biasdef} are
sensitive to a stochastic bias; however, restricted to two-point
statistics, they do not discriminate between a non-linear
deterministic bias and a stochastic bias. The only exceptions are
linear scales where \mbox{$k\ll1\,h\,\rm Mpc^{-1}$} and density
fluctuations $\tilde{\delta}(\vec{k})$ are small
\citep{2001MNRAS.320..289S,1999ApJ...520...24D}. For example, on
linear scales \mbox{$r(k)<1$} indicates a stochastic scatter
$\epsilon_k$ in the relation
\mbox{$\tilde{\delta}_{\rm g}(\vec{k})=b(k)\,\tilde{\delta}_{\rm
    m}(\vec{k})+\epsilon_k$} between galaxy density contrast,
$\tilde{\delta}_{\rm g}$, and matter density contrast,
$\tilde{\delta}_{\rm m}$, whereas for large density fluctuations
\mbox{$r(k)<1$} can also indicate
\mbox{$\tilde{\delta}_{\rm g}(\vec{k})=b_k[\tilde{\delta}_{\rm
    m}(\vec{k})]+\epsilon_k$} with a non-linear mapping $b_k[x]$ and
smaller (or vanishing) $\epsilon_k$. Studies of galaxy clustering show
that both non-linearity and stochasticity increase towards smaller
scales, and galaxy bias is essentially linear and deterministic
($\epsilon_k=0$) at \mbox{$2\pi\,k^{-1}\gtrsim30\,h^{-1}\rm Mpc$}
\citep[e.g.][]{2018PhRvD..98b3508F, 2018PhRvD..98b3507G,
  2018MNRAS.479.1240D, 2016A&A...594A..62D, 2016MNRAS.460.1310P,
  2011ApJ...731..102K, 2008MNRAS.385.1635S, 2005MNRAS.356..247W,
  1999ApJ...518L..69T}.  Despite the limitations, these measures of
galaxy biasing highlight differences between galaxy populations and
their relation to the matter density field, they are easily measured
in simulations \citep{2018MNRAS.475..676S,2004ApJ...601....1W}, and
are hence a useful quantity to test galaxy models.

Perfectly suited to measuring scale-dependent galaxy bias is the
analysis of the weak gravitational-lensing effect
\citep{schneider2006gravitational}. This exploits small coherent
distortions of distant galaxy images (sources) by the large-scale
matter-density field in the foreground. For second-order statistics
the analysis correlates shear distortions with lens galaxy positions
to obtain, apart from an amplitude normalisation, galaxy bias in
projection. This can be normalised or deprojected if the redshift
distributions of sources and lens galaxies, as well as the background
cosmology, mainly $\Omega_{\rm m}$, is known
\citep{1998A&A...334....1V, 1998ApJ...498...43S, 2001ApJ...558L..11H}.
Several previous studies have successfully applied this (or a similar)
lensing analysis \citep{2018MNRAS.479.1240D,2016arXiv160908167P,
  2013MNRAS.433.1146C, 2012ApJ...750...37J, 2007A&A...461..861S,
  2003MNRAS.346..994P, 2001ApJ...558L..11H}, but they neglect the effects
of lensing magnification and source intrinsic alignments
\citep[e.g.][]{2004PhRvD..70f3526H,2019arXiv191006400U}.
In particular, intrinsic alignments (IA) have to be included in the
lensing signal to reach an accuracy of several per cent for $b(k)$ and
$r(k)$, as demonstrated in \cite{2017arXiv171102677S} (hereafter SH18)
by using mock galaxy and shear catalogues that are similar to ours
here. We apply the SH18 procedure with some improvements to measure
the functions in Eqs. \Ref{eq:biasdef}.

This lensing technique probes galaxy biasing with a minimum of
assumptions and is also valid in the non-linear regime. This sets it
apart from other techniques that, for instance, only apply in the
weakly linear regime; require a theoretical reference model for the
clustering of matter as a function of scale; or depend on a detailed
physical model for the clustering of matter and galaxies
\citep[e.g.][]{2013MNRAS.430..767C, 2013A&A...557A..17M,
  2018MNRAS.479.1240D, 2020arXiv200715632H}. However, our technique
employs flexible model functions for $b(k)$ and $r(k)$ to stabilise
the deprojection of the projected galaxy bias. Although based on a
halo model prescription, their purpose notably is not an accurate
interpretation of the projected galaxy bias in terms of a physical
model, but rather an accurate deprojection of the model-free biasing
functions.  The completely generic templates in SH18 could be used
equally well, albeit with less accuracy. To highlight our focus on the
biasing functions, we refer to the stabilising model functions in the
reconstruction as `templates'.

The outline of this paper is as follows. Section 2 describes our
\cfhtlens~shear catalogue and the selection of galaxy (foreground)
samples for the analysis. For these we estimate the redshift
distributions (lenses and sources) and the mean number density of the
lenses; both are used in the analysis. For the model predictions, we
select similar galaxy samples from the SAMs in Sect. 3, accounting for
the stellar-mass errors in \cfhtlens.  Section 4 summarises the
template-based deprojection technique for $b(k)$ and $r(k)$.  Our new
analysis improves on that of SH18 with regard to the transformation
bias in the aperture statistics (Sect. 4.4), the prior for the
template parameters of the biasing functions (Appendix
\ref{app:templateprior}), and an allowed mismatch between matter and
satellite galaxy distribution inside dark matter halos (Sect. 4.6). We
present raw (i.e. not normalised) \cfhtlens~measurements for the
projected galaxy bias already in this section. We compare our main
results for $b(k)$ and $r(k)$ to the H15 and L12 predictions in
Sect. 5, along with some physical interpretation, and discuss the
results in Sect. 6.

The fiducial background cosmology defines the projection of galaxy
bias and the lensing kernel; a small additional dependence is
introduced by the shape of a fiducial matter power spectrum that
impacts the smoothing in the projection.  Our results adopt a flat
$\Lambda\rm CDM$ cosmology with a matter density parameter of
$\Omega_{\rm m}=0.306$, a Hubble constant of
$H_0=h\,100\,\rm km\,s^{-1}\,Mpc^{-1}$ and $h=0.683$, a baryon matter
density of $\Omega_{\rm b}=0.0479$, and a primordial spectral index of
$n_{\rm s}=0.965$,in agreement with \cite{2016A&A...594A..13P}
(Sect. 4.1). Systematic errors in the amplitude of the
\cfhtlens~biasing functions due to uncertainties in the fiducial
cosmological model and galaxy redshift distributions are explored in
Sect.  \ref{sect:normerrors}.

\section{Canada-France-Hawaii Lensing Survey}

\begin{table}
  \begin{center}
    \caption{\label{tab:omeff} Effective survey area
      $\Omega_{\rm eff}$ for individual patches and all patches
      combined (\texttt{mask}=0). Only $129$ MegaCam fields that are
      flagged as suitable for a cosmic-shear analysis are included.}
 \begin{tabular}{lcccc|c}
   \hline\hline
    & W1 & W2 & W3 & W4 & W1 to W4\\
    \hline & & & & &\\
    $\Omega_{\rm eff}/{\rm deg^2}$ & $42.42$ & $11.72$ & $25.43$ & $12.55$ & $92.12$  
  \end{tabular}
\end{center}
\end{table}

The \cfhtlens\footnote{\url{http://cfhtlens.org}} is a multi-colour
lensing survey based on the Wide part of the Canada-France-Hawaii
Survey (CFHTLS) taken with the $2{\rm k}\times4{\rm k}$
\mbox{MegaPrime} camera (\mbox{MegaCam}) between 2003 and 2008
\citep{2013MNRAS.433.2545E}. The pixel size of the camera CCD chips is
$0.187\, \rm arcsec$. The survey consists of four patches W1
(\mbox{$\sim63.8\,\rm deg^2$}), W2 (\mbox{$\sim22.6\,\rm deg^2$}), W3
(\mbox{$\sim44.2\,\rm deg^2$}), and W4 (\mbox{$\sim23.3\,\rm deg^2$})
with $154\,\rm deg^2$ total sky area. The patches are mosaics of $171$
\mbox{$1\times1\,\rm deg^2$} \mbox{MegaCam} fields, observed with the
five filters $u^\ast g^\prime r^\prime i^\prime z^\prime$ to the
limiting AB magnitudes of \mbox{$z^\prime\lesssim23.5\,\rm mag$} and
between $24.5$ and $25.5\,\rm mag$ for the other filters ($5\sigma$
detection limit in a $2\,\rm arcsec$ aperture). All filters were
obtained under sub-arcsec seeing conditions of which $i^\prime$, used
for shear estimates, has a median seeing of $0.68\,\rm arcsec$. After
a break of the $i^\prime$ filter in 2008, $33$ fields were observed
with the slightly different $y^\prime$ filter. The survey combines
weak-lensing data processing with \texttt{THELI}
\citep{2013MNRAS.433.2545E}, shear measurement with \emph{lens}fit
\citep{2013MNRAS.429.2858M}, and photometric redshift measurement with
PSF-matched photometry \citep{2012MNRAS.421.2355H}. A full
systematic-error analysis of the shear measurements in combination
with the photometric redshifts is presented in
\cite{2012MNRAS.427..146H}, with additional error analyses of the
photometric redshift measurements presented in
\cite{2013MNRAS.431.1547B}.

The survey data product is the object and lensing catalogue
v1.1\footnote{\url{http://www.cadc-ccda.hia-iha.nrc-cnrc.gc.ca/en/community/CFHTLens/query.html}},
for which galaxies were selected from the $i^\prime$-band stacks with
the \texttt{SExtractor} software \citep{1996A&AS..117..393B}. The
revised photometry of these objects yields a root mean square (RMS)
uncertainty of \mbox{$\sigma\approx0.01-0.03\,\rm mag$} for the
photometric calibration in all five bands
\citep{2012MNRAS.421.2355H}. From {these data}, photometric
redshifts $z_{\rm p}$ as galaxy distance estimators were inferred with
the Bayesian Photometric Redshift Code
\citep[\texttt{BPZ};][]{2000ApJ...536..571B}.

From the catalogue we used only the $129$ science-ready fields that
are considered suitable for cosmic shear studies in
\cite{2012MNRAS.427..146H}. To estimate the effective unmasked area of
this subset of fields, we counted the number of pixels in the mask
files of the co-added images with \mbox{\texttt{mask}=0}; only source
and lens galaxies from these regions are selected for the analysis.
Table \ref{tab:omeff} lists the effective survey area of all patches
W1 to W4, amounting in total to an effective area of
$92.12\,\rm deg^2$. We summarise here the details of our lens and
source samples.

\begin{table*}
  \begin{center}
    \caption{\label{tab:lenssamples} Overview of lens samples and
      selection criteria in addition to \mbox{\texttt{star\_flag}<1},
      \mbox{\texttt{T\_B}>-99}, \mbox{\texttt{Flag}<3}, and
      $i^\prime\in[17.5,22.5)$.}
 \begin{tabular}{lccl}
   \hline\hline
    Sample & $N_{\rm g}$\tablefootmark{a} & $\bar{z}\pm\sigma_{\rm z}$\tablefootmark{b} & Selection\tablefootmark{c}\\
    \hline\\
    SM1 low-$z$ & $4.96\times10^{4}$ & $0.34\pm0.11$ & $z_{\rm p}\in[0.2,0.44]$, $M_\ast\in[0.5,1]$ 
    \\
    SM2 low-$z$ & $3.98\times10^{4}$ & $0.34\pm0.11$ & $z_{\rm p}\in[0.2,0.44]$, $M_\ast\in[1,2]$ 
    \\
    SM3 low-$z$ & $3.01\times10^{4}$ & $0.35\pm0.10$ & $z_{\rm p}\in[0.2,0.44]$, $M_\ast\in[2,4]$ 
    \\
    SM4 low-$z$ & $2.02\times10^{4}$ & $0.36\pm0.08$ & $z_{\rm p}\in[0.2,0.44]$, $M_\ast\in[4,8]$ 
    \\
    SM5 low-$z$ & $9.20\times10^{3}$ & $0.37\pm0.07$ & $z_{\rm p}\in[0.2,0.44]$, $M_\ast\in[8,16]$ 
    \\
    SM6 low-$z$ & $2.46\times10^{3}$ & $0.38\pm0.07$ & $z_{\rm p}\in[0.2,0.44]$, $M_\ast\in[16,32]$ 
    \\
    RED low-$z$ & $7.15\times10^{4}$ & $0.35\pm0.08$ & $z_{\rm p}\in[0.2,0.44]$, $M_\ast\in[0.5,32]$, $u^\ast-r^\prime>1.93\,z_{\rm p}+1.85$ 
    \\
    BLUE low-$z$ & $7.98\times10^{4}$ & $0.34\pm0.12$ & $z_{\rm p}\in[0.2,0.44]$, $M_\ast\in[0.5,32]$, $u^\ast-r^\prime\le1.93\,z_{\rm p}+1.85$ 
    \\\\
    SM1 high-$z$ & $4.80\times10^{4}$ & $0.51\pm0.10$ & $z_{\rm p}\in[0.44,0.6]$, $M_\ast\in[0.5,1]$ 
    \\
    SM2 high-$z$ & $5.74\times10^{4}$ & $0.51\pm0.09$ & $z_{\rm p}\in[0.44,0.6]$, $M_\ast\in[1,2]$ 
    \\
    SM3 high-$z$ & $5.37\times10^{4}$ & $0.52\pm0.08$ & $z_{\rm p}\in[0.44,0.6]$, $M_\ast\in[2,4]$ 
    \\
    SM4 high-$z$ & $3.78\times10^{4}$ & $0.52\pm0.08$ & $z_{\rm p}\in[0.44,0.6]$, $M_\ast\in[4,8]$ 
    \\
    SM5 high-$z$ & $1.68\times10^{4}$ & $0.52\pm0.07$ & $z_{\rm p}\in[0.44,0.6]$, $M_\ast\in[8,16]$ 
    \\
    SM6 high-$z$ & $3.61\times10^{3}$ & $0.52\pm0.07$ & $z_{\rm p}\in[0.44,0.6]$, $M_\ast\in[16,32]$ 
    \\
    RED high-$z$ & $8.84\times10^{4}$ & $0.51\pm0.07$ & $z_{\rm p}\in[0.44,0.6]$, $M_\ast\in[0.5,32]$, $u^\ast-r^\prime>1.93\,z_{\rm p}+1.85$ 
    \\
    BLUE high-$z$ & $1.29\times10^{5}$ & $0.52\pm0.09$ & $z_{\rm p}\in[0.44,0.6]$, $M_\ast\in[0.5,32]$, $u^\ast-r^\prime\le1.93\,z_{\rm p}+1.85$ 
  \end{tabular}
 \tablefoot{
  \tablefoottext{a}{total number of objects for W1 to W4 combined (effective $92.12$ $\rm deg^2$)};
  \tablefoottext{b}{mean and dispersion of maximum-likelihood redshift distribution};
  \tablefoottext{c}{photometric redshift $z_{\rm p}$, stellar mass
  $M_\ast$ in units of $h_{70}^{-2}10^{10}\,{\rm M}_\odot$,
  apparent AB \cfhtlens~magnitudes $u^\ast$ and $r^\prime$} }
\end{center}
\end{table*}

\subsection{Galaxy samples}
\label{sect:lenses}

We measure galaxy bias for foreground galaxies that {were} binned
in stellar mass (SM1 to SM6) or colour (RED or BLUE), and additionally
in redshift (low-$z$ or high-$z$).  Table \ref{tab:lenssamples} is an
overview of all $2\times8=16$ samples and their selection criteria.

All samples select only bright objects
(\mbox{$17.5\le i^\prime<22.5$}) within regions of
\mbox{\texttt{mask}=0}. At this brightness the error on the redshift
estimators is typically \mbox{$\sigma_{\rm z}=(1+z)\,0.04$} with a few
per cent outliers.  We {used} stellar-mass estimates in the
\cfhtlens~catalogue. These estimators apply stellar-synthesis
modelling to the photometric data in the five bands, assuming a
\cite{2003PASP..115..763C} initial stellar mass-function, the stellar
population synthesis model by \cite{2003MNRAS.344.1000B}, the dust
extinction law in \cite{2000ApJ...533..682C}, and a flat
$\Lambda\rm CDM$ model with Hubble parameter of $h=0.7$ and
$\Omega_{\rm m}=0.27$ for the luminosity distance, which is similar to
our fiducial cosmology in Table \ref{tab:fidcosmo}. For the stellar
masses, we assume the RMS error of $0.27\,\rm dex$ in
\citet{2014MNRAS.437.2111V}.

As in SH18 we split the galaxy samples into six stellar mass bins
between $5\times10^9$ and
$3.2\times10^{11}\,h_{70}^{-2}\,{\rm M}_\odot$, and into two colour
bins along the line $u^\ast-r^\prime=1.93\,z_{\rm p}+1.85$ for the
full stellar mass range. We further {divided} these eight samples
in photometric redshift to obtain low-$z$ samples with
\mbox{$0.2\le z_{\rm p}<0.44$} ($\bar{z}\approx0.35$) and high-$z$
samples with \mbox{$0.44\le z_{\rm p}<0.6$}
($\bar{z}\approx0.51$). For each sample, the total number of objects
is in the range \mbox{$10^3-10^5$}, which corresponds to a mean number
density of \mbox{$0.003-0.3\,\rm arcmin^{-2}$} on the sky.

\begin{table*}
  \begin{center}
    \caption{\label{tab:sourcesamples} Overview of source samples and
      selection criteria in addition to \mbox{\texttt{star\_flag}<1},
      \mbox{\texttt{FLAG}<3}, \mbox{\texttt{fitsclass=0}}, and
      $i^\prime\in[17.5,24.7)$.}
 \begin{tabular}{lccccl}
   \hline\hline
    Sample & $N_{\rm g}$\tablefootmark{a} & $N_{\rm eff}$\tablefootmark{b} & $\bar{z}\pm\sigma_{\rm z}$\tablefootmark{c} & 2D $\sigma_\epsilon$\tablefootmark{d} & Selection\tablefootmark{e}\\
    \hline\\
   BACK low-$z$ & $2.36\times10^{6}$ & $1.94\times10^{6}$ & $0.97\pm0.28$ & $0.413$ & $z_{\rm p}\in[0.65,1.30]$
   \\
   BACKa low-$z$ & $1.29\times10^{6}$ & $1.08\times10^{6}$ & $0.86\pm0.26$ & $0.408$ & $z_{\rm p}\in[0.65,0.90]$
   \\
   BACKb low-$z$ & $1.06\times10^{6}$ & $8.63\times10^{5}$ & $1.12\pm0.22$ & $0.419$ & $z_{\rm p}\in[0.90,1.30]$
   \\\\
   BACK high-$z$ & $1.80\times10^{6}$ & $1.47\times10^{6}$ & $1.03\pm0.25$ & $0.413$ & $z_{\rm p}\in[0.75,1.30]$
   \\
   BACKa high-$z$ & $8.97\times10^{5}$ & $7.43\times10^{5}$ & $0.93\pm0.23$ & $0.407$ & $z_{\rm p}\in[0.75,0.95]$
   \\
   BACKb high-$z$ & $9.03\times10^{5}$ & $7.32\times10^{5}$ & $1.14\pm0.21$ & $0.420$ & $z_{\rm p}\in[0.95,1.30]$
  \end{tabular}
 \tablefoot{
 \tablefoottext{a}{total number of objects for W1 to W4 combined (effective $92.12$ $\rm deg^2$)};
 \tablefoottext{b}{effective number $(\sum_i w_i)^2/\sum_i w_i^2$ of sources};
 \tablefoottext{c}{mean and dispersion of binned redshift distribution}; 
 \tablefoottext{d}{weighted ellipticity dispersion $\ave{|\epsilon|^2}^{1/2}$ of sources};
 \tablefoottext{e}{photometric redshift $z_{\rm p}$}.
 }
  \end{center}
\end{table*}

\subsection{Shear catalogues}
\label{sect:sources}

As background sources for the lensing analysis, we use resolved
galaxies with reliable photo-$z$ estimates and
\mbox{$i^\prime<24.7$}. Each source $i$ has an estimated
\mbox{$\epsilon_i=\epsilon_{i1}+\i\,\epsilon_{i2}$} gravitational
shear $\gamma_i$ in the source line-of-sight direction and a
statistical weight \mbox{$w_i>0$}, which depends on the variance of
its posterior ellipticity and the ellipticity distribution in the
entire source galaxy population. The additional calibration factor
\mbox{$m_i>-1$}, determined with simulated \cfhtlens~ images,
addresses the multiplicative shear bias. We {accounted} for the
shear calibration in our analysis, equivalent to the procedure
described in \cite{2013MNRAS.429.2858M}, by rescaling the source
ellipticities \mbox{$\epsilon_i\to\epsilon_i\,(1+m_i)^{-1}$} and
weights \mbox{$w_i\to w_i\,(1+m_i)$}. Furthermore, we {subtracted}
from $\epsilon_{i2}$ the additive shear bias $c_2$, provided by the
catalogue.

We then {constructed} six source samples according to Table
\ref{tab:sourcesamples}. For the full analysis, sources are in two
samples, BACK low-$z$ and BACK high-$z$, which have little redshift
overlap with the low-$z$ and high-$z$ foreground samples in Sect.
\ref{sect:lenses}, but aim at a large number density of sources on the
sky. For low-$z$ we {applied} the cut
\mbox{$0.65\le z_{\rm p}<1.3$}, and for high-$z$ the cut
\mbox{$0.75\le z_{\rm p}<1.3$}. Additionally, for robustness tests of
our results later on, we {subdivided} each source bin into the two
complementary photo-$z$ bins BACKa and BACKb. In total, there are
roughly $2\times10^6$ sources in each sample BACK low-$z$ and
high-$z$, which corresponds to a number density of
$6\,\rm arcmin^{-2}$. However, due to the source weights, the
effective source number $N_{\rm eff}:=(\sum_i\,w_i)^2/\sum_i w_i^2$ is
roughly $20\%$ smaller. The mean, $w$-weighted redshift of BACK
low-$z$ (high-$z$) is $\bar{z}=0.97\,(1.03)$, and the weighted
ellipticity dispersion (i.e. the source shape noise) in all samples is
\mbox{$\ave{|\epsilon|^2}^{1/2}\approx0.41$}, or $0.29$ for each
component individually.

\subsection{Redshift distributions}

The average \texttt{BPZ} redshift posterior of the sample galaxies is
the estimated redshift distribution of a sample (which is the sum of
all posteriors followed by a normalisation). According to
\cite{2013MNRAS.431.1547B}, these estimates provide a sufficiently
accurate model for the actual redshift distribution $p(z)$ of
\cfhtlens~galaxies at \mbox{$z<1.3$}.

However, the \texttt{BPZ} posterior is binned with bin width
\mbox{$\Delta z=0.05$} and therefore artificially broadened. While
this has a small effect for cosmic shear or galaxy-galaxy lensing
studies, which are mostly sensitive to the mean of the distribution,
it is potentially problematic for predictions of the angular
clustering of galaxies and for the galaxy-bias normalisation. As
correction of the broadening, we {deconvolved} the binned, average
posterior for the lens samples as described in Appendix
\ref{sect:pofz_method}. For the source samples, where the broadening
is not relevant, we directly {used} the \texttt{BPZ} posteriors.

\begin{figure}
  \begin{center}
    \includegraphics[width=92mm]{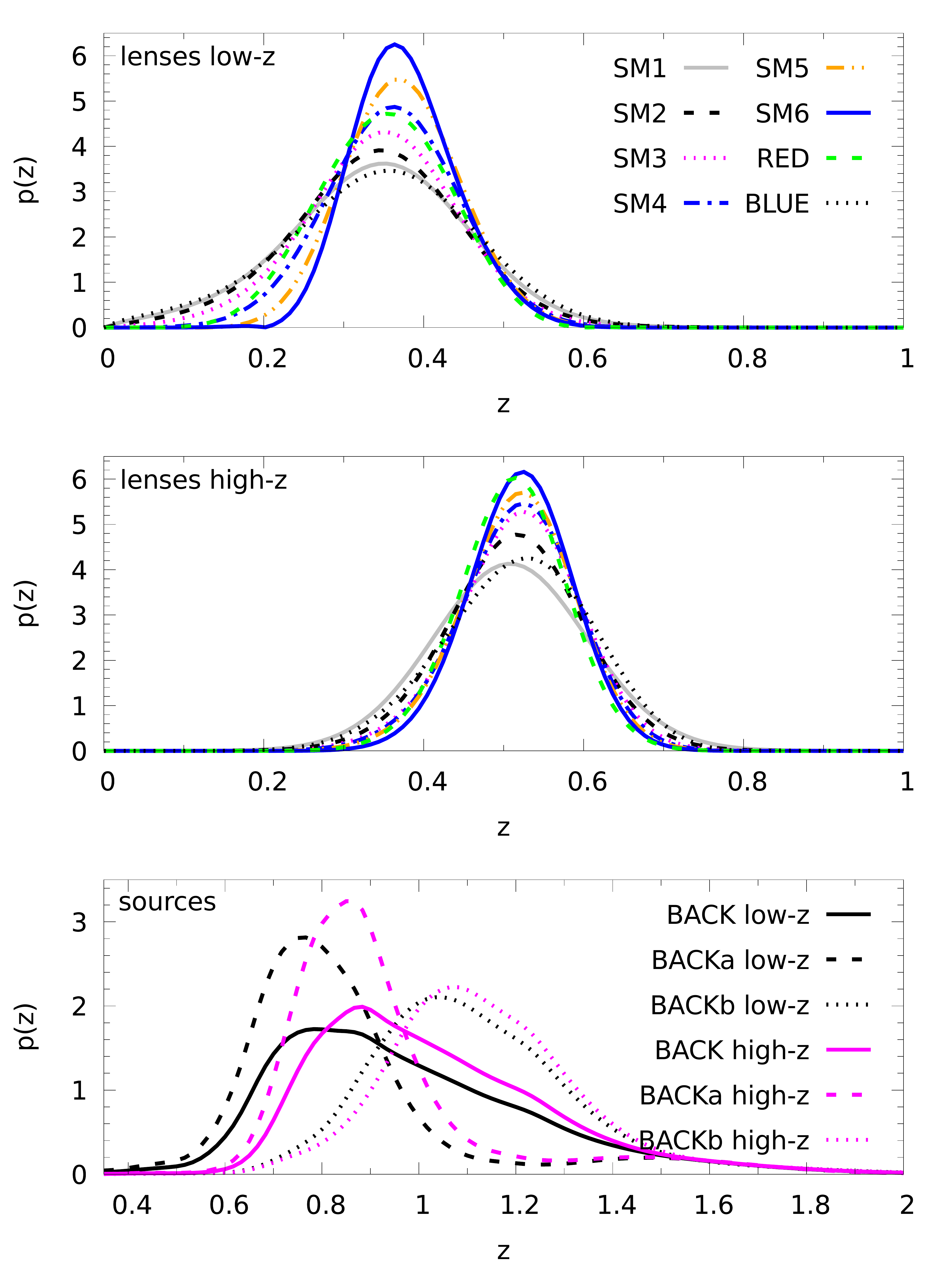}
  \end{center}
  \caption{\label{fig:pofz} Redshift distributions $p(z)$ of all
    galaxy samples. The two top panels display the lens samples,
    low-$z$ and high-$z$, the bottom panel displays the two source
    samples (solid lines) and their two subsamples (dotted or dashed
    lines).}
\end{figure}

The two top panels in Fig. \ref{fig:pofz} show the deconvolved $p(z)$
for all our lens samples. Despite the same photo-$z$ cuts, the differences
in the redshift distributions between the lens samples is evident:
bluer samples or samples with less stellar mass tend to be more spread
out in redshift. The bottom panel plots the average \texttt{BPZ}
posteriors (weighted) of all source samples.

\subsection{Average number densities}
\label{sect:ng}

\begin{figure}
  \begin{center}
    \includegraphics[width=95mm]{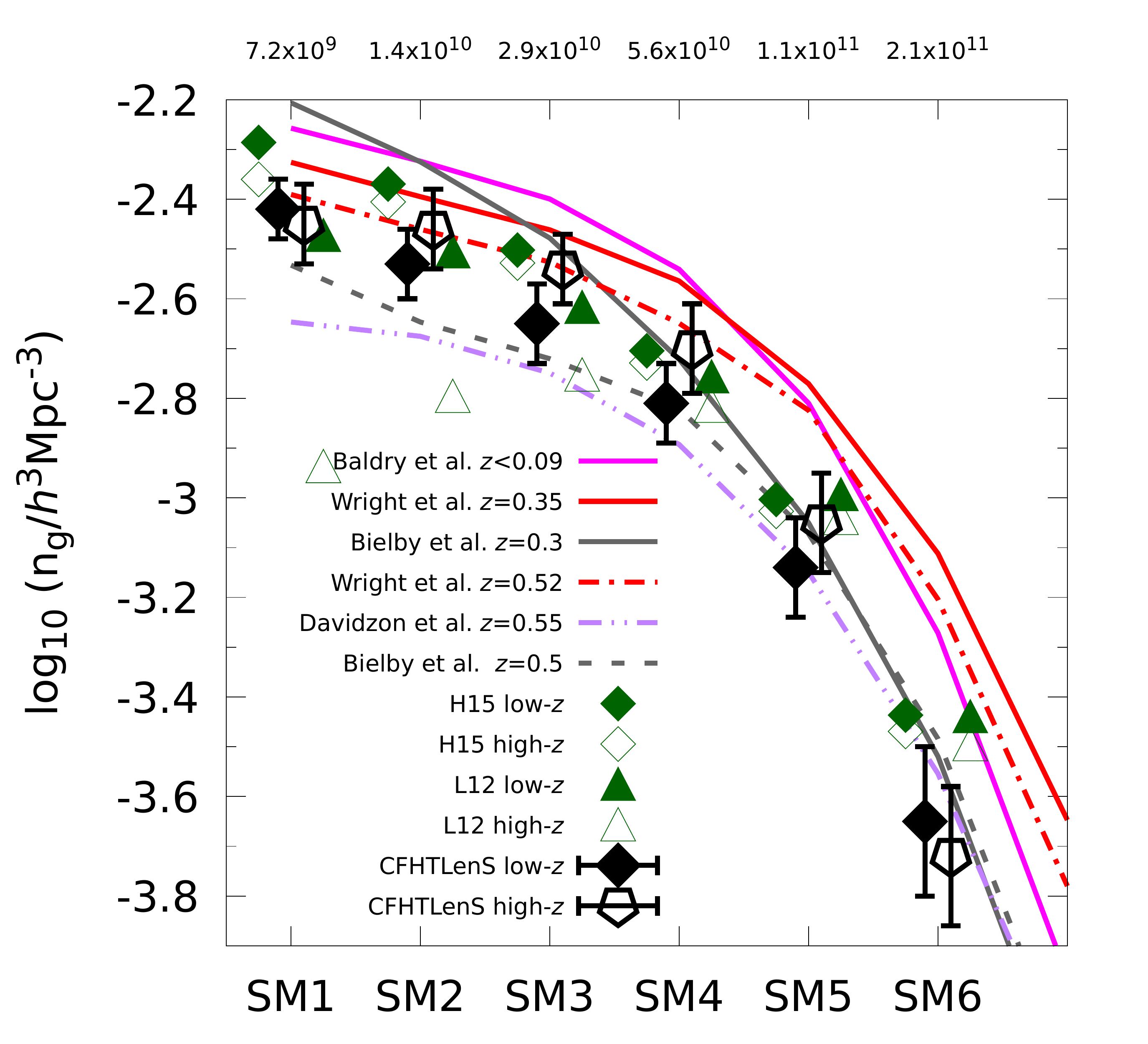}
  \end{center}
  \caption{\label{fig:ng} Comoving number density of galaxies for
    \cfhtlens~(black data points) and the SAMs (green data points) in
    our stellar-mass bins for low-$z$ (filled symbols) and high-$z$
    (open symbols). The sample labels SM1 to SM6 are indicated at the
    bottom; the small numbers at the top are the mean stellar masses
    inside the bins in units of $h_{70}^{-2}\msol$. The coloured lines
    are the galaxy SMFs by
    \citet[][GAMA/G10-COSMOS/3D-HST]{2018MNRAS.480.3491W},
    \citet[][WIRDS/CFHTLS]{2012A&A...545A..23B},
    \citet[][VIPERS/CFHTLS]{2013A&A...558A..23D}, and
    \citet[][GAMA]{2012MNRAS.421..621B}. The SMFs are, before
    integrating over our stellar-mass bins, convolved with a
    stellar-mass error of $0.27\,\rm dex$ to account for the Eddington
    bias in our data.}
\end{figure}

We obtain the mean (comoving) number density of lens galaxies with our
modified $V_{\rm max}$-estimator, described in Appendix
\ref{sect:vmaxdetails}. Originally, $V_{\rm max}$ assumed an exactly
determined maximum (comoving) volume $V_i$ inside which a galaxy $i$
is observable under the adopted selection criteria; the estimator of
the number density was then $\bar{n}_{\rm g}=\sum_{i=1}^n\,V_i^{-1}$
for all $n$ galaxies in the sample
\citep{2000ApJS..129....1T,1968ApJ...151..393S}. In our case with
photometric redshifts, however, the volume $V_i$ is uncertain. Our
modified $V_{\rm max}$ estimator therefore accounts for this
uncertainty, and we find values for $\bar{n}_{\rm g}$ that are then a
few per cent larger than the standard $V_{\rm max}$ based on the
best-fit photo-$z$. Our estimates safely ignore a density correction
for inhomogeneous galaxy samples because our broad redshift bins at
$z>0.1$ in combination with the \mbox{$\sim100\,\rm deg^2$} survey
area span a sufficiently large and homogeneous volume
\citep[e.g.][]{2012MNRAS.421..621B}.

Using the $V_{\rm max}$ densities for \cfhtlens~in Fig. \ref{fig:ng}
(black data points with error bars) as priors improves the convergence
of our Monte Carlo analysis because they break parameter degeneracies
in the galaxy-bias templates. The priors are log-normal distributions
with RMS errors for $\log_{10}{\bar{n}_{\rm g}}$ based on jackknife
re-sampling \citep{knight1999mathematical}; we removed one
\mbox{MegaCam} field at a time for each jackknife sample. The number
densities typically range between
$2\times10^{-4}-4\times10^{-3}\,h^3\,\rm Mpc^{-3}$ and have a clear
trend towards fewer galaxies for higher stellar masses.

Figure \ref{fig:ng} also compares our estimates for $\bar{n}_{\rm g}$
to measured stellar-mass functions (SMFs) from selected studies, which
are shown as lines. For this comparison, we {convolved} the
best-fitting SMFs in \citet{2018MNRAS.480.3491W},
\citet{2012A&A...545A..23B}, \citet{2013A&A...558A..23D}, and
\citet{2012MNRAS.421..621B} with our RMS error of $0.27\,\rm dex$ for
stellar-mass estimates (log-normal error model), and we then
{integrated} the convolved functions over the stellar-mass ranges
in Table \ref{tab:lenssamples}. The convolution accounts for the
Eddington bias, which softens the drop in galaxy counts beyond the
characteristic mass in the Schechter function around
$5\times10^{10}\,h_{70}^{-2}\msol$ and which suppresses the counts
below this cutoff. The top axis of the figure indicates, averaged for
low-$z$ and high-$z$, the mean stellar masses of our samples. All SMFs
in this comparison make assumptions for the stellar population
synthesis and the stellar initial mass-function that are consistent
with the \cfhtlens~stellar masses, and they employ, with the exception
of \cite{2018MNRAS.480.3491W} and partly \cite{2013A&A...558A..23D},
the same dust extinction law.

That there are still differences between the surveys is probably a
sign of inconsistent stellar-mass estimates or galaxy selections.
Closest to our selection is the \mbox{VIPERS} study by
\citet{2013A&A...558A..23D}, which identifies galaxies in the patches
W1 and W4 of \mbox{CFHTLS} (at that time $\sim15\,\rm deg^2$) with our
flux limit \mbox{$i_{\rm AB}<22.5$}. This sample was, superior to our
photometric redshift estimates, followed up with the VIMOS
multi-object spectrograph at the ESO VLT to obtain precise estimates
of stellar mass and redshift with \mbox{HYPERZMASS}
\citep{2010A&A...524A..76B,2000A&A...363..476B}. In addition to the
$i_{\rm AB}$-band flux limit, they applied a colour cut to exclude
low-redshift objects ($z<0.5$). The resulting SMF in their lowest
redshift bin \mbox{$z\in[0.5,0.6]$} agrees well with our high-$z$
galaxy counts SM5 and SM6, but is \mbox{$\sim40\%$} lower for the less
massive galaxies SM1 to SM3. In another study,
\citet{2012A&A...545A..23B} combine deep \mbox{CFHTLS}
($\sim2\,\rm deg^2$) photometry with near-infrared observations by the
WIRcam Deep Survey (WIRDS) to attain the galaxy SMF at various
redshifts, in particular also around our low-$z$ and high-$z$. Again,
there is consistency for SM5 and SM6, but our galaxy counts are lower
(higher) for low-$z$ (high-$z$) in the other stellar-mass bins. This
discrepancy for less massive galaxies may be related to their
near-infrared colour cuts, which were used to separate stars from
galaxies, and the fainter flux-limit of \mbox{$i=25.5$}. Finally, the
studies by \citet{2018MNRAS.480.3491W} and \citet{2012MNRAS.421..621B}
are based on the Galaxy And Mass Assembly (GAMA) survey and, in the
former case, additionally on the deeper observations in the Cosmos
Evolution Survey (COSMOS) and 3D-HST. They consistently find higher
galaxy counts for low-$z$, whereas now high-$z$ SM1 to SM4 are in good
agreement with our results.

In conclusion, our measurements for $\log_{10}\bar{n}_g$ are broadly
supported within $0.2-0.3\,\rm dex$ by other studies. Mismatches are
presumably a mixture of inconsistent selections and stellar mass
errors. A clear explanation probably requires a completely consistent
selection of galaxies, beyond the scope of this study.

\section{Semi-analytic models of galaxies}

Semi-analytic galaxy models (SAMs) populate matter halos in purely
dark matter simulations with galaxies, reducing galaxies and halos to
entities with a few global properties that follow a simplified set of
physical and empirical laws. By applying prescriptions for a stellar
population synthesis and dust extinction the SAM results are, with
increased uncertainty, converted into galaxy spectral
energy-distributions, and thus mock galaxy catalogues with multi-band
photometry. A detailed description of the semi-analytic approach can
be found in \cite{10.1093/mnras/stw1888}, among others. We use the
Munich model H15 and the Durham model L12.

\subsection{Simulation and galaxy sample selection}
\label{sect:MR}

Our galaxy mocks are based on the Millennium Simulation
\citep{2005Natur.435..629S}. This $N$-body simulation of the dark
matter density field has a comoving spatial resolution of
\mbox{$\sim5\,h^{-1}\,\rm kpc$} over a cubic volume with
\mbox{$500\,h^{-1}\,\rm Mpc$} side length, sampled by
\mbox{$\sim10^{10}$} mass particles. The fiducial cosmology of the
simulation ($\Omega_{\rm m}=1-\Omega_\Lambda=0.25$,
$\Omega_{\rm b}=0.045$, $\sigma_8=0.9$, $h=0.73$, $n_{\rm s}=1.0$) is
different from our fiducial \emph{Planck} cosmology in Table
\ref{tab:fidcosmo}.  As shown in \cite{2017MNRAS.469.2626H}, however,
this has only a mild impact on the galaxy clustering because the
combination of a higher $\sigma_8$ and a lower $\Omega_{\rm m}$ gives
a similar structure growth. We therefore do not expect large
differences between the galaxy models and the \cfhtlens~galaxy bias
from this side.

To compute predictions of the galaxy biasing functions, we combine the
galaxy and dark matter power spectra, and their cross-power spectra at
the Millennium Simulation snapshots $z=0.3623,\,0.5086$ for low-$z$
and high-$z$, exactly as described in Section 2.3 of SH18. To prepare
the various lens samples in these snapshots, we {selected} SAM
galaxies according to Table \ref{tab:lenssamples}, after adding a
random RMS error of $0.27\,\rm dex$ to the SAM stellar masses
$\log_{10}{M_\ast}$ and after converting SDSS magnitudes of the SAMs
to \cfhtlens~magnitudes according to Section 3.2 in
\cite{2016arXiv160808629S}. By SAM stellar mass we specifically mean
that we do not apply \texttt{BPZ} to the SAM photometry to emulate
\cfhtlens, rather we utilise the randomised SAM stellar masses. For
the colour cut, we identify $z_{\rm p}$ with the snapshot $z$.

\subsection{Model differences}

The H15 and L12 models follow somewhat differing SAM approaches and
hence produce distinct predictions for the galaxy bias. A major
difference is their calibration: H15 fits stellar-mass functions and
the red-galaxy fractions as a function of $M_\ast$ to observations for
$z\le1$; L12 calibrates the model with local luminosity functions
(near-infrared and blue bands) and the observed correlation between
galactic bulge mass and central black hole mass. In addition, the
identification of subhalos in the Millennium Simulation differs and so
does the treatment of subhalos and satellite galaxies. L12 satellites,
for instance, can merge with a central galaxy while they still have
resolved subhalos, whereas H15 satellites have to disrupt their halos
first. This directly affects the galaxy clustering on small scales and
the radial distribution of satellites inside galaxy clusters
\citep{2013MNRAS.432.2717C}. Another difference, affecting the
photometry of the model galaxies, are the initial mass functions, the
implemented models for the stellar population synthesis, and the dust
extinction law.

More can be seen in the stellar mass function (SMF) of the models. The
green data points in Fig. \ref{fig:ng} show the number density of the
SAM galaxy samples for the various stellar masses (filled data points:
low-$z$; open data points: high-$z$). We obtained these data by
dividing the total number of galaxies at the snapshot redshifts by the
comoving simulation volume of $(500\,h^{-1}\,\rm Mpc)^3$. Within
$0.1\,\rm dex$ these data points closely follow those of
\cfhtlens~with two exceptions. Firstly, the L12 high-$z$ samples (open
triangles) have up to $0.5\,\rm dex$ fewer galaxies for SM1 and SM2,
due to a strong evolution of the stellar-mass function at low stellar
masses. This strong evolution is not present in H15. Secondly, the
SAMs predict approximately $0.2\,\rm dex$ more SM6 galaxies for both
redshifts.

For the colour-selected samples we find values broadly compatible with
\cfhtlens,~ but also differences between H15 and L12: for RED low-$z$
in H15 (L12), we have
$\log_{10}{(\bar{n}_{\rm g}/h^3\,\rm Mpc^{-3})}=-2.20\,(-2.11)$ and
for RED high-$z$ $-2.37(-2.37)$. The corresponding figures for BLUE
low-$z$ and high-$z$ are $-2.02(-2.37)$ and $-1.99(-2.51)$,
respectively. L12 thus leans towards a smaller number of blue (SM1 and
SM2) galaxies, especially at high-$z$.

\section{Aperture statistics and reconstruction of biasing functions}
\label{sect:method}

This section briefly summarises the SH18 formalism for galaxy-bias
reconstruction and reports the \cfhtlens~aperture statistics to which
we apply this technique. The basic idea is to measure the fluctuations
of galaxy numbers and lensing convergence inside circular apertures
for a range of angular scales. {The ratios of their (co-)variances}
directly probe the projected scale-dependent galaxy bias on the
sky. By then forward-fitting templates to the observed ratio
statistics we reconstruct the average biasing functions $b(k)$ and
$r(k)$ for a given fiducial cosmology and galaxy redshift
distributions.

\subsection{Fiducial cosmology}

\begin{table}
  \caption{\label{tab:fidcosmo} Fiducial $\Lambda\rm CDM$ cosmological
    model for the galaxy-bias analysis. The parameters are those from
    \cite{2016A&A...594A..13P} for the TT, TE, and EE CMB power
    spectrum combined with constraints from baryon acoustic
    oscillations of galaxies. We changed $\sigma_8$ to a lower value,
    albeit irrelevant for the bias analysis, to be consistent with the
    cosmic shear constraints in \cfhtlens.}
  \begin{tabular}{ccccccc}
    $\Omega_{\rm m}$\tablefootmark{a} & 
    $\Omega_\Lambda$\tablefootmark{b} & 
    $\Omega_{\rm b}$\tablefootmark{c} & 
    $\sigma_8$\tablefootmark{d}      & 
    $h$\tablefootmark{e}             & 
    $n_{\rm s}$\tablefootmark{f}      & 
    $A_{\rm ia}$\tablefootmark{g}
    \\\hline\hline\\
    $0.306$ & $1-\Omega_{\rm m}$ & $0.0479$ & $0.714$ & $0.683$ & $0.965$ & $-0.48$
  \end{tabular}
  \tablefoot{
    \tablefoottext{a}{Density parameter of dust-like matter}
    \tablefoottext{b}{Density parameter of the cosmological constant (chosen for a spatially flat universe)}
    \tablefoottext{c}{Density parameter of baryonic matter}
    \tablefoottext{d}{Normalisation of the linear matter power spectrum}
    \tablefoottext{e}{Hubble parameter $H_0$ in units of $100\,\rm km\,s^{-1}\,Mpc^{-1}$}
    \tablefoottext{f}{Spectral index of the primordial matter power spectrum}
    \tablefoottext{g}{Correlation amplitude of the intrinsic alignment of source galaxies, taken from the \cfhtlens~analysis by \citet{2013MNRAS.432.2433H}}}
\end{table}

The baseline cosmology in the analysis are the parameters listed in
Table \ref{tab:fidcosmo}, taken from the recent CMB analysis by the
\cite{2016A&A...594A..13P}. In this baseline, we changed the
normalisation $\sigma_8$ of the matter power spectrum to the smaller
value of \mbox{$\sigma_8=0.714$} in order to be consistent with
cosmic-shear constraints
\mbox{$\sigma_8(\Omega_{\rm m}/0.27)^{0.5}\approx0.76$} in \cfhtlens~
\citep{2017MNRAS.465.2033J,2014MNRAS.441.2725F,2013MNRAS.432.2433H}.
For the following analysis of the ratio aperture-statistics, however,
this smaller value has no practical relevance. This is reflected by
the insensitivity of the ratio statistics with respect to any
rescaling of the matter power spectrum amplitude.

Nevertheless, the ratio statistics depends weakly on the shape of the
matter power spectrum, if it is not a power law in $k$
\citep{1998A&A...334....1V}.  In our analysis, the power spectrum
$P_{\rm m}(k,\chi)$ at comoving distance $\chi$ is the updated
\texttt{Halofit} \citep{2012ApJ...761..152T,Smith03}, as it is
implemented in the latest version of
\texttt{Nicea}\footnote{\url{https://github.com/CosmoStat/nicaea}}
\citep{2009A&A...497..677K}.  In addition, we modify the power
spectrum
\mbox{$P_{\rm m}(k,\chi)\mapsto F(k,\chi)\,P_{\rm m}(k,\chi)$} in the
non-linear regime by the transfer function $F(k,\chi)$ in
\cite{2018arXiv180108559C} to correct for AGN feedback, star
formation, and adiabatic contraction of matter halos due to gas
cooling \citep{2015JCAP...12..049S}.  Since $F(k,\chi)$ is not well
known, we randomly vary $F(k,z)-1$ later on to include systematic
errors by baryons.

\subsection{Aperture statistics}

This section is a summary of the relation between power spectra and
aperture statistics.  For a circular aperture of size
$\theta_{\rm ap}$ at position $\vec{\theta}$, the density fluctuation
of galaxy numbers inside the aperture is
\begin{equation}
  \label{eq:apcount}
  {\cal N}(\theta_{\rm ap};\vec{\theta})=
  \int\d^2\theta^\prime\;U(|\vec{\theta}^\prime|;\theta_{\rm ap})\,
  \kappa_{\rm g}(\vec{\theta}^\prime+\vec{\theta})\;.
\end{equation}
This aperture number count convolves the galaxy number-density
contrast
$\kappa_{\rm g}(\vec{\theta})=N(\vec{\theta})/\bar{N}_{\rm g}-1$ of
galaxy density $N(\vec{\theta})$ and mean density $\bar{N}_{\rm g}$ on
the sky with the compensated polynomial filter
\begin{equation}
  \label{eq:apfilter}
  U(\theta;\theta_{\rm ap})=
  \frac{1}{\theta_{\rm ap}^2}\,u(\theta\,\theta_{\rm ap}^{-1})~;~
  u(x)=\frac{9}{\pi}\,(1-x^2)\,\left(\frac{1}{3}-x^2\right)\,{\rm H}(1-x)\;,
\end{equation}
where ${\rm H}(x)$ is the Heaviside step function.  The discreteness
of galaxies at positions $\vec{\theta}_i$ (i.e.
$N(\vec{\theta})=\sum_i\delta_{\rm D}(\vec{\theta}-\vec{\theta}_i)$
for Dirac delta functions $\delta_{\rm D}(\vec{\theta})$) gives rise
to shot-noise in ${\cal N}(\theta_{\rm ap};\vec{\theta}),$ which is
corrected for in the statistical moments of $\cal N$ below. The
aperture number count ${\cal N}$ compares to the aperture mass (i.e.
the fluctuation of the lensing convergence $\kappa(\vec{\theta})$
inside the same aperture):
\begin{equation}
  \label{eq:apmass}
  M_{\rm ap}(\theta_{\rm ap};\vec{\theta})=
  \int\d^2\theta^\prime\;U(|\vec{\theta}^\prime|;\theta_{\rm ap})\,
  \kappa(\vec{\theta}^\prime+\vec{\theta})\;.
\end{equation}

Our statistical analysis utilises second-order moments of the joint
distribution of ${\cal N}$ and $M_{\rm ap}$ over all positions
$\vec{\theta}$. These second-order moments are measures of angular
power spectra $P(\ell)$,
\begin{eqnarray}
  \label{eq:n2b}
  \ave{{\cal N}^2}_{\rm th}(\theta_{\rm ap};b)&=&
  2\pi\int\limits_0^\infty\d\ell\,\ell\,P_{\rm n}(\ell;b)\,\left[I(\ell\theta_{\rm
      ap})\right]^2\;,\\
    \label{eq:nmapbr}
    \ave{{\cal N}M_{\rm ap}}_{\rm th}(\theta_{\rm ap};b,r)&=&
  2\pi\int\limits_0^\infty\d\ell\,\ell\,P_{{\rm n}\kappa}(\ell;b,r)\,\left[I(\ell\theta_{\rm
    ap})\right]^2\;,\\
  \label{eq:mapsq}
    \ave{M_{\rm ap}^2}_{\rm th}(\theta_{\rm ap})&=&
  2\pi\int\limits_0^\infty\d\ell\,\ell\,P_\kappa(\ell)\,\left[I(\ell\theta_{\rm
    ap})\right]^2\;,
\end{eqnarray}
for the angular band-pass filter
\begin{equation}
  I(x)=\frac{12}{\pi}\frac{{\rm J}_4(x)}{x^2}\;.
\end{equation}
We estimate these statistics from observations (see
Sect. \ref{sect:estimators}), and denote them just by
$\ave{{\cal N}^2(\theta_{\rm ap})}$,
$\ave{{\cal N}M_{\rm ap}(\theta_{\rm ap})}$, and
$\ave{M_{\rm ap}^2(\theta_{\rm ap})}$ without the subscript `th' (for
theory) and arguments `$b$' or `$b,r$'.

To model the aperture statistics, we use  Eqs.
\Ref{eq:n2b}-\Ref{eq:mapsq} and the following three power-spectra
models: $P_{\rm n}(\ell;b)$, $P_{{\rm n}\kappa}(\ell;b,r)$, and
$P_\kappa(\ell)$. In the models the sources have the probability
distribution $p_{\rm s}(\chi)\,\d\chi$ in comoving distance $\chi$,
and lens galaxies have the distribution $p_{\rm d}(\chi)\,\d\chi$. The
specific biasing functions of the lens galaxies are given by $b(k)$
and $r(k)$ as indicated by the arguments `$b$' or `$b,r$'. There is
the angular galaxy power-spectrum
\begin{equation}
  \label{eq:pn}
  P_{\rm n}(\ell;b)=
  \int\limits_0^{\chi_{\rm h}}\frac{\d\chi\;p_{\rm d}^2(\chi)}{f_K^2(\chi)}\,
  b^2(k_\ell^\chi)\,P_{\rm m}\left(k_\ell^\chi;\chi\right)\;,
\end{equation}
the galaxy-convergence cross-power
\begin{multline}
  \label{eq:pnkappa}
  P_{{\rm n}\kappa}(\ell;b,r)=\\
  \frac{3H_0^2\,\Omega_{\rm m}}{2c^2}
  \int\limits_0^{\chi_{\rm h}}
  \frac{\d\chi\;p_{\rm d}(\chi)\,g_{\rm s}(\chi)}
  {a(\chi)\,f_K(\chi)}\,
  b(k_\ell^\chi)\,r(k_\ell^\chi)\,P_{\rm m}\left(k_\ell^\chi;\chi\right)\;,
\end{multline}
and the convergence power-spectrum
\begin{equation}
  \label{eq:pkappa}
  P_\kappa(\ell)=
  \frac{9H_0^4\,\Omega_{\rm m}^2}{4c^4}
  \int\limits_0^{\chi_{\rm h}}
  \frac{\d\chi\,\;g_{\rm s}^2(\chi)}{a^2(\chi)}\,
  P_{\rm m}\left(k_\ell^\chi;\chi\right)\;,
\end{equation}
where $f_K(\chi)$ is the comoving angular diameter distance,
\begin{equation}
  g_{\rm s}(\chi):=
  \int\limits_\chi^{\chi_{\rm h}}\d\chi^\prime\;p_{\rm s}(\chi^\prime)
  \,\frac{f_K(\chi^\prime-\chi)}{f_K(\chi^\prime)}
\end{equation}
is a lensing kernel, \mbox{$k_\ell^\chi:=(\ell+0.5)/f_K(\chi)$} is a
wave number at $\chi$, $a(\chi)$ is the scale factor, $\chi_{\rm h}$
is the maximum distance to a source, and ${\rm J}_n(x)$ is the
$n$th-order Bessel function of the first kind.

\subsection{Intrinsic alignment and magnification bias}

We add corrections to the theoretical power spectra, Eqs. \Ref{eq:n2b}
to \Ref{eq:mapsq}, to account for the intrinsic source alignments
(IAs) and the magnification of lenses. To this end, we employ the
approximate `non-linear linear' model for IA
\citep{2004PhRvD..70f3526H,2007NJPh....9..444B,
  2011A&A...527A..26J}. Specifically, instead of $P_\kappa(\ell)$ in
the integral Eq. \Ref{eq:mapsq}, we integrate over
\begin{equation}
  \label{eq:iii}
  P_\kappa^\prime(\ell)=
  P_\kappa(\ell)+P_\kappa^{\rm II}(\ell)+P_\kappa^{\rm GI}(\ell)\;,
\end{equation}
where the additional intrinsic-intrinsic (II) and shear-intrinsic (GI)
terms are\footnote{In \cite{2017arXiv171102677S} there is a missing
  minus sign for $F_{\rm ia}(\chi)$ in {their} Eq. (31) and a typo
  in the pre-factor of
  $P^{\rm mb}_{{\rm n}\kappa}(\ell)\equiv P^{(2)}_{{\rm
      n}\kappa}(\ell)$ in {their} Eq. (34), which have both been
  corrected here. The pre-factor of the first integral in {their}
  Eq. (36) should be $9H_0^4c^{-4}\Omega^2_{\rm m}$.}
\begin{eqnarray}
  P_\kappa^{\rm II}(\ell)&=&
  \int_0^{\chi_{\rm h}}\frac{\d\chi\;p^2_{\rm s}(\chi)}{f^2_K(\chi)}
  \,F_{\rm ia}^2(\chi)\,P_{\rm m}(k^\chi_\ell;\chi)\;;
  \\
  P_\kappa^{\rm GI}(\ell)&=&
  \frac{3H_0^2\,\Omega_{\rm m}}{c^2}\,
  \int_0^{\chi_{\rm h}}\frac{\d\chi\;p_{\rm s}(\chi)\,g_{\rm s}(\chi)}
  {a(\chi)\,f_K(\chi)}
  \,F_{\rm ia}(\chi)\,P_{\rm m}(k^\chi_\ell;\chi)
,\end{eqnarray}
and 
\begin{equation}
  F_{\rm ia}(\chi)\approx
  -2.4\times10^{-2}
  \left(\frac{A_{\rm ia}}{3.0}\right)\,
  \left(\frac{\Omega_{\rm m}}{0.3}\right)\,
  \left(\frac{D_+(\chi)}{0.5}\right)^{-1}
\end{equation}
uses the linear growth factor of structure $D_+(\chi)$, normalised by
$D_+(\chi=0)$ \citep{peebles80}. The parameter $A_{\rm ia}$ defines
the amplitude of the intrinsic alignments in the model. In our
Bayesian analysis we use as prior $A_{\rm ia}=-0.48^{+0.75}_{-0.87}$
the constraints from \cite{2013MNRAS.432.2433H} ($68\%$ credible
interval, {hereafter CI}).

Likewise, we correct for the magnification of the lens number density
by foreground structure in the power spectrum
$P_{{\rm n}\kappa}(\ell;b,r)$, Eq. \Ref{eq:nmapbr}, through
\begin{equation}
P_{{\rm n}\kappa}^\prime(\ell;b,r)=
P_{{\rm n}\kappa}(\ell;b,r)+P^{\rm GI}_{{\rm n}\kappa}(\ell;b,r)+P_{{\rm
    n}\kappa}^{\rm mb}(\ell)
\end{equation}
with
\begin{equation}
  P^{\rm GI}_{{\rm n}\kappa}(\ell;b,r)=
  \int_0^{\chi_{\rm h}}\frac{\d\chi\;p_{\rm s}(\chi)\,p_{\rm d}(\chi)}{f^2_K(\chi)}\,
  b(k_\ell^\chi)\,r(k_\ell^\chi)\,F_{\rm ia}(\chi)\,P_{\rm m}(k_\ell^\chi;\chi)
\end{equation}
and
\begin{equation}
P_{{\rm
    n}\kappa}^{\rm mb}(\ell)=
-\frac{9H_0^4\,\Omega_{\rm m}^2}{2c^4}
\int_0^{\chi_{\rm h}}\d\chi\;
\frac{g_{\rm s}(\chi)\,g_{\rm d}(\chi)}{a^2(\chi)}\,
P_{\rm m}(k^\chi_\ell;\chi)\;,
\end{equation}
where
\begin{equation}
  g_{\rm d}(\chi):=
  \int_\chi^{\chi_{\rm h}}\d\chi^\prime\;p_{\rm d}(\chi^\prime)\,
  \frac{f_K(\chi^\prime-\chi)}{f_K(\chi^\prime)}
\end{equation}
is the lensing kernel of the foreground structure
\citep{2019arXiv191006400U,2009A&A...499...31H,2009PhDThesisHartlap}.
The expression $P_{{\rm n}\kappa}^{\rm mb}(\ell)$ assumes a
volume-complete sample of lens galaxies which is approximately correct
for our stellar-mass limited samples. The flux limit of
$i^\prime<22.5$ mostly affects the samples SM1 and SM2 for low-$z$ and
high-$z$ (Section 3.2 in \citealt{2013MNRAS.430.2476S}). The estimated
magnification bias for $\ave{{\cal N}M_{\rm ap}}$ can reach levels of
several per cent for lenses with low clustering amplitude and high
redshifts (\mbox{$z\sim0.5$}) and should therefore be modelled. The
impact of magnification on $\ave{{\cal N}^2}$, on other hand, is
negligible but nevertheless included here (using Equation 36 in SH18).

\subsection{Estimators and data covariance}
\label{sect:estimators}

The second-order moments of the aperture statistics are linear
transformations of the angular correlations functions for the lens
clustering $\omega(\theta)$, mean tangential shear around lenses
$\bar{\gamma}_{\rm t}(\theta)$, and the cosmic shear
$\xi_\pm(\theta)$, which all can be obtained relatively
straightforwardly from observational data
\citep[e.g.][]{2007A&A...461..861S}. To perform this transformation we
measure with
\texttt{Athena}\footnote{\url{http://www.cosmostat.org/software/athena}}
the correlation functions for angular separations between
$9\,\rm arcsec$ and $3\,\rm deg$ in $8000$ bins in the four patches W1
to W4 and combine the measurements \citep{2014ascl.soft02026K}. The
result is statistics for $n=8$ aperture radii $\theta_{\rm ap}$
between $2$ and $90\,\rm arcmin$, equally spaced on a logarithmic
scale.

The largest aperture size is our choice to retain a reasonably large
number of jackknife samples; the resampling scheme removes subfields
that stretch the full angular range of the correlation functions. The
aim of the smallest aperture size, on the other hand, is to keep the
bias in the estimated statistics below $10\%$ at a few arcmin. The
bias arises due to the blending of galaxy images which compromises
\emph{lens}fit below $9\, \rm arcsec$, our smallest galaxy separation,
and hence produces a gap in the correlation functions
\citep{2013MNRAS.429.2858M,2012MNRAS.427..146H}. Previous studies set
the correlation signal to zero inside the gap and produced a
`transformation bias' in the aperture statistics. We follow a revised,
even though not perfect, approach with less transformation bias.

Specifically, we extrapolate the correlations below $9\,\rm arcsec$ in
two different ways, depending on which correlation function is
used. For $\bar{\gamma}_{\rm t}(\theta)$ and $\xi_\pm(\theta)$, which
are respectively converted into $\ave{{\cal N}M_{\rm ap}}$ and
$\ave{M_{\rm ap}^2}$, we extend the lowest angular bin to
$1\,\rm arcsec$, which means that we assume a constant signal equal to
that at $9\,\rm arcsec$. For $\omega(\theta)$, which is converted into
$\ave{{\cal N}^2}$, we fit a power law
\mbox{$\omega(\theta)\propto\theta^{-\delta}$} to the data points
between \mbox{$9\,\rm arcsec\le\theta\le1\,\rm arcmin$}, and we then
extrapolate the best fit to
\mbox{$0.1\,\rm arcsec\le\theta\le9\,\rm arcsec$} in the
transformation. A power-law correlation function for galaxy clustering
is a sensible approximation on small scales
\citep[e.g.][]{2005ApJ...630....1Z}. To quantify the remaining
transformation bias in the estimated aperture statistics after
application of these extrapolations, we use mock galaxy (SM1-SM6, RED,
and BLUE low-$z$ and high-$z$) and shear catalogues that were
constructed from the H15 data and previously used for the work in
SH18. With these mocks we compare the estimated aperture statistics
(with extrapolation) to the true aperture statistics (with no
truncation at $9\,\rm arcsec$). In summary, the transformation bias
for the aperture moments $\ave{{\cal N}M_{\rm ap}}$ and
$\ave{M_{\rm ap}^2}$ is typically below $1\%$ at
\mbox{$\theta_{\rm ap}\approx2\,\rm arcmin$}. For $\ave{{\cal N}^2}$
it is about $10\%$ at $\theta_{\rm ap}\approx2\,\rm arcmin$ (biased
low), but quickly falls below a few per cent beyond several arcmin.

For estimates of the errors in the aperture statistics, we perform
jackknife resampling of the data
\citep{2016arXiv161100752S,2009MNRAS.396...19N,knight1999mathematical}. The
resulting jackknife covariance matrix,
\begin{equation}
  \mat{C}=
  \frac{n_{\rm jk}-1}{n_{\rm jk}}
  \sum_{i=1}^{n_{\rm jk}}
  \left(\vec{d}_i-\overline{\vec{d}}\right)
  \left(\vec{d}_i-\overline{\vec{d}}\right)^{\rm
    t}~;~
  \overline{\vec{d}}=
  \frac{1}{n_{\rm jk}}\sum_{i=1}^{n_{\rm jk}}\vec{d}_i\;,
\end{equation}
for $i=1\ldots n_{\rm jk}$ jackknife samples,
\begin{equation}
  \label{eq:datavec}
  \vec{d}_i=\left(\vec{d}^{(1)}_i,\vec{d}^{(2)}_i,\vec{d}^{(3)}_i\right)\;,
\end{equation}
comprises the error covariances between different angular scales and
types of the aperture statistics, namely
\mbox{$d^{(1)}_{ij}=\ave{{\cal N}^2}_i(\theta_{{\rm ap},j})$},
\mbox{$d^{(2)}_{ij}=\ave{{\cal N}M_{\rm ap}}_i(\theta_{{\rm ap},j})$},
and \mbox{$d^{(3)}_{ij}=\ave{M_{\rm ap}^2}_i(\theta_{{\rm ap},j})$}.
For each jackknife sample, we remove two adjacent \mbox{MegaCam}
pointings from one field, redo the measurement, and finally combine
the results from all four patches W1 to W4 into $\vec{d}_i$. In three
samples, where the fields have an odd number of pointings, we remove
three pointings. In total, we thereby obtain $n_{\rm jk}=63$ jackknife
samples to estimate the error covariance.

\begin{figure}
  \begin{center}
    \includegraphics[width=92mm]{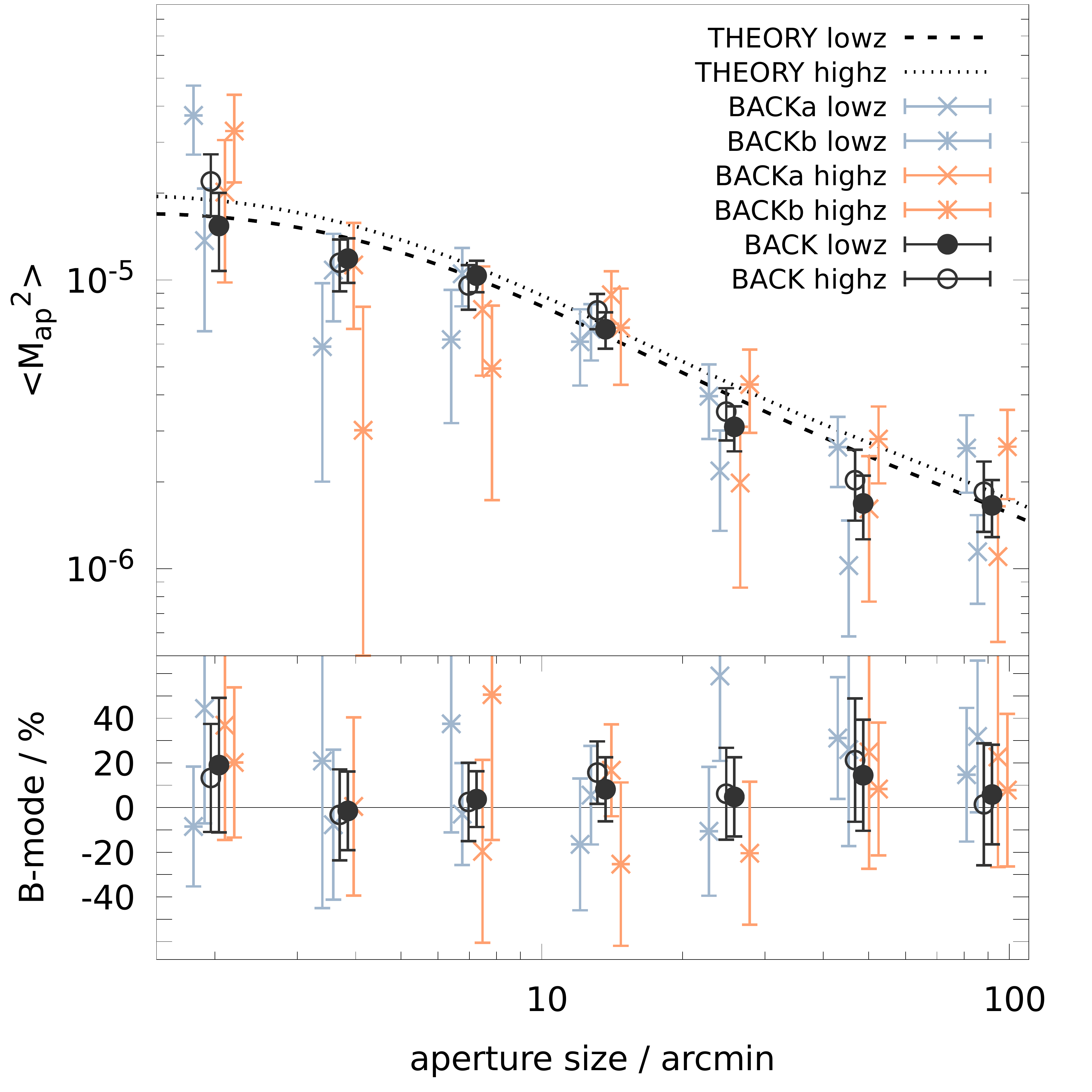}
  \end{center}
  \caption{\label{fig:mapsqd} Variance $\ave{M^2_{\rm ap}}$ as
    function of aperture size for our source samples. The top panel
    shows the E-mode and the  predictions (lines) for the BACK
    low-$z$ and BACK high-$z$ data points (solid and open
    circles). The crosses and stars correspond to the (noisier)
    signals of the four subdivisions $a$ and $b$ of the source
    samples. The bottom panel shows the B-mode relative to the E-mode
    signal (in percentage). This plot uses seven evenly spaced aperture
    radii instead of the eight used in the analysis.}
\end{figure}

Figure \ref{fig:mapsqd} shows our measurements for
$\ave{M^2_{\rm ap}}$ and the jackknife error; the top panel is the
E-mode signal, the bottom panel the B-mode signal, which is consistent
with zero, relative to the E-mode. In our analysis the measurements
for BACK low-$z$ and BACK high-$z$ {are plotted as circles}. The
noisier data of the source-sample subdivisions $a$ and $b$ are
depicted as crosses and stars. Importantly, the circle data points are
the weighted average of crosses, stars, and the cross-correlation
signal of the subdivisions (not shown here). The dashed and dotted
lines are predictions based the cosmology in Table \ref{tab:fidcosmo}.

\subsection{Ratio statistics and projected galaxy bias}

\begin{figure*}
  \begin{center}
    \includegraphics[width=180mm]{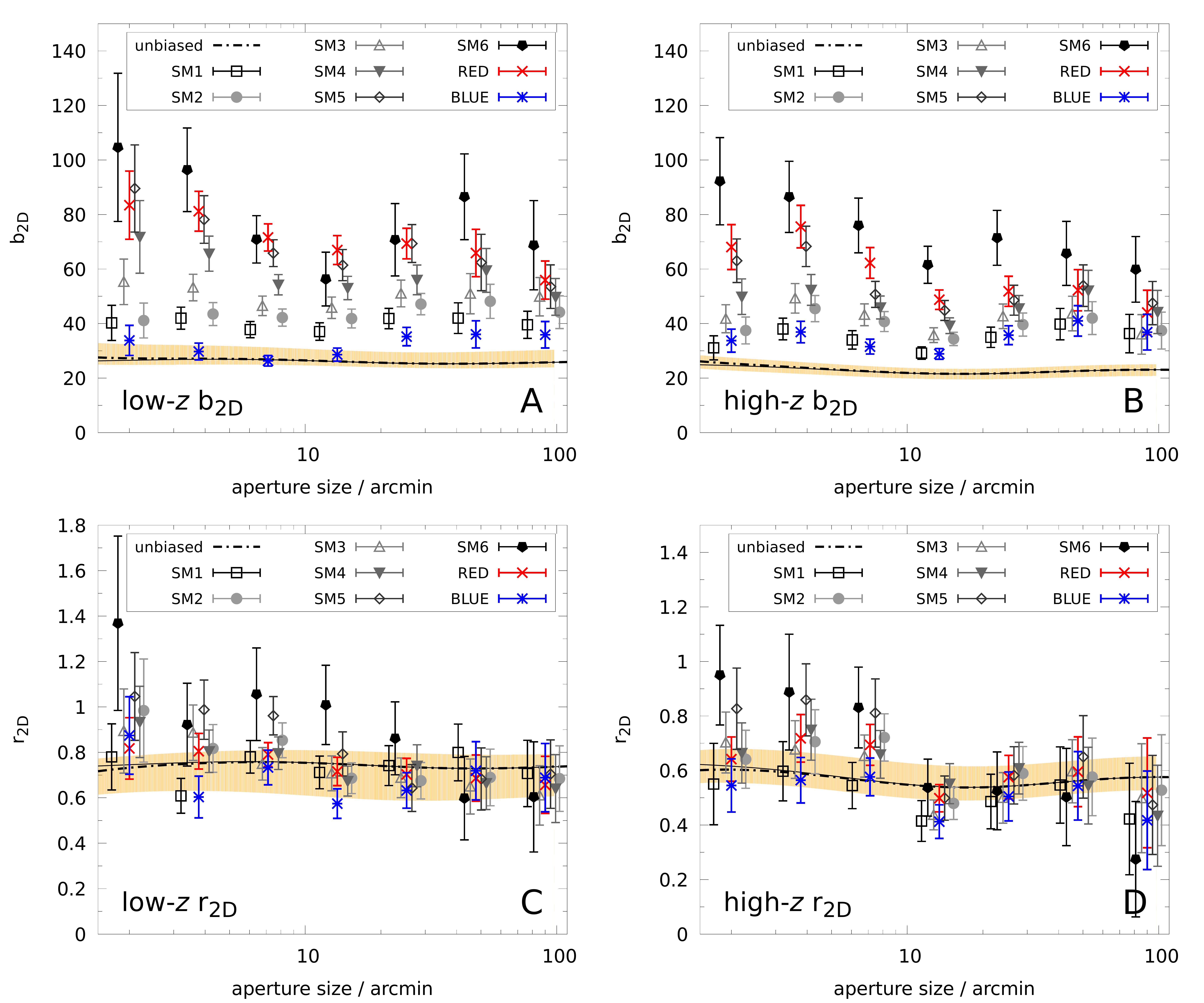}
  \end{center}
  \caption{\label{fig:br2d_results} Observed ratio statistics with
    jackknife errors for the \cfhtlens~lens samples in the low-$z$ bin
    (panels A and C) and the high-$z$ bin (panels B and D). The source
    samples are BACK low-$z$ and BACK high-$z$, respectively.  The
    data points are slightly shifted along the $x$-axis for better
    visibility. The dashed lines are predictions for unbiased galaxies
    (the galaxy-bias normalisation) with SM3 $p_{\rm d}(z)$.  The
    shaded regions bracket the prediction variations using the
    $p_{\rm d}(z)$ of the other lens samples. \emph{Top panels:}
    Relative clustering $\hat{b}_{\rm 2D}(\theta_{\rm ap})$ of
    galaxies and matter. \emph{Bottom panels:} Correlation
    $\hat{r}_{\rm 2D}(\theta_{\rm ap})$ between the (projected) galaxy
    and matter distribution. This plot uses seven evenly spaced
    aperture sizes instead of the eight used in the analysis.}
\end{figure*}

The ratio statistics
\begin{eqnarray}
\nonumber
  \hat{b}_{\rm 2D}(\theta_{\rm ap})
  &=&
      \sqrt{
      \frac{\ave{{\cal N}^2}(\theta_{\rm ap})}
     {\ave{M_{\rm ap}^2}(\theta_{\rm ap})}}\;;
  \\
  \label{eq:ratioobs}
  \hat{r}_{\rm 2D}(\theta_{\rm ap})
  &=&
      \frac{\ave{{\cal N}M_{\rm ap}}(\theta_{\rm ap})}
      {\sqrt{\ave{{\cal N}^2}(\theta_{\rm ap})\,\ave{M_{\rm ap}^2}(\theta_{\rm
      ap})}}
\end{eqnarray}
is the observed projection of the biasing functions $b(k)$ and $r(k)$
on the sky. These raw data are plotted in Fig. \ref{fig:br2d_results},
as function of aperture size, for \cfhtlens~with the bias factor
$\hat{b}_{\rm 2D}$ in the top panels A and B, and the correlation
coefficient $\hat{r}_{\rm 2D}$ in the bottom panels C and D. The error
bars indicate symmetric jackknife errors, whereas the analysis below
accounts for asymmetric errors in the ratio statistics through a more
complex likelihood model. The likelihood model also accounts for the
correlation between the data points at different aperture sizes,
encoded in the jackknife covariance. Moreover, statistical errors for
different lens samples are correlated because they share the same
shear data.

The lines and shaded areas inside the panels are predictions for
unbiased galaxies and the fiducial cosmology in Table
\ref{tab:fidcosmo}. They are traditionally used as normalisation for
the projected galaxy bias to correct, among other things, for the
distinct radial projection kernels of galaxies and matter
\citep{2002ApJ...577..604H}. Specifically, the dashed lines are the
normalisation for SM3 low-$z$ (left panels) and high-$z$ (right
panels). The normalisation varies weakly with aperture size, and
therefore essentially rescales the amplitude of the inferred galaxy
bias. This is also the case for the other samples.  Across all lens
samples in the same photo-$z$ bin, however, the normalisation
amplitude varies by about $10\%$ within the shaded regions owing to
differences among the $p_{\rm d}(z)$ (Fig. \ref{fig:pofz}). The
predictable uncertainties in the normalisation amplitude for a
specific galaxy sample are estimators for the systematic error in the
reconstructed biasing functions. This is exploited below.

Coming back to the \cfhtlens~data points, our samples are usually more
strongly clustered than unbiased galaxies (i.e. $\hat{b}_{\rm 2D}$ is
larger). In addition, our galaxy samples are strongly correlated with
the matter-density field, which means that $\hat{r}_{\rm 2D}$ is
consistent with unbiased galaxies. Inside large apertures the
correlation is similar for all lens samples, whereas there is some
up-scatter below $20\,\rm arcmin$. This hints at sample differences in
the galaxy biasing at small scales. To investigate this in detail we
deproject the aperture statistics.

\subsection{Template models for deprojection}
\label{sect:templatemodels}

To reconstruct spatial biasing functions from the projected ratio
statistics in Fig. \ref{fig:br2d_results} we fit the models
\begin{eqnarray}
  \nonumber
  \hat{b}_{\rm 2D}(\theta_{\rm ap};b)
  &=&
      \sqrt{
      \frac{\ave{{\cal N}^2}_{\rm th}(\theta_{\rm ap};b)}
      {\ave{M_{\rm ap}^2}_{\rm th}(\theta_{\rm ap})}}\;;
  \\
  \label{eq:br2dmodel}
  \hat{r}_{\rm 2D}(\theta_{\rm ap};b,r)
  &=&
      \frac{\ave{{\cal N}M_{\rm ap}}_{\rm th}(\theta_{\rm ap};b,r)}
      {\sqrt{\ave{{\cal N}^2}_{\rm th}(\theta_{\rm ap};b)\,\ave{M_{\rm
      ap}^2}_{\rm th}(\theta_{\rm
      ap})}}
\end{eqnarray}
to the measurements by using templates for $b(k)$ and
$r(k)$. Templates are useful here because, due to the smoothing in the
projections \Ref{eq:br2dmodel}, a deprojection into $b(k)$ and $r(k)$
for a given $\hat{b}_{\rm 2D}$ and $\hat{r}_{\rm 2D}$ without
regularisation is unstable. We denote the templates
$b(k;\vec{\Theta})$ and $r(k;\vec{\Theta})$ to indicate the set
$\vec{\Theta}$ of parameters they are based on.

\renewcommand{\arraystretch}{1.1}
\begin{table}
  \caption{\label{tab:model} List of template parameters}
\begin{center}
  \begin{tabular}{ll}
    \hline\hline
    Param & Description\\
    \hline\\
    $b_{\rm ls}$   & large-scale bias factor\\
    $r_{\rm ls}$   & large-scale correlation factor\\
    $b(m)$       & mean biasing function (interp.)\\
    $V(m)$       & normalised excess-variance (interp.)\\
    $m_{\rm piv}$  & $\ave{N|m_{\rm piv}}=1$; pivotal halo mass \\
    $f_{\rm cen}$  & halo fraction open for central galaxies\\
    $\zeta$      & concentration parameter of satellites\\
  \end{tabular}
  \tablefoot{The HOD functions $b(m)$, related to $\ave{N|m}$, and
    $V(m)$, related to $\ave{N(N-1)|m}$, cover the mass range
    $10^4-10^{16}\,h^{-1}\,\msol$ and are interpolated with $22$
     points each.}
\end{center}
\end{table}
\renewcommand{\arraystretch}{1.0}

We employ templates that {are} thoroughly tested with mock galaxy
survey data, similar to our \cfhtlens~data, in SH18. They are based on
the halo model in \cite{2000MNRAS.318..203S} but with more freedom for
the halo-occupation distribution (HOD) of galaxies and a model-free
galaxy bias on linear scales. This halo model ignores relevant
physics, such as halo exclusion or assembly bias or {galactic
  subhalos}, and it has a crude treatment of central
galaxies. Nevertheless, the templates produce an accurate deprojection
of the biasing functions which is our primary interest here.  We
outline the template construction in Appendix
\ref{app:templateconstruction} and restrict ourselves here to a brief
description of the template parameters, listed in Table
\ref{tab:model}.

The templates have $49$ free parameters
\mbox{$\vec{\Theta}=(\ldots,b_i,\ldots,V_i,\ldots,\zeta,m_{\rm
    piv},b_{\rm ls},r_{\rm ls},f_{\rm cen})$}. The large number of
parameters is due to the fine sampling of the HOD mean number of
galaxies, $\ave{N|m}$, and galaxy pairs, $\ave{N(N-1)|m}$, with $22$
interpolation points each that cover the broad halo mass range
\mbox{$10^4\le m\le10^{16}\,h^{-1}\,\msol$}. Our intention is to keep
the HOD as flexible as possible and to avoid a model bias in the
deprojection with a too restrictive HOD.  The meaning of the
parameters is as follows:
\begin{itemize}
\item The two parameters \mbox{$b_{\rm ls}=b(k\to0)$} and
  \mbox{$r_{\rm ls}=r(k\to0)$} quantify the linear stochastic bias on
  large (linear) scales. They are completely independent from the
  galaxy bias on small scales, i.e. independent from the template HOD;
\item The parameter \mbox{$f_{\rm cen}\in[0,1]$} is the fraction of
  halos that permit a central galaxy in the selected galaxy
  population. This fraction is not to be confused with the fraction of
  central galaxies in a sample;
\item A concentration parameter \mbox{$\zeta\ne1$} indicates biased
  satellite galaxies that have, averaged over all halos, a different
  distribution from the dark matter inside halos. For \mbox{$\zeta>1$}
  satellites have a more diffuse distribution, and for
  \mbox{$\zeta<1$} a more concentrated distribution;
\item The pivotal mass $m_{\rm piv}$ defines the halo mass scale,
  where \mbox{$\ave{N|m_{\rm piv}}=1$} and $N$ comprises both central
  and satellite galaxies;
\item The mean number of galaxies in a halo of mass $m$,
  \mbox{$\ave{N|m}=b(m)\,m/(b(m_{\rm piv})\,m_{\rm piv})$}, is
  expressed by the mean biasing function $b(m)$ and its value at
  $m_{\rm piv}$. The $b(m)$ is linearly interpolated on a logarithmic
  $m$-scale with $22$ interpolation points, and it has to satisfy the
  normalisation condition in Eq. \Ref{eq:bmnorm}.
\item The mean number of galaxy pairs, expressed by the normalised
  excess variance $V(m)$ via
  $\ave{N(N-1)|m}=\ave{N|m}\,[\ave{N|m}+V(m)]$, is sensitive to the
  variance of galaxy numbers inside halos, which means for
  \mbox{$V(m)=0$} that galaxy numbers have a Poisson variance, for
  \mbox{$V(m)>0$} a super-Poisson variance, and for \mbox{$V(m)<0$} a
  sub-Poisson variance. Similar to $b(m)$, $V(m)$ is interpolated with
  $22$ points.  To be physically meaningful, the function $V(m)$ has
  to be larger than \mbox{${\rm max}\{-1,-\ave{N|m}\}$}, which is
  accounted for in the template fitting.
\end{itemize}
Unbiased galaxies have $b_{\rm ls}=r_{\rm ls}=\zeta=1$,
$f_{\rm cen}=0$, $b(m)=1$, and $V(m)=0$ for all halo masses $m$.

The template parameter space is highly degenerate, which may pose
numerical problems when sampling the space during our deprojection
technique. Therefore, we also fit the relation
\begin{equation}
  \label{eq:nbarg}
  \bar{n}_{\rm g}(\vec{\Theta})=
  \int_0^\infty\d m\;n(m)\,\ave{N|m}=
  \frac{\Omega_{\rm m}\bar{\rho}_{\rm crit}}{m_{\rm piv}\,b(m_{\rm piv})}
\end{equation}
to our observed $\bar{n}_{\rm g}$ in Fig. \ref{fig:ng} to better
confine $m_{\rm piv}$ and to break parameter degeneracies.\footnote{An
  alternative route without the extra parameter $m_{\rm piv}$ would be
  to use the relation
  \mbox{$\ave{N|m}=m\,b(m)\,\bar{n}_{\rm g}\,\bar{\rho}_{\rm
      m}^{-1}$},
  $\bar{\rho}_{\rm m}=\Omega_{\rm m}\bar{\rho}_{\rm crit}$, and the
  measured $\bar{n}_{\rm g}$ with marginalisation over the statistical
  error in $\bar{n}_{\rm g}$.}  Here,
$\bar{\rho}_{\rm crit}=3H_0^2/(8\pi G_{\rm N})$ denotes the critical
matter density at $z=0$. Equation \Ref{eq:nbarg} follows from the
definition of $b(m)$ and the condition Eq. \Ref{eq:bmnorm}.

\subsection{Bayesian reconstruction of biasing functions}

We forward-fit in a Bayesian analysis the templates
$b(k;\vec{\Theta})$ and $r(k;\vec{\Theta})$ to the ratio statistics
(Eq. \ref{eq:ratioobs}); the vector $\vec{\Theta}$ comprises all
template parameters in Table \ref{tab:model}. For the likelihood
function of the data we assume a Gaussian multivariate error
distribution of the data,
\begin{equation}
  \vec{d} =
  \left(\ldots,\ave{{\cal N}}(\theta_{{\rm ap},i}),
    \dots,\ave{{\cal N}M_{\rm ap}}(\theta_{{\rm ap},i}),
    \dots,\ave{M_{\rm ap}^2}(\theta_{{\rm ap},i}),\ldots\right)\;,    
\end{equation}
and a lognormal error distribution of the galaxy number density
$\bar{n}_{\rm g}$.  Specifically, the probability distribution
function $p_\delta$ of statistical errors $\delta\vec{d}$ and
$\delta_n$ in $\vec{d}$ and $\log_{10}{\bar{n}_{\rm g}}$,
respectively, is
\begin{multline}
  \label{eq:likelihood}
  -2\,\ln{p_\delta(\delta\vec{d},\delta_n)}= \\ {\rm const}+n_{\rm
    jk}\,\ln{\left(1+\frac{\delta\vec{d}^{\rm
          T}\mat{C}^{-1}\delta\vec{d}}{n_{\rm jk}-1}\right)}
  +\frac{\delta_n^2}{\sigma_{\rm logn}^2}\;,
\end{multline} 
for the error covariance $\mat{C}$ of $\vec{d}$
(Sect. \ref{sect:estimators}) and an error variance
$\sigma_{\rm logn}$ of $\log_{10}{\bar{n}_{\rm g}}$ (error bars in
Fig. \ref{fig:ng}). This noise model also marginalises over
statistical errors in our estimate of $\mat{C}$, which is constructed
from $n_{\rm jk}$ jackknife samples \citep{2016MNRAS.456L.132S}. As
discussed in SH18, the posterior density of the template parameters
given the ratio statistics in Eqs. \Ref{eq:ratioobs} is then a
marginalisation of $p_\delta$ with respect to
\mbox{$\vec{x}:=(\ldots,\ave{M^2_{\rm ap}}(\theta_{{\rm
      ap},i}),\ldots)$},
\begin{equation}
  \label{eq:posterior}
  p(\vec{\Theta}|\vec{d},\bar{n}_{\rm g})\propto
  p_\Theta(\vec{\Theta})\,
  \int\d\vec{x}\;
  p_\delta\Big(\vec{d}-\vec{m}(\vec{\Theta},\vec{x}),\log_{10}{\frac{\bar{n}_{\rm
    g}}{\bar{n}_{\rm g}(\vec{\Theta})}}\Big)
  \,p_x(\vec{x})\;,
\end{equation}
where
$\vec{m}(\vec{\Theta},\vec{x})=\left(\vec{m}^{(1)},\vec{m}^{(2)},\vec{m}^{(3)}\right)$
and
\begin{eqnarray}
  m^{(1)}_i&=&\hat{b}^2_{\rm 2D}(\theta_{{\rm ap},i};b)\,x_i\;;
  \\
  m^{(2)}_i&=&\hat{b}_{\rm 2D}(\theta_{{\rm ap},i};b)\,\hat{r}_{\rm
  2D}(\theta_{{\rm ap},i};b,r)\,x_i\;;
  \\
  m_i^{(3)}&=&x_i\;.
\end{eqnarray}
The integral kernel $p_\delta(\ldots)$ inside \Ref{eq:posterior} is
the likelihood of $\vec{d}$ given $\vec{\Theta}$, while
$p_\Theta(\vec{\Theta})$ and $p_x(\vec{x})$ are prior densities for
$\vec{\Theta}$ and $x_i$. The prior densities are detailed in the
following section.

Since we seek constraints on the biasing functions rather than
$\vec{\Theta}$, we convert the posterior density of $\vec{\Theta}$
into a posterior distribution for $b(k)$ and $r(k)$ by drawing
\mbox{$\vec{\Theta}_i\sim p(\vec{\Theta}|\vec{d},\bar{n}_{\rm g})$}
from \Ref{eq:posterior} in a Monte Carlo process and by computing the
mean and variance of $\{b(k;\vec{\Theta}_i)\}$ and
$\{r(k;\vec{\Theta}_i)\}$ for a range of $k$.

\subsection{Bayesian priors}

For the prior densities
$p_\Theta(\vec{\Theta})=\prod_i p_{\Theta_i}(\Theta_i)$ and
$p_{\rm x}(\vec{x})=\prod_i p_{x_i}(x_i)$ in Eq. \Ref{eq:posterior},
we assert, similar to the analysis in SH18, uniform top-hat priors and
the boundaries $V(m)\in[\max{\{-1,-\ave{N|m}\}},1]$,
$\zeta\in[0.1,2]$, $\log_{10}{(m_{\rm piv}\,h\,\msol^{-1})}\in[4,16]$,
$b_{\rm ls}\in[0,5]$, $r_{\rm ls}\in[0,2]$, and $f_{\rm cen}\in[0,1]$.

An exception here is the prior density for $b_i:=b(m_i)$, which
compared to SH18 now has the less informative density
\mbox{$p_{{\rm b}_i}(b_i)\propto1/b_i$} or, equivalently,
\mbox{$p_{{\rm b}_i}(\ln{b_i})=\rm const$}.  This change is needed to
avoid a bias in the $r(k)$ reconstruction for
\mbox{$k\gtrsim3\,h^{-1}\,\rm Mpc^{-1}$} due to the noisier \cfhtlens~
data (for more details on this problem, see Appendix
\ref{app:templateprior}). Likewise, the prior density for $x_i$ is now
\mbox{$p_{\rm x}(\vec{x})\propto1/\prod_i x_i$}.

\subsection{Systematic error of the galaxy-bias amplitude}
\label{sect:normerrors}

\begin{table}
  \caption{\label{tab:errorbase} Baseline of assumed systematic errors
    for the reconstruction of biasing functions. The resulting
    systematic RMS uncertainties of $b(k)$ and $r(k)$ amplitudes are
    within $[6.8\%,13.5\%]$ and $[5.2\%,8.5\%]$, respectively (see
    text).}
\begin{center}
  \begin{tabular}{ll}
    \hline\hline
    Source & RMS error\\
    \hline\\
    redshift bias   (sources and lenses) & 2.0\%\\
    width of $p(z)$ (sources and lenses) & 5.0\%\\
    \\
    $\Omega_{\rm m}$  & 4.3\%\\
    $\Omega_{\rm b}$  & 4.8\%\\
    $w$              & 6.2\%\\
    $H_0$            & 2.3\%\\
    $n_{\rm s}$       & 0.5\%\\
    $\sigma_8$       & 2.6\%\\
    \\
    $A_{\rm ia}$      & 170\%\\
    \\
    $F(k,z)$         & 30.0\%
  \end{tabular}
  \tablefoot{Correlated errors on cosmological parameters
    $\Omega_{\rm m}$ to $\sigma_8$ are the posterior constraints from
    \cite{2016A&A...594A..13P} for a flat $w\rm CDM$ model. The IA
    parameter is \mbox{$A_{\rm ia}=-0.48^{+0.75}_{-0.87}$} from
    \cite{2013MNRAS.432.2433H}. The baryon transfer function $F(k,z)$
    is from \cite{2018arXiv180108559C}; we use as fiducial values
    their Figure 11, but randomly perturb $F(k,z)$ by
    $\delta F(k,z)=\delta\times(F(k,z)-1)$ with a normally distributed
    amplitude $\delta$ of the quoted RMS variance.}
\end{center}
\end{table}

Errors in the fiducial cosmology or the redshift distributions of
lenses and sources (i.e. in the projection kernel) cause systematic
errors in the amplitude of the reconstructed biasing functions $b(k)$
and $r(k)$. Since changes in the projection kernel produce variations
in the amplitudes of $\hat{b}_{\rm 2D}$ and $\hat{r}_{\rm 2D}$ with
only weak dependence on angular scale, the $k$-dependence of errors in
the deprojected biasing functions is also weak, and the amplitude
errors in the deprojection are of similar magnitude. We therefore
estimate errors in $b(k)$ and $r(k)$ by propagating variations in the
lensing kernel parameters to $\hat{b}_{\rm 2D}$ and $\hat{r}_{\rm 2D}$
for unbiased galaxies (i.e. the normalisation for the projected bias),
and we average their RMS variations over
\mbox{$1^\prime\lesssim\theta_{\rm ap}\lesssim2^\circ$}.

As baseline for the error model we assume the RMS values in Table
\ref{tab:errorbase} for (i) the mean and width of the redshift
distributions, (ii) the set of fiducial cosmological parameters, (iii)
the baryon physics in the non-linear matter power spectrum, and (iv)
the amplitude \mbox{$A_{\rm ia}=-0.48^{+0.75}_{-0.87}$} of the IA,
based on the \cfhtlens~constraints in \cite{2013MNRAS.432.2433H}.  The
error distribution of cosmological parameters for (ii) is the
posterior distribution of the \emph{Planck} TT, TE, and EE cosmology
constraints combined with data from BAO experiments; the reference
cosmology is a flat \mbox{$w\rm CDM$ model}
\citep{2016A&A...594A..13P}.\footnote{We use the MCMC posterior
  samples for \texttt{base\_w\_plikHM\_TTTEEE\_lowTEB\_BAO} at
  \url{http://pla.esac.esa.int/pla/#cosmology}.} For the fiducial
redshift distribution of the lenses in (i), we take that of SM1, SM4,
SM6, RED, and BLUE, and we then average the results. The redshift
distribution of the sources are either BACK low-$z$ or BACK high-$z$.

We perform thousands of random variations in (i) to (iv) and average
for all variations the quadratic means in the interval
\mbox{$1^\prime\lesssim\theta_{\rm ap}\lesssim2^\circ$} for
$\delta b=\hat{b}_{\rm 2D}/\hat{b}_{\rm 2D}^{\rm fid}-1$ and
$\delta r=\hat{r}_{\rm 2D}/\hat{r}_{\rm 2D}^{\rm fid}-1$ relative to
our fiducial models $\hat{b}_{\rm 2D}^{\rm fid}$ and
$\hat{r}_{\rm 2D}^{\rm fid}$ in Table \ref{tab:fidcosmo} and the
redshift distributions in Fig. \ref{fig:pofz}. For every new variation
in the cosmological model we draw a set of cosmological parameters
from the published \emph{Planck} MCMC posterior, thereby accounting
for the correlations of their errors. Variations in the mean
($\delta z$) and RMS width ($\delta\sigma$) of redshift distributions
of lenses and sources, the IA amplitude ($\delta A_{\rm ia}$) and the
baryon transfer function ($\delta F$) are independently drawn from a
Gaussian distribution with the variances given in Table
\ref{tab:errorbase}. To change the mean of a redshift distribution
$p(z)$, we map \mbox{$p(z)\to p(z[1+\delta z])$}, whereas
\mbox{$p(z)\to p(z)^{1/(1-\delta\sigma)^2}$} varies the width of the
distribution (Section 7.3 in SH18). We perform this analysis for a
scenario where all errors (i)-(iv) are present simultaneously but are
statistically independent, and scenarios where only one error source
(i) to (iv) is switched on at a time, giving the isolated RMS error
$\sigma_i$. The detailed error distributions of the bias amplitudes
can be found in Appendix \ref{sect:normerror}.

In summary, the RMS error $\sigma_{\rm tot}$ for $b(k)$ ranges between
$6.1\%$ and $11.6\%$, and between $4.3\%$ and $6.9\%$ for $r(k)$.  For
these values, the lower limits adopt uncorrelated systematic errors
(i) to (iv) (the optimistic scenario, i.e.
$\sigma_{\rm tot}^2=\sum_i\sigma_i^2$), while the upper limits assume
that systematic errors (i) to (iv) conspire to maximise the amplitude
error (the conservative scenario, i.e.
$\sigma_{\rm tot}=\sum_i\sigma_i$).

However, these are not all the systematic errors that need to be
considered.  There may be additional systematic errors from the
methodology in Sect. \ref{sect:method}, for instance model bias by the
templates or the likelihood, and its specific implementation in a
computer code.  According to SH18, these errors fall between $3\%$ and
$7\%$ for $b(k)$ and between $3\%$ and $5\%$ for $r(k)$. These
estimates were obtained from galaxy-bias reconstructions with mock
lensing data (H15 based) that have exactly known galaxy bias, redshift
distributions, and cosmological and IA parameters. Since they are
related to methodology, they are uncorrelated to uncertainties in the
galaxy bias normalisation.

Combining all the systematic errors above, the accuracy for $b(k)$
ranges somewhere between $6.8\%$ (optimistic) and $13.5\%$
(conservative), and for $r(k)$ between $5.2\%$ and $8.5\%$. A
systematic error would be similar on all scales, and therefore offset
the inferred biasing functions.

\section{Results}

\subsection{Biasing functions}

Our $b(k)$ and $r(k)$ results are shown on the following pages, after
the references, as Figs. \ref{fig:bk_lowz} to \ref{fig:rk_highz} in
the Appendix. The orange regions depict the $68\%$ and $95\%$ {CIs} of
the reconstructed biasing functions around the median (dashed
lines). The reconstructions assume the fiducial parameters in Table
\ref{tab:fidcosmo} and the galaxy redshift distributions in Fig.
\ref{fig:pofz}. There is one panel for each galaxy sample, indicated
by the in-panel labels at the top left. For most \cfhtlens~samples,
the bias factor $b(k)$ clearly changes with $k$, while the high
galaxy-matter correlation, \mbox{$r(k)\approx1$}, has a moderate scale
dependence.

The green data points show the SAM predictions in comparison: filled
diamonds are for H15, and filled triangles are those for L12. In
comparison to the models, the relative changes of the \cfhtlens~$b(k)$
with $k$ is broadly similar to the models, but real galaxies are
clearly more biased for \mbox{$k\lesssim0.5\,h\,\rm
  Mpc^{-1}$}. Towards the small spatial scales at higher $k$, the
biasing of galaxies in both \cfhtlens~and the models increases.  Due
to a steeper increase of L12 for SM2 to SM5, however, the Durham model
typically agrees better with \cfhtlens~than H15. Still, the models
have a too small bias for SM1 and BLUE on all scales. With regard to
the matter-galaxy correlation, the value of $r(k)$ evidently increases
for \mbox{$k\gtrsim1\,h\,\rm Mpc^{-1}$} in the \cfhtlens~samples SM3
to SM6. This reflects, despite the noise, the scale-dependence in the
SAMs. The model agreement is somewhat worse for high-$z$ where the
high value of $r(k)$ at \mbox{$k\approx10\,h\,\rm Mpc^{-1}$} is not
supported by the \cfhtlens~samples SM1, SM2, RED, and BLUE.

For a test of signal robustness in our data we split the source
catalogues into two subsamples of distinct mean redshifts $\bar{z}$:
BACK low-$z$ ($\bar{z}\approx0.97$) into BACKa low-$z$
($\bar{z}\approx0.86$) and BACKb low-$z$ ($\bar{z}\approx1.12$); BACK
high-$z$ ($\bar{z}\approx1.03$) into BACKa high-$z$
($\bar{z}\approx0.93$) and BACKb high-$z$ ($\bar{z}\approx1.14$). Then
we repeat every galaxy-bias reconstruction twice using those
subsamples. The overlaid transparent grey regions in the figures are
the 68\% posterior CIs about the median for these
reconstructions. Since they constrain the same biasing functions as
the combined source sample, they should overlap with the filled
dark-orange region, and they should overlap with each other. This is
indeed the case for most reconstructions. A few exceptions are found
near \mbox{$k\approx2\,h\,\rm Mpc^{-1}$} for $b(k)$ or at
\mbox{$k\lesssim0.1\,h\,\rm Mpc^{-1}$} for $r(k)$, such as for RED
low-$z$ in both cases. Another disagreement is visible at small scales
for SM4 high-$z$ where the two extra reconstructions prefer smaller
$b(k)$ values. On the whole, however, and with some disagreement to be
expected due to random chance, we consider the reconstruction overall
robust. The next section tests the adequacy of the templates in
describing the ratio statistics.

\subsection{Posterior predictive check of bias reconstruction}

\begin{figure*}
  \centering \includegraphics[scale=0.36,clip=false,angle=0]{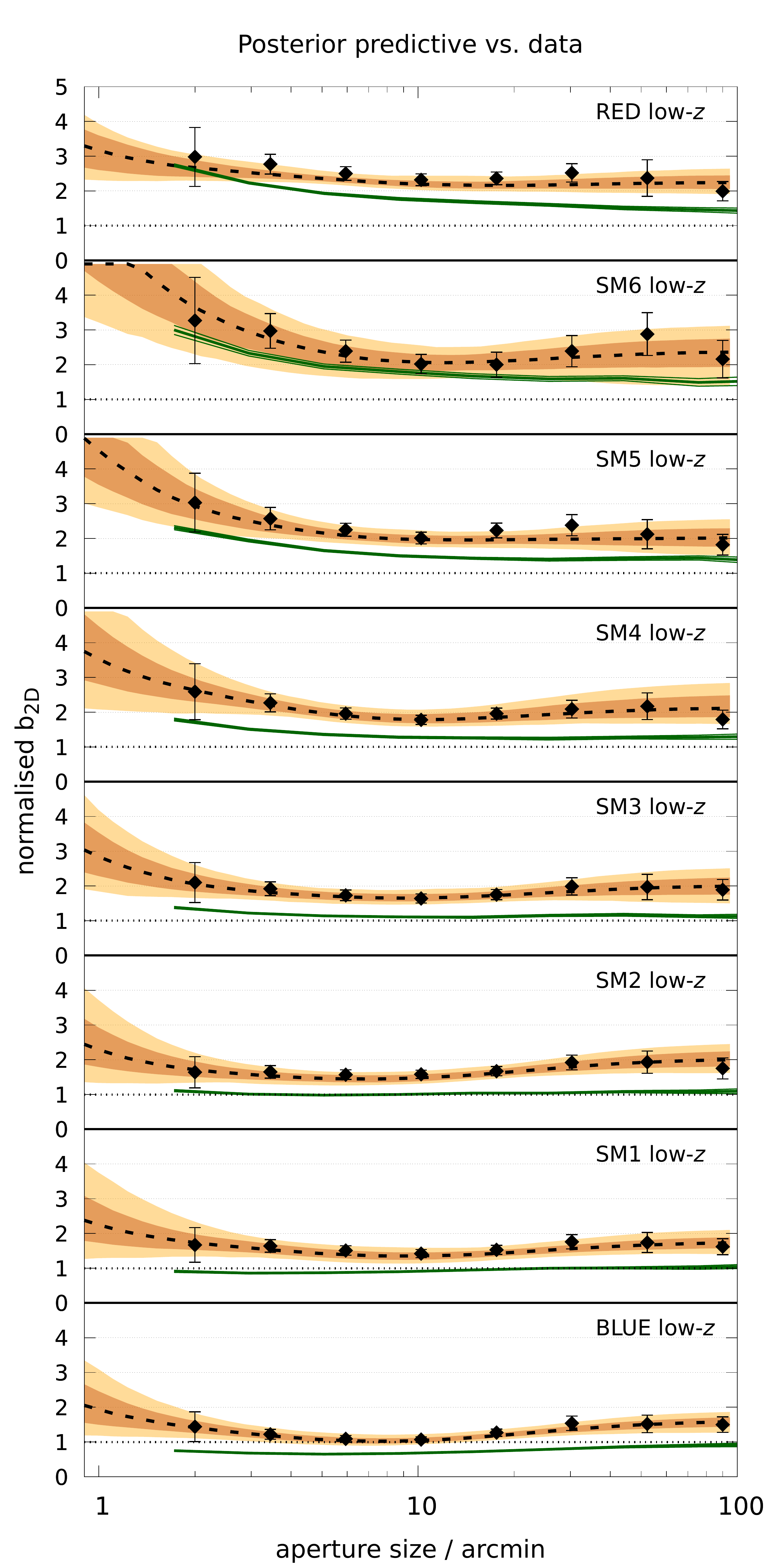}\includegraphics[scale=0.36,clip=false,angle=0]{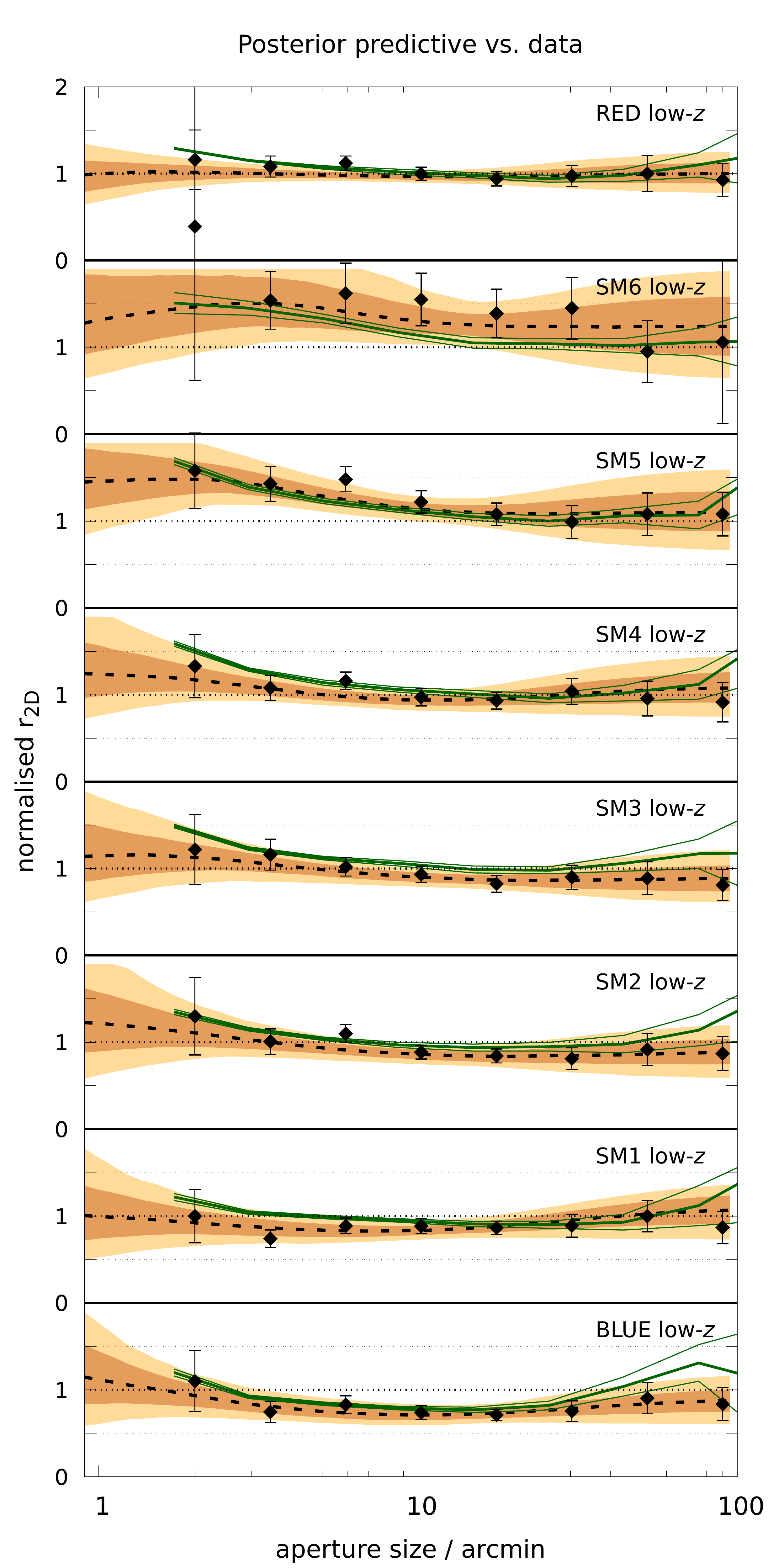}
  \caption{\label{fig:postpredlowz} Posterior predictive checks for
    the low-$z$ \cfhtlens~samples. The data points are normalised bias
    measurements $b_{\rm 2D}$ (left panel) and $r_{\rm 2D}$ (right
    panel) with jackknife errors as a function of aperture size. The
    orange regions are the posterior predictive distributions of our
    templates with $68\%$ and $95\%$ CIs; the dashed lines are the
    median. The green solid lines depict the normalised bias and
    $68\%$ confidence intervals for H15 mock data in SH18 without
    shape noise.}
\end{figure*}

\begin{figure*}
  \centering
  \includegraphics[scale=0.36,clip=false,angle=0]{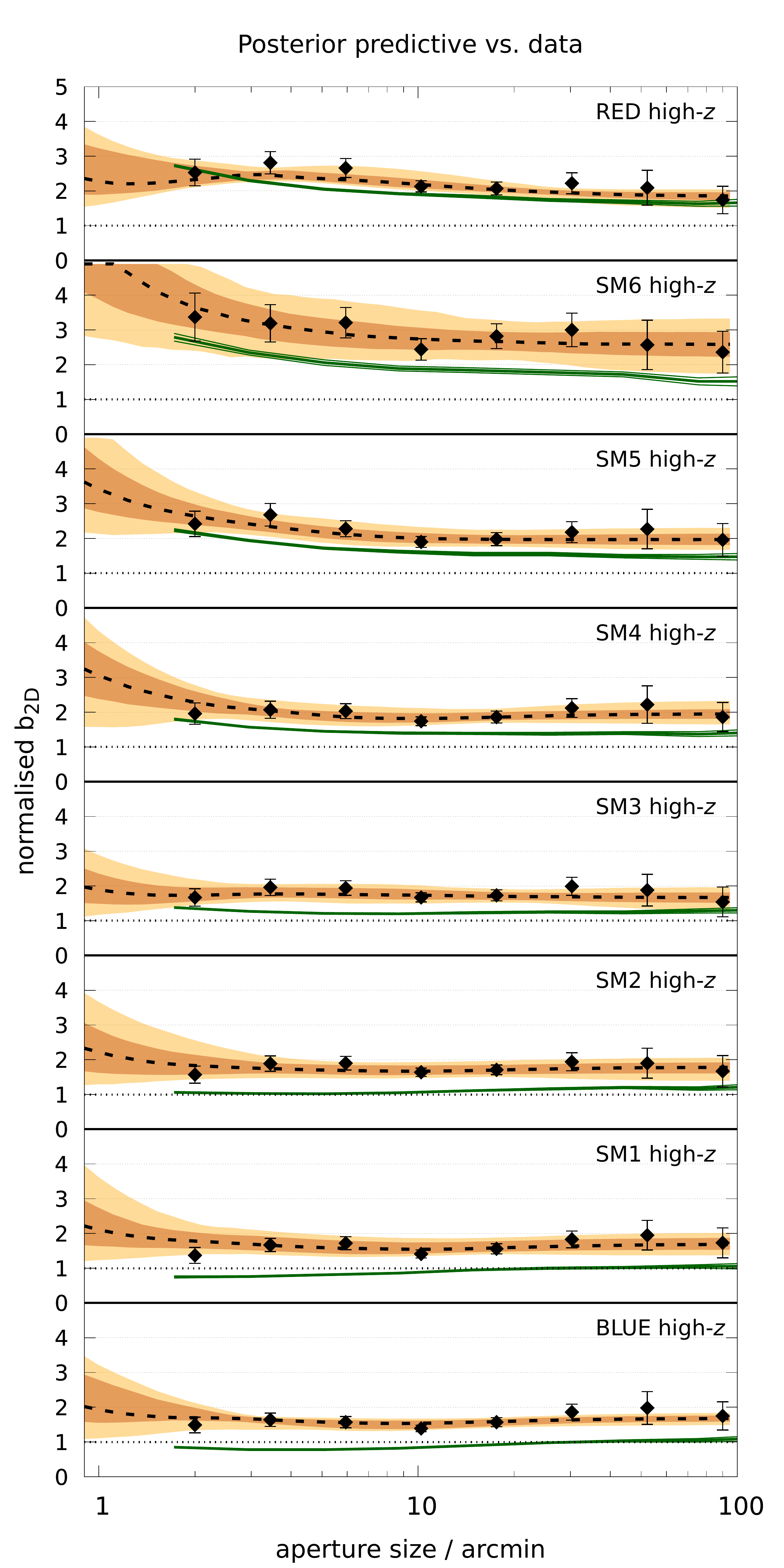}\includegraphics[scale=0.36,clip=false,angle=0]{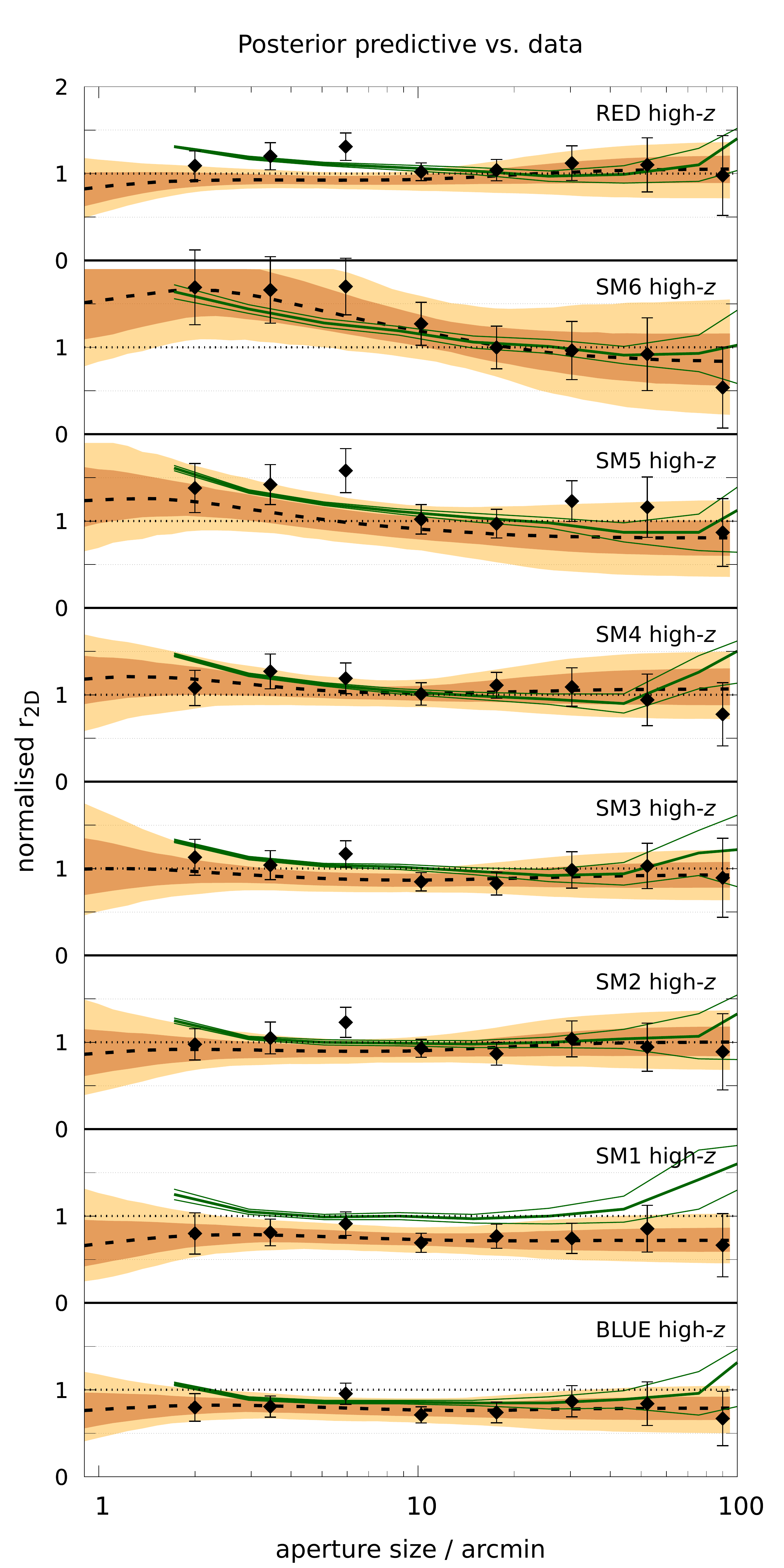}
  \caption{\label{fig:postpredhighz} As in Fig. \ref{fig:postpredlowz},
    but   for the high-$z$ samples. There are signs of a model
    misfit to $r_{\rm 2D}$ at around $6\,\rm arcmin$, especially for
    RED and SM5. There is also weak evidence in favour of a stochastic
    bias, \mbox{$r_{\rm 2D}<1$}, for SM1 and BLUE at degree scales
    (bottom right panels).}
\end{figure*}

A posterior predictive check validates the template models in our
Bayesian deprojection \citep{gelman2003bayesian}.  We construct the
posterior predictive, plotted in Figs. \ref{fig:postpredlowz} and
\ref{fig:postpredhighz}, by drawing values
$\vec{\Theta}\sim p(\vec{\Theta}|\vec{d},\bar{n}_{\rm g})$ from the
posterior distribution of template parameters in
Eq. \Ref{eq:posterior} and by comparing the scatter of their
$b_{\rm 2D}(\theta_{\rm ap};b)$ (left panels) and
$r_{\rm 2D}(\theta_{\rm ap};b,r)$ (right panels) to the \cfhtlens~data
points of the ratio statistics. The ratio statistics is normalised by
the prediction for unbiased galaxies using the fiducial cosmology in
Table \ref{tab:fidcosmo}. The $68\%$ and $95\%$ CIs are indicated by
the shaded regions. We plot, for reference,
\mbox{$b_{\rm 2D}=r_{\rm 2D}=1$} as dotted lines.  With the exception
of $r_{\rm 2D}$ at \mbox{$\theta_{\rm ap}\approx6\,\rm arcmin$} for
high-$z$, the replicated and \cfhtlens~data agree very well for the
(normalised) ratio statistics $b_{\rm 2D}$ and $r_{\rm 2D}$. The
origin of the $r_{\rm 2D}$ misfit is unclear, but it may be a bias of
an overly smooth template or a systematic error in the shear data at
this particular angular scale.

The projected galaxy bias in Figs. \ref{fig:postpredlowz} and
\ref{fig:postpredhighz} nicely illustrates that the conflict with the
SAM galaxy bias is already present before our deprojection of the
ratio statistics; it has nothing to do with the specific templates in
our analysis. For comparison with the \cfhtlens~data points, we plot
the normalised statistics and their $68\%$ confidence intervals for
the H15-based mock galaxy and shear catalogues without shape noise,
used in SH18, as green solid lines (cosmological parameters as in the
Millennium Simulation).  Throughout, the H15 data for $b_{\rm 2D}$ are
systematically lower than \cfhtlens~on all angular scales. The
correlation factor, $r_{\rm 2D}$, on the other hand, is broadly
similar to \cfhtlens, with the notable exception of SM1
high-$z$. Similar data for L12 are not available to us. However, we do
not expect their predictions to be significantly different to H15 due
to their similar biasing functions for $k\lesssim5\,h\,\rm Mpc^{-1}$
(cf. green data points in Figs. \ref{fig:bk_lowz} to
\ref{fig:rk_highz}).

\subsection{Galaxy bias on linear scales}
\label{sect:gallin}

\begin{figure*}
  \begin{center}
    \includegraphics[width=175mm]{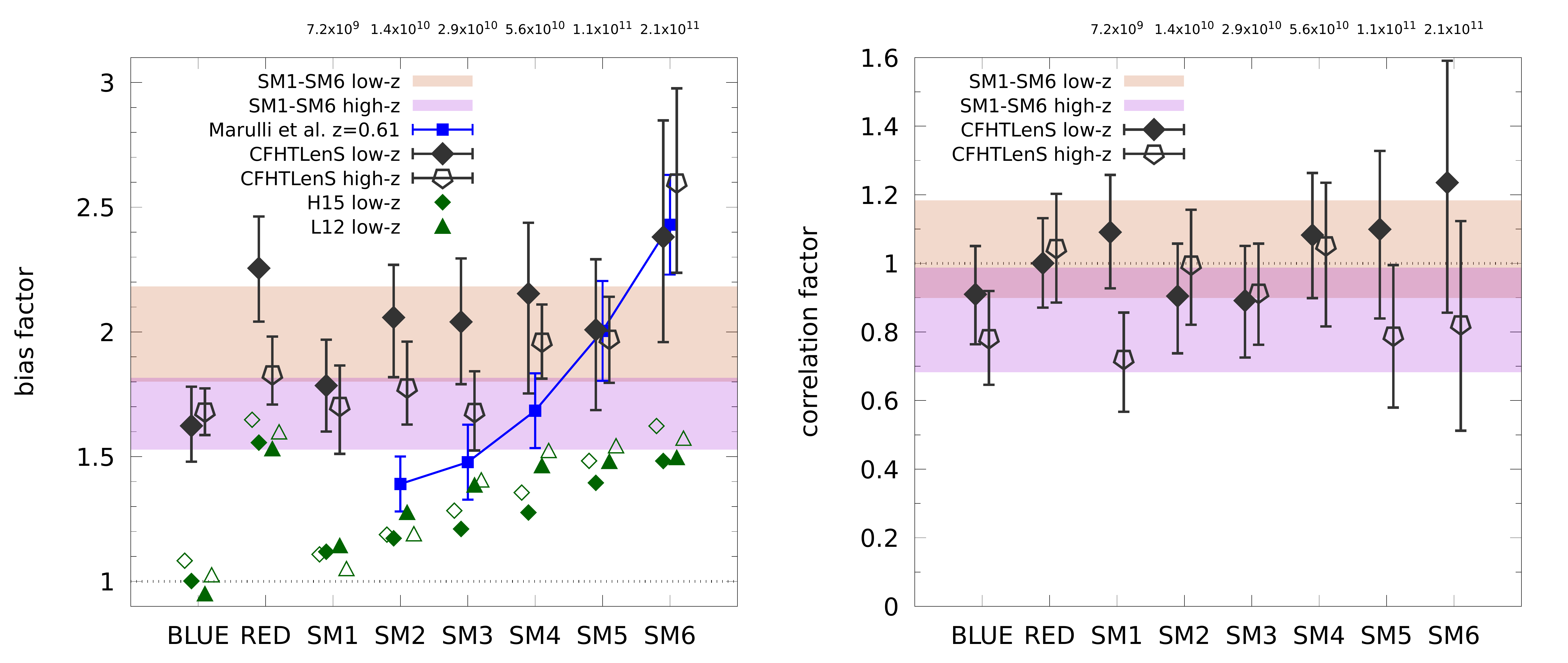}
  \end{center}
  \caption{\label{fig:brls_results} Large-scale bias $b_{\rm ls}$
    (left panel) and correlation $r_{\rm ls}$ (right panel) for the
    galaxy populations indicated on the $x$-axis. The black data
    points are \cfhtlens~results (medians and $68\%$ CIs), the green
    data points in the left panel are SAM predictions (diamonds for
    H15, triangles for L12, filled symbols for low-$z$, and open
    symbols for high-$z$). The shaded areas are $68\%$ CIs for
    \cfhtlens, merging SM1 to SM6 in an extra analysis. The blue
    squares are \mbox{VIPERS} $b_{\rm ls}$ for stellar-mass binned
    galaxies with \mbox{$i_{\rm AB}<22.5$} and median $\bar{z}=0.61$
    by \citet{2013A&A...557A..17M}.}
\end{figure*}

\begin{figure*}
  \begin{center}
    \includegraphics[width=175mm]{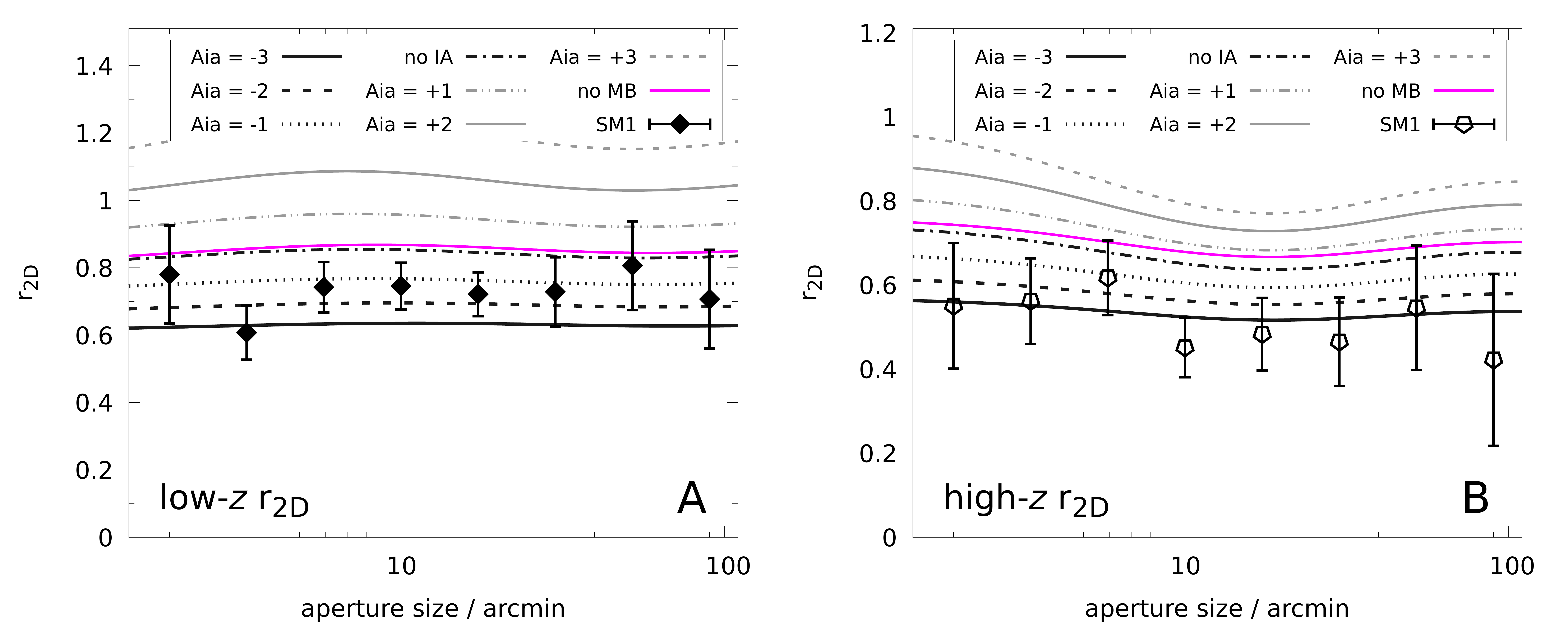}
  \end{center}
  \caption{\label{fig:rlin_ai} Ratio statistics $\hat{r}_{\rm 2D}$ as
    a function of aperture size. The data points with error bars
    (jackknife $68\%$ errors) are for SM1 low-$z$ (left panel A) and
    SM1 high-$z$ (right panel B). The lines are predictions for
    unbiased galaxies with varying values of
    \mbox{$A_{\rm ia}\in[-3,+3]$}. Our fiducial value is
    \mbox{$A_{\rm ia}=-0.48$}. The solid magenta line `no MB' is a
    model for \mbox{$A_{\rm ia}=0$} and no magnification.}
\end{figure*}

Our technique separates galaxy bias in the non-linear (one-halo)
regime from the linear stochastic bias $b_{\rm ls}$ and $r_{\rm ls}$
on large scales. The \cfhtlens~ results and model predictions are
plotted in Fig. \ref{fig:brls_results}. Diamonds with error bars in
the left panel are \cfhtlens~ posterior medians of $b_{\rm ls}$ with
$68\%$ CIs or $r_{\rm ls}$ in the right panel. Filled data points
denote the low-$z$ samples and open data points the high-$z$
samples. The corresponding galaxy samples are shown at the bottom of
each panel and the mean stellar masses at the top. The shaded
rectangular regions indicate the $68\%$ intervals of the average bias,
obtained from merged samples SM1-SM6 in an extra analysis.

On the one hand, for $b_{\rm ls}$ there are clear differences from the
model predictions, which are the green diamonds (H15) and triangles
(L12): model galaxies are less clustered relative to matter. On the
other hand, the relative bias $b_i/b_j$ for any two samples $i$ and
$j$ is similar for \cfhtlens~and the models. This can be seen when
multiplying the H15 or L12 values by roughly $1.5$, bringing them into
broad agreement with \cfhtlens. Furthermore, the models predict an
evolution of biasing from low-$z$ (open symbols) to high-$z$ (filled
symbols) that is too small for the \cfhtlens~errors.  Nevertheless,
our merged samples SM1-SM6 weakly indicate an evolution but with the
opposite trend to in the models, namely a decline in $b_{\rm ls}$
($68\%$ credibility) from $1.98^{+0.20}_{-0.18}$ at low-$z$ to
$1.65^{+0.17}_{-0.12}$ at high-$z$.

The blue squares in the left panel are non-lensing measurements of
$b_{\rm ls}$ by \citet{2013A&A...557A..17M} in the \mbox{VIMOS} Public
Extragalactic Redshift Survey (\mbox{VIPERS}) for galaxy samples
similar to ours at high-$z$ (Sect. \ref{sect:ng}). For their estimate
of $b_{\rm ls}$, the \mbox{VIPERS} samples were divided into three
stellar-mass bins, encompassing $10^{9.8}\,h_{70}^{-2}\,\msol$ to
$10^{11.8}\,h_{70}^{-2}\,\msol$, and the galaxy-clustering correlation
function between $1$ and $10\,h^{-1}\,\rm Mpc$ was normalised by the
theoretical clustering of dark matter for a flat $\Lambda\rm CDM$
universe ($\Omega_{\rm m}=0.25$, $\sigma_8=0.8$, $n_{\rm s}=1$). To
make these results comparable to ours, we interpolate the
\mbox{VIPERS} data points (their Table 4) to the mean stellar masses
of SM2 to SM6 by using a best-fitting second-order polynomial. Then,
except for our higher SM2 data point, the \mbox{VIPERS} results agree
well with our lensing-based high-$z$ measurements (open diamonds) and
conflict with the SAMs for the highest stellar masses SM5 and SM6.

The linear correlation factors in the right panel are $68\%$
consistent with a deterministic bias \mbox{$r_{\rm ls}=1$} (and the
SAMs), except for two cases: BLUE and, in particular, SM1 high-$z$
significantly fall below this value.  Since SM1 high-$z$ is an
interesting case, Fig. \ref{fig:rlin_ai} (right panel) plots its
$\hat{r}_{\rm 2D}$ as function of aperture size. The lines are
predictions for unbiased galaxies with IA amplitudes varying between
\mbox{$-3\le A_{\rm ia}\le3$}. Importantly, all \cfhtlens~ galaxies
with \mbox{$r_{\rm ls}=1$}, bias or unbiased, should for large
apertures be consistent with our fiducial value
\mbox{$A_{\rm ia}=-0.48$}. While this is clearly the case for SM1
low-$z$ (left panel), the data points of SM1 high-$z$ in the right
panel fall even below \mbox{$A_{\rm ia}=-3$} for
\mbox{$\theta_{\rm ap}\gtrsim30\,\rm arcmin$}, reflecting our low
value for $r_{\rm ls}$ in the Bayesian interpretation of the aperture
statistics.

\subsection{Template parameters}

Our main results are the reconstructed biasing functions for
\cfhtlens~in Figs. \ref{fig:bk_lowz} to
\ref{fig:rk_highz}. Nevertheless, marginal constraints for the
template parameters in the one-halo regime (i.e. $b(m)$, $V(m)$,
$f_{\rm cen}$, $\zeta$, and $m_{\rm piv}$) are also reported in
Appendix \ref{app:template_prms} and are briefly summarised
here. These inform us about physical galaxy parameters, albeit with
significantly lower accuracy compared to the reconstructed $b(k)$,
$r(k)$, and their asymptotic values $b_{\rm ls}$ and $r_{\rm ls}$ on
large scales.

Typically, we find only weak constraints for HOD-related template
parameters in the \cfhtlens~ data, yet differences to the SAMs are
visible for $b(m)$: \cfhtlens~ galaxies of same stellar mass seemingly
prefer to populate halos of higher mass when compared to the true
$b(m)$ of the SAMs. However, this shift of the inferred \cfhtlens~
$b(m)$ towards higher halo masses is at least partly an artefact of
the crude template modelling because it is also visible in the
simulated reconstruction of biasing functions in SH18, using H15-based
mock data.  We also note here that the specific results for $b(m)$
(and $V(m)$) rely on the adopted halo-mass spectrum $n(m)$ which
differs for \cfhtlens~ and the Millennium Simulation due to different
fiducial cosmologies. Furthermore, the normalised excess variance
$V(m)$ is mostly consistent with a Poisson variance of galaxy numbers
inside halos ($V(m)=0$) or possibly a sub-Poisson variance ($V(m)<0$)
around $10^{13}\lesssim m_{\rm piv}\lesssim10^{14}\,h^{-1}\,\rm\msol$
in some cases, SM6 high-$z$ for instance. In comparison, the $V(m)$
for the SAMs also indicates a sub-Poisson variance but again offset to
somewhat lower halo masses. For high halo masses
$m\gtrsim10^{14}\,h^{-1}\,\msol$ and stellar masses below SM5 the
variance of galaxy numbers $N(m)$ becomes super-Poisson ($V(m)>0$) for
H15. This is neither seen in L12 nor visible in our \cfhtlens~
results.  Then the distribution of satellites is consistent with that
of matter inside halos ($\zeta\sim1$) within typical errors of
roundabout $\pm0.45$ ($68\%$ CI). And, finally, the mixing parameter
of templates with and without central galaxies, $f_{\rm cen}$, falls
broadly within $20\%$ and $80\%$, including the high values of
$70\%\lesssim f_{\rm cen}\lesssim90\%$ in the SAMs.

\section{Discussion}

\begin{figure}
  \begin{center}
    \includegraphics[width=95mm]{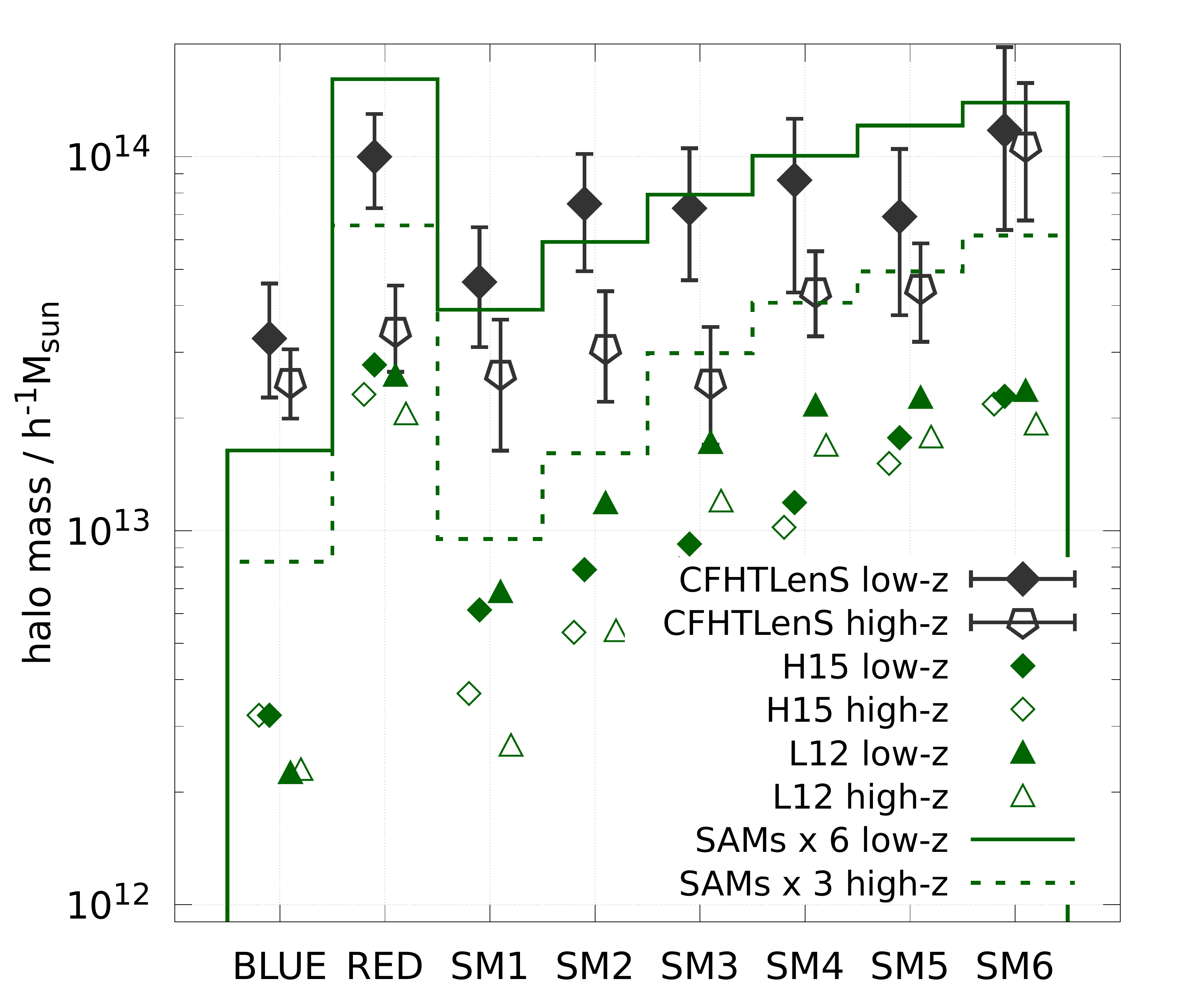}
  \end{center}
  \caption{\label{fig:mhalo} Characteristic halo mass $m_\ast$ that
    corresponds to the large-scale bias $b_{\rm ls}$ in
    Fig. \ref{fig:brls_results} for the halo bias model by
    \cite{2005ApJ...631...41T} and the Millennium Simulation cosmology
    as reference (Sect.  \ref{sect:MR}). Data points with error bars
    indicate the \cfhtlens~galaxy samples ($x$-axis) and the green
    points are the SAM predictions. The green lines average the SAM
    values and rescale them by a factor of $6$ (solid line for
    low-$z$) or $3$ (dashed line for high-$z$).}
\end{figure}

Our results for the galaxy bias in Fig. \ref{fig:brls_results} and
Figs. \ref{fig:bk_lowz}-\ref{fig:rk_highz} (Appendix) clearly show a
conflict with the SAM predictions by H15 (Munich model) and L12
(Durham model). This conflict hints at inaccuracies in the SAM
modelling of galaxy physics through its connection with galaxy bias
\citep[e.g.][]{2004ApJ...601....1W,1999ApJ...522..590B}. There is no
clear preference for any model by \cfhtlens, despite the inter-model
differences on small scales, most prominent for $b(k)$ in SM3 to SM5.
We discuss our findings and possible systematic errors below.

The linear bias factor $b_{\rm ls}$ is overall $50\%$ higher relative
to the models due to a stronger clustering of galaxies in
\cfhtlens. This may highlight modelling problems for
\mbox{$z\gtrsim0.3$} that affect all stellar masses.  We measure for
$b(k)$ and $b_{\rm ls}$ the amplitude ratio of galaxy and matter
clustering in the data. Comparing the amplitudes of $\ave{{\cal N}^2}$
in \cfhtlens~ on scales
\mbox{$4^\prime\le\theta_{\rm ap}\le40^\prime$} to the H15 mock data
in SH18, we find typical values in \cfhtlens~ that are $20\%$ to
$40\%$ higher. This conflict is further increased in $b_{\rm ls}$ by a
weaker clustering of matter in \cfhtlens,
\mbox{$\Sigma_8=\sigma_8\sqrt{\Omega_{\rm m}/0.27}\approx0.76$}, relative
to that in the Millennium Simulation, \mbox{$\Sigma_8\approx0.87$}.
Problems with galaxy clustering in the Durham and Munich models are
known for low-mass galaxies at $z\lesssim0.1$ on small scales of
\mbox{$\sim1\,h^{-1}\,\rm Mpc$} \citep{2017MNRAS.469.2626H,
  2016MNRAS.458..934V, 2015MNRAS.454.2120F, 2014MNRAS.437.3385K}.
According to Fig. \ref{fig:brls_results}, supported by the
\mbox{VIPERS} data for SM3-SM6, this problem deepens in the range
\mbox{$0.3\lesssim z\lesssim0.6$} and then also affects linear scales
and stellar masses as high as
\mbox{$M_\ast\sim10^{11}\,h_{70}^{-2}\,\msol$}. In a halo-model
picture the lower $b_{\rm ls}$ suggests that H15 or L12 galaxies tend
to reside inside matter halos of systematically low mass, decreasing
the clustering amplitude on linear scales
\citep{2002ApJ...575..587B}. At first sight this seems to be supported
by the constraints of our template parameters for
\mbox{$b(m)\propto\ave{N|m}/m$} (Fig.  \ref{fig:summary_bm}), but we
caution here that the templates, which are employed for a different
purpose, have not been tested for their ability to truthfully
reconstruct $b(m)$.  Clearly visible in Fig. \ref{fig:ng}, however, is
that the SMF of the SAMs in the two redshift bins closely matches that
in \cfhtlens. First of all, this supports a selection of galaxy
populations compatible with \cfhtlens; secondly, if there are
differences in the SAMs' $\ave{N|m}$, then they have to be such that
we nevertheless obtain similar galaxy number densities
$\bar{n}_{\rm g}=\int_0^\infty\d m\,n(m)\,\ave{N|m}$. In other words,
variations $\delta\ave{N|m}$ between the SAMs and \cfhtlens~ have to
satisfy $\int_0^\infty\d m\,n(m)\,\delta\ave{N|m}\approx0$. The
large-scale bias factor, on the other hand, can be changed by
$\delta b_{\rm ls}=\bar{n}_{\rm g}^{-1}\int_0^\infty\d m\,n(m)\,b_{\rm
  h}(m)\,\delta\ave{N|m}$ while keeping $\bar{n}_{\rm g}$ fixed. To
identify HOD variations between \cfhtlens~ and the SAMs, permitted by
this $\delta\ave{N|m}$ constraint, let us crudely assume that all
galaxies in a sample live inside halos of characteristic mass
$m_\ast$, hence exhibiting the large-scale bias
$b_{\rm ls}=b_{\rm h}(m_\ast)$. In Fig. \ref{fig:mhalo}, we explore
the relative $m_\ast$ differences needed to explain the $b_{\rm ls}$
of observational data and models, using the same reference cosmology
for both to make them comparable (see that of the Millennium
Simulation, Sect.  \ref{sect:MR}). For the halo bias factor
$b_{\rm h}(m)$, we employ \cite{2005ApJ...631...41T} at the mean
redshift of a galaxy sample.  On average, the $m_\ast$ in \cfhtlens~
is broadly a factor of $6(3)$ higher than the average SAM at low-$z$
(high-$z$), as indicated by the solid (dashed) line; in the SAMs
$m_\ast$ increases from around $4\times10^{12}\,h^{-1}\,\msol$ for SM1
to $2\times10^{13}\,h^{-1}\,\msol$ for SM6. In conclusion, our results
hint at problems in H15 and L12 for $z\gtrsim0.3$. These problems may
be related to inaccuracies in their HOD, preferring parent halos of
$3$ to $6$ times lower mass than in reality. This view, however,
assumes well-controlled systematic errors in our galaxy selection and
the reconstruction of galaxy bias.

In this respect it is conceivable that the conflicting results might
also indicate large systematic errors in our galaxy bias amplitude
combined with an inconsistent selection of SAM galaxies, such as
systematically low \cfhtlens~ stellar masses.  The estimated
systematic RMS error in $b_{\rm ls}$ is somewhere within
$[6.8\%,13.5\%]$ (Sect. \ref{sect:normerrors}). This is well below the
$b_{\rm ls}$ conflict of \mbox{$\sim50\%$}, and therefore cosmology or
redshift errors alone probably do not explain the model conflict. But
two other systematic errors in the galaxy sample selection can also
bias our results.  Firstly, our flux limit of \mbox{$i^\prime<22.5$}
might remove too many faint weakly clustered galaxies in \cfhtlens~ if
the SAM $i^\prime$-band magnitudes are too faint relative to
\cfhtlens~ (the flux limit is also applied to the SAMs). Even then the
SAM conflict for the massive (luminous) galaxies
\mbox{$M_\ast\gtrsim10^{11}\,h_{70}^{-2}\,\msol$} is hard to explain
where the flux limit is irrelevant. In addition, the SAM luminosity
functions give good fits to observations for \mbox{$z\lesssim1$}
\citep[e.g.][]{2006MNRAS.370..645B}; the Durham model even uses
luminosity functions in two bands for the model calibration. We note
that the low-mass samples SM1 and SM2 are especially affected by the
flux limit, implying a higher value for $b_{\rm ls}$ compared to
volume-limited stellar-mass samples because the bias factor increases
with galaxy luminosity
\citep[e.g.][]{2005ApJ...630....1Z,2002MNRAS.332..827N}.  More
relevant could be a second selection bias: stellar-mass estimates
might be systematically low in \cfhtlens. Shifting the data points in
the left panel of Fig. \ref{fig:brls_results} by
\mbox{$\Delta\log_{10}M_\ast=+0.6\,\rm dex$} gives a better match to
the SAM predictions for $b_{\rm ls}$ (e.g. SM1 to SM3, SM2 to SM4).
On the other hand, a stellar-mass error that large is hard to justify
considering previous error estimates \citep[Section 2.1
of][]{2014MNRAS.437.2111V}, and considering that the \mbox{VIPERS}
results for both $b_{\rm ls}$ and $\bar{n}_{\rm g}$, using full
spectroscopic information, are comparable to ours in the high-$z$ bin
(Marulli et al. in Fig. \ref{fig:brls_results}; Davidzon et al.  in
Fig. \ref{fig:ng}). In addition, the stellar-mass functions of the
SAMs and \cfhtlens~ in Fig. \ref{fig:ng} are good matches, which is
unlikely for systematically different \cfhtlens~ stellar
masses. Nevertheless, a large yet realistic stellar-mass error of
\mbox{$\sim0.3\,\rm dex$} combined with a very large systematic error
of \mbox{$\sim25\%$} for $b_{\rm ls}$ could in principle resolve the
SAM conflict for the large-scale galaxy clustering. A systematic error
of such magnitude in $b_{\rm ls}$, however, would imply that all our
results for \mbox{$r_{\rm ls}\approx1$} in Fig. \ref{fig:brls_results}
have to be offset by about $15\%$, which {would then mildly
  conflict with a deterministic galaxy bias on large scales};
systematic errors in $r_{\rm ls}$ are typically $50\%$ to $70\%$ of
those in $b_{\rm ls}$ (Appendix \ref{sect:normerror}).

For systematic errors within the estimated range, our results for
$r_{\rm ls}$ agree with the SAM picture of a deterministic galaxy
bias. Only the high-$z$ samples SM1 and BLUE are weak evidence for a
stochastic linear bias, which will be tested with upcoming survey
data. Although the biasing functions $b(k)$ and $r(k)$ generally do
not distinguish non-linear deterministic bias from a stochastic bias,
the linear regime is an exception, where
\mbox{$\tilde{\delta}_{\rm g}\approx b_{\rm ls}\,\tilde{\delta}_{\rm
    m}+\epsilon$} and $r_{\rm ls}$ is only sensitive to the shot-noise
corrected stochastic scatter $\epsilon$
\citep{1998ApJ...500L..79T,1999ApJ...520...24D}.  For this regime, and
after the IA (and magnification) correction, the data points in the
right panel of Fig. \ref{fig:brls_results} establish, in agreement
with the SAMs, a deterministic bias \mbox{$r_{\rm ls}=1$} for
\cfhtlens~ galaxies on linear scales. More interestingly, and contrary
to the SAMs, the samples SM1 and BLUE high-$z$ prefer an average
stochastic bias
\mbox{$r_{\rm ls}\approx0.75\pm0.14\,{\rm (stat.)}\pm0.06\,{\rm
    (sys.)}$} at $68\%$ credibility.  Some stochasticity of blue
galaxies, \mbox{$r\approx0.9$} on scales of
\mbox{$15\,h^{-1}\,\rm Mpc$}, is supported by other observations, but
it is unclear if this is still present on linear scales and for our
redshift regime
\citep{2016MNRAS.460.1310P,2008MNRAS.385.1635S,2005MNRAS.356..247W,2000ApJ...544...63B}. At
least for the CMB lensing results of $i$-band limited galaxies by
\cite{2016MNRAS.456.3213G}, a linear stochastic bias is one
interpretation (Section 7.4). Unfortunately, our new evidence for
stochasticity ($2\sigma$) is further weakened by the systematic
amplitude uncertainty of $[5.2\%,8.5\%]$, quoted above as `sys.'  for
the pessimistic $8.5\%$ (Sect. \ref{sect:normerrors}). In the near
future we anticipate better constraints from the Kilo Degree Survey
\citep{2013ExA....35...25D}, the Hyper Suprime-Cam Survey
\citep{2018PASJ...70S...4A}, or the Dark Energy Survey
\citep{2016PhRvD..94b2001A} which will have ten times the survey area
of \cfhtlens~ and a better control of galaxy redshift distributions,
which are, {in addition to IA}, the major source for the systematic
error in $r_{\rm ls}$ (bottom panels in Fig. \ref{fig:normerror}).

Turning to the non-linear regime, \cfhtlens~ galaxy biasing clearly
varies with scale, stellar mass, and colour, but differently from the
SAMs and with no clear preference for either H15 or L12. Nevertheless,
there is a qualitative agreement with the models insofar as the bias
factor $b(k)$ increases on non-linear scales
\mbox{$k\sim10\,h\,\rm Mpc^{-1}$} relative to linear scales by up to a
factor of two (Figs. \ref{fig:bk_lowz} and \ref{fig:bk_highz}). The
relative increase of $r(k)$ is weaker, typically a factor of $1.2$
(Figs. \ref{fig:rk_lowz} and \ref{fig:rk_highz}). These relative
changes are insensitive to systematic errors in our method which
offset the biasing functions as a whole. Physically the $b(k)$
increase is probably related to central galaxies inside halos that
have \mbox{$\ave{N|m}\lesssim1$} and dominate the signal for
sufficiently large $k$ (Section 5.3 in SH18). Central galaxies affect
$r(k)$ in a similar way, like probably the stellar mass samples of
\cfhtlens~ at low-$z$. For other samples, however, where the $r(k)$
increase is small or absent in contrast to the SAMs, this
interpretation is less clear, RED and BLUE high-$z$ for instance. For
a physical interpretation of these observations, an accurate
halo-model fit \citep[e.g.][]{2018MNRAS.479.1240D,2005PhRvD..71d3511S}
or a study of $b(k)$ and $r(k)$ variations with respect to SAM
parameters is needed.

Finally, in a broader context, the possibility of a stochastic galaxy
bias, IA distortions, or magnification effects on linear scales is
also relevant for the popular cosmological probe $E_{\rm G}$ for
gravity \citep{2010Natur.464..256R}. This probe relies on a
deterministic bias, whereas a stochastic bias changes the value of
$E_{\rm G}$ to $r_{\rm ls}\,E_{\rm G}$. Additionally, lensing
magnification changes $E_{\rm G}$ through the galaxy-galaxy lensing
(GGL) amplitude by up to several per cent, depending on type and mean
redshift of lens galaxies (Figure 4 in SH18). Likewise, GI
correlations bias the GGL signal if the distributions of lenses and
sources overlap in radial direction \citep{2004PhRvD..70f3526H}. SH18
estimate the GI bias of GGL, and thus of $E_{\rm G}$, to up to $5\%$
for \mbox{$-2\lesssim A_{\rm ia}\lesssim-1$} and lens, source samples
similar to ours (Figure 3 in SH18). While this GI contamination can in
principle be reduced to zero by avoiding any overlap, the
magnification bias is always present and has to be accounted for by
applying the method in \cite{2019arXiv191006400U}, for
example. Importantly, our work suggests not to use BLUE (star-forming)
or SM1-like galaxies as lenses for $E_{\rm G}$ due to the possibility
of a stochastic bias.

As online material attached to this paper, we provide Monte Carlo
realisations of the posterior distributions of $b(k)$ and $r(k)$ for
every galaxy sample in our analysis. The scatter among the
realisations reflects the distribution of their statistical errors,
and the full set of realisations can be utilised to propagate this
uncertainty in future studies.

\section*{Acknowledgements}

We thank Christopher Bonnett for fruitful discussions at an early
stage of this project, and we thank Peter Schneider and Catherine
Heymans for comments on the paper manuscript. This work has been
supported by Collaborative Research Center TR33 `The Dark Universe'
and by the Deutsche Forschungsgemeinschaft through the project SI
1769/1-1.

This work is based on observations obtained with MegaPrime/MegaCam, a
joint project of CFHT and CEA/DAPNIA, at the Canada-France-Hawaii
Telescope (CFHT) which is operated by the National Research Council
(NRC) of Canada, the Institut National des Sciences de l'Univers of
the Centre National de la Recherche Scientifique (CNRS) of France, and
the University of Hawaii. This research used the facilities of the
Canadian Astronomy Data Centre operated by the National Research
Council of Canada with the support of the Canadian Space Agency.
CFHTLenS data processing was made possible thanks to significant
computing support from the NSERC Research Tools and Instruments grant
program.

\bibliographystyle{aa}
\bibliography{cfhtlens-galbias}

\appendix
\counterwithin{figure}{section}

\section{Reconstructed biasing functions}

\begin{landscape}
\begin{figure}
  \centering
  \includegraphics[scale=0.22,clip=false,angle=0]{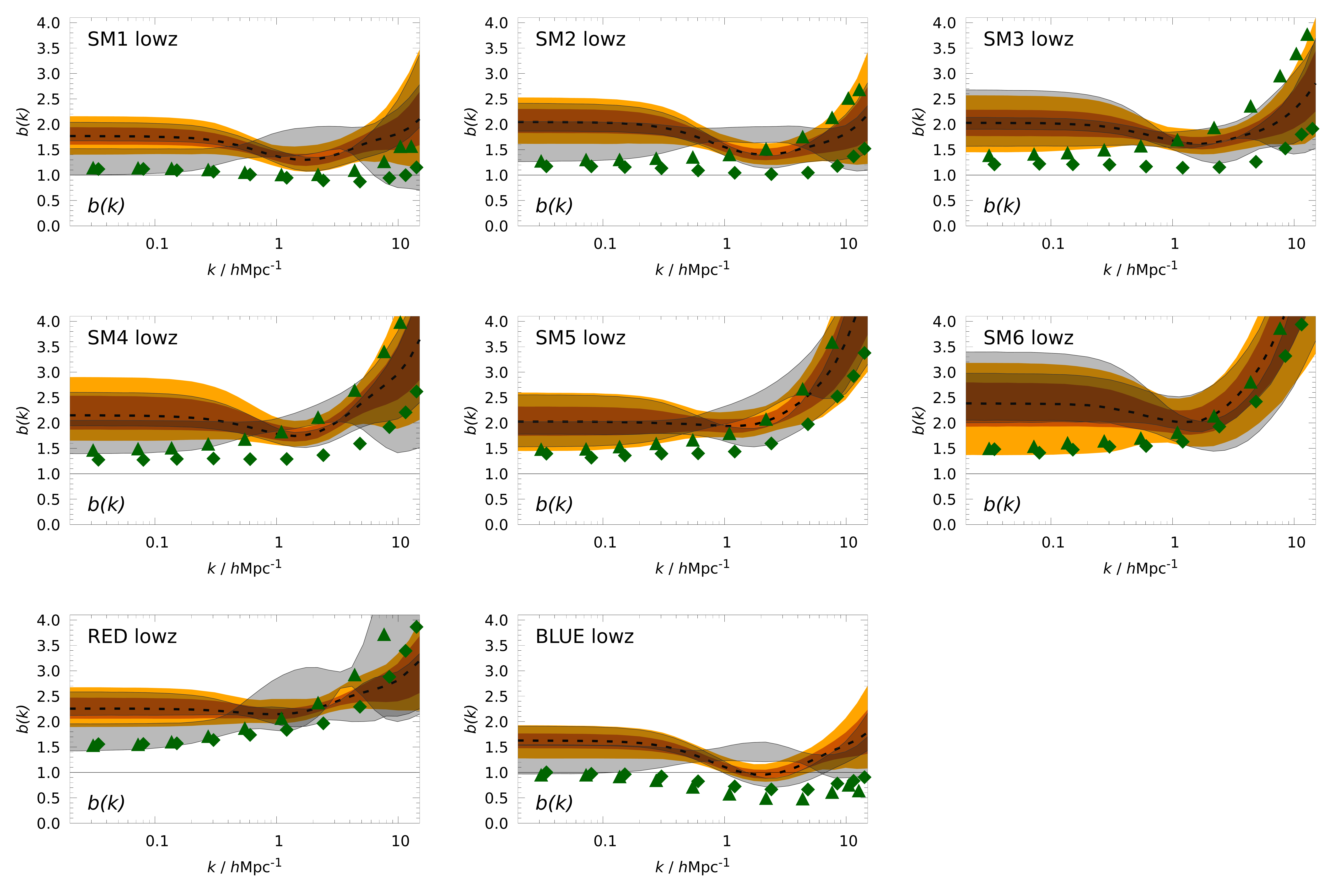}
  \caption{\label{fig:bk_lowz} Posterior scale-dependent bias $b(k)$
    for galaxy samples at redshift $\bar{z}\approx0.35$. The galaxy
    sample is indicated in the top left corner. The orange filled regions
    and the dashed lines show the median, 68\%, and 95\% credible
    region for the \cfhtlens~measurement using the full source sample
    BACK low-$z$. The overlapping transparent grey regions show 68\%
    regions that are based on the source subsamples BACKa and BACKb
    low-$z$. These noisier measurements at mean source redshifts
    $\bar{z}\approx0.86$ and $\bar{z}\approx1.12$ test the robustness
    of the measurement. The filled green diamonds and triangles
    indicate the SAM predictions by H15 and L12,
    respectively.}
\end{figure}
\end{landscape}

\begin{landscape}
\begin{figure}
  \centering
  \includegraphics[scale=0.22,clip=false,angle=0]{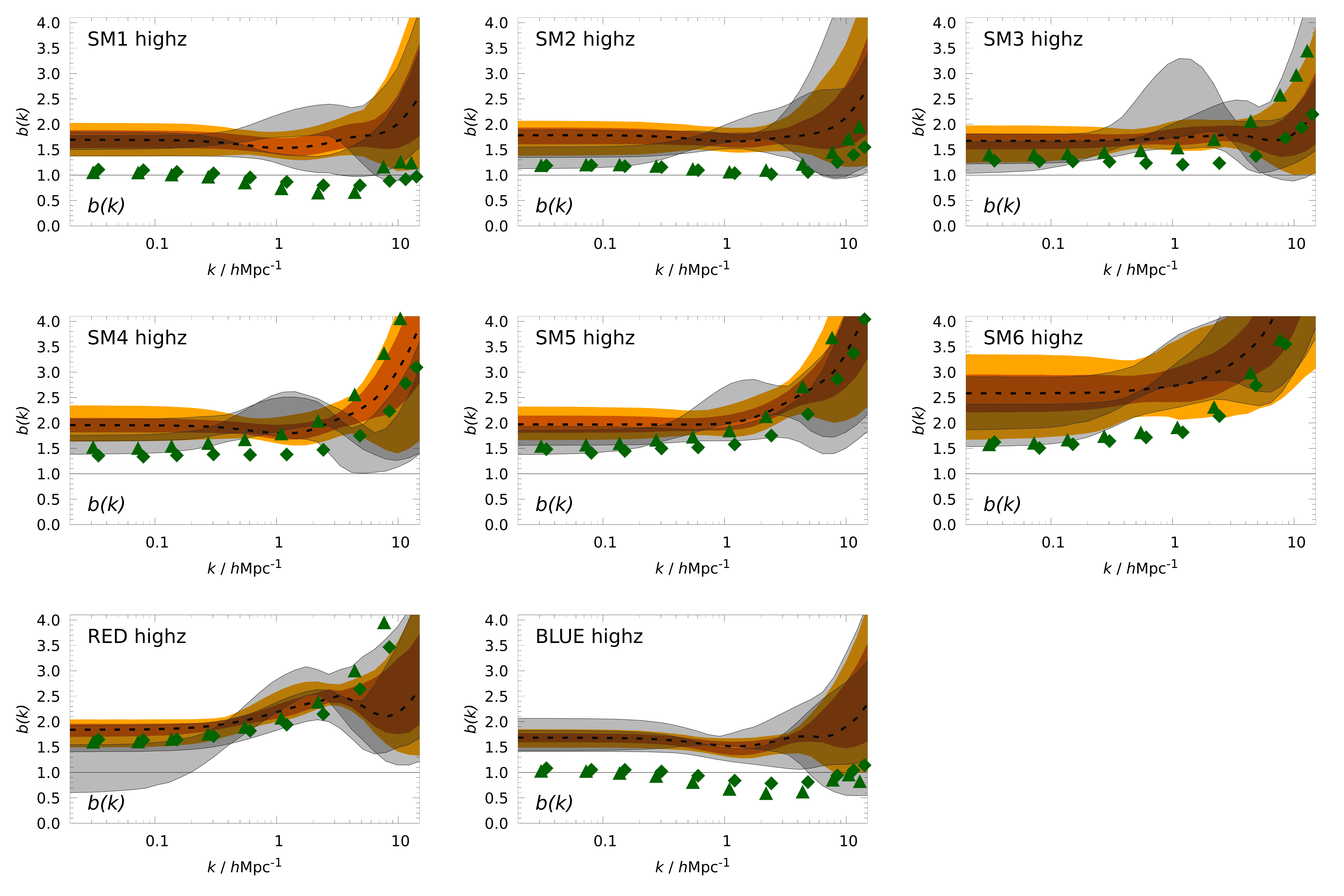}
  \caption{\label{fig:bk_highz} As in Fig. \ref{fig:bk_lowz}, but   for
    the scale-dependent bias $b(k)$ of the high-$z$ galaxy samples
    ($\bar{z}\approx0.52$). The grey regions depict 68\% credible
    regions, based on the source subsamples BACKa and BACKb high-$z$
    ($\bar{z}\approx0.93,1.14$). The green data points
    are SAM predictions.}
\end{figure}
\end{landscape}

\begin{landscape}
\begin{figure}
  \centering
  \includegraphics[scale=0.22,clip=false,angle=0]{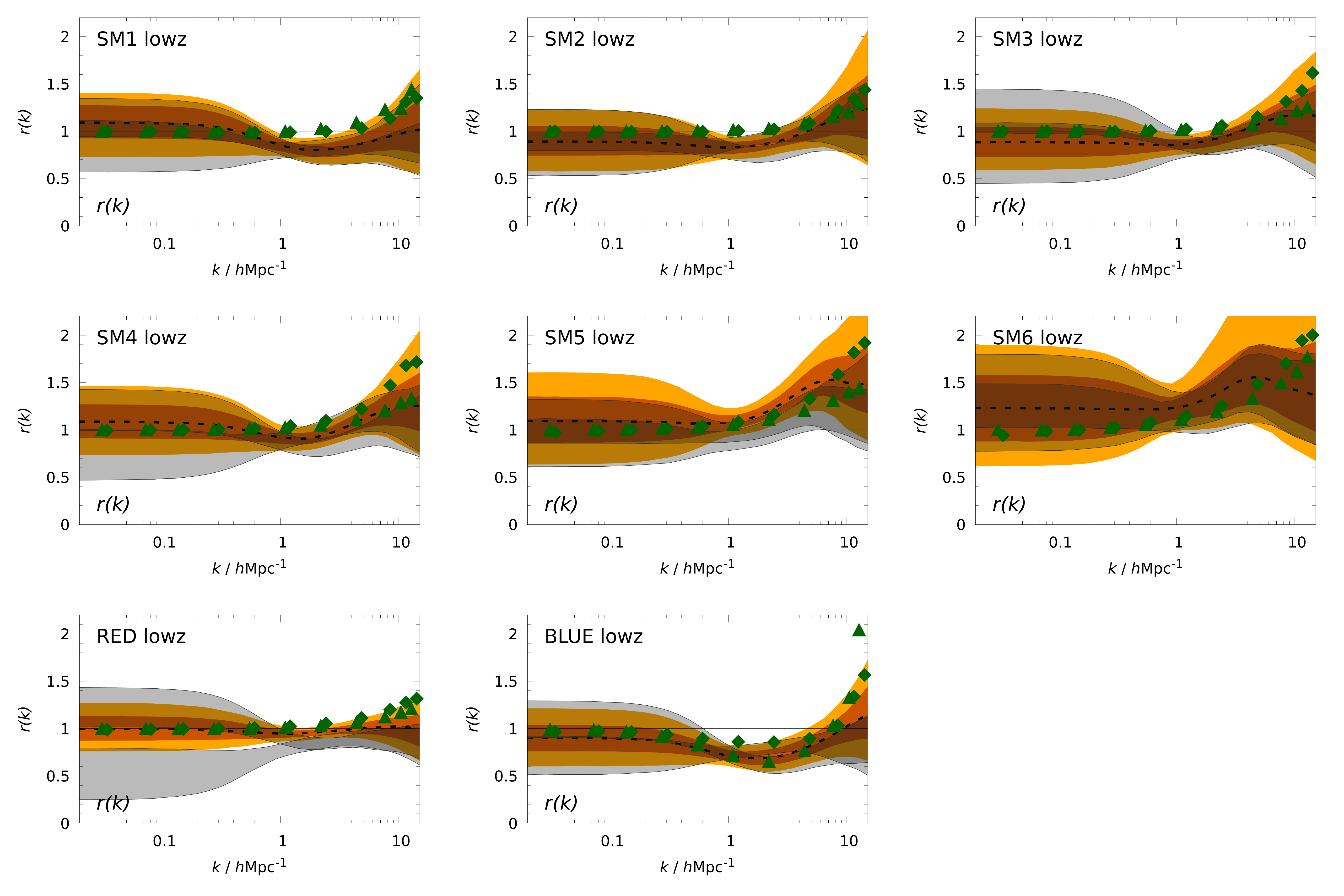}
  \caption{\label{fig:rk_lowz} As in Fig. \ref{fig:bk_lowz}, but   for
    the scale-dependent correlation factor $r(k)$ of the low-$z$
    galaxy samples ($\bar{z}\approx0.35$). The grey 68\% regions are
    based on two source subsamples. The green data points are SAM
    predictions.}
\end{figure}
\end{landscape}

\begin{landscape}
\begin{figure}
  \centering
  \includegraphics[scale=0.22,clip=false,angle=0]{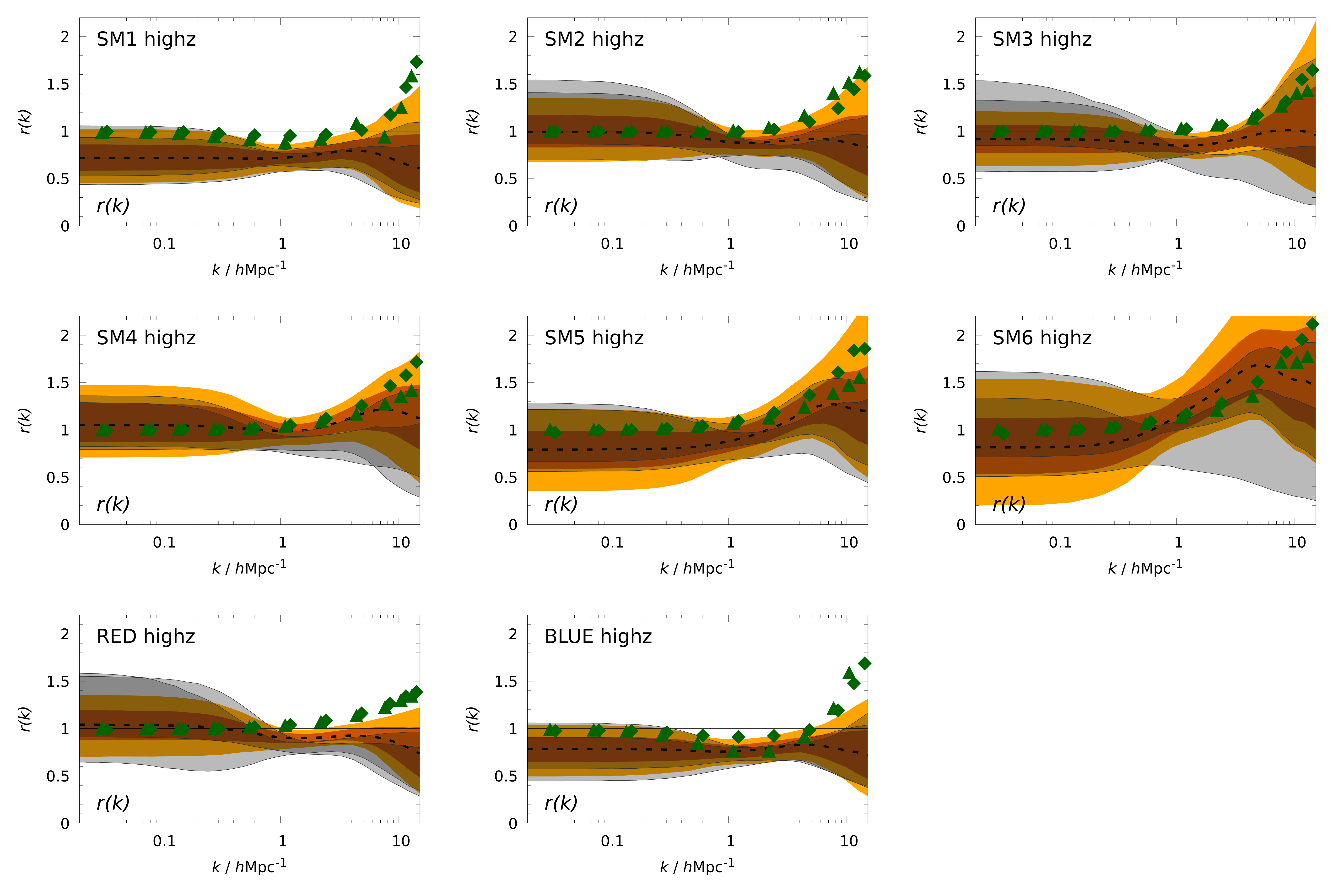}
  \caption{\label{fig:rk_highz} As in Fig. \ref{fig:bk_lowz}, but   for
    the scale-dependent correlation factor $r(k)$ of the high-$z$
    samples ($\bar{z}\approx0.52$). The grey 68\% regions are based on
    two source subsamples. The green data points are SAM predictions.}
\end{figure}
\end{landscape}

\section{Deconvolution of binned redshift distributions}
\label{sect:pofz_method}

To estimate the redshift distribution of our lens galaxies from the
binned \texttt{BPZ} posterior distribution we discuss here a
maximum-likelihood technique. Generally, the binning of a probability
density $p(z)\,\d z$ into \mbox{$i=0\ldots (N_{\rm bin}-1)$} bins,
\begin{equation}
  p_i:=\int\limits_{i\,\Delta z}^{(i+1)\,\Delta z}\d z\;p(z)
,\end{equation}
is not invertible with respect to $p(z)$. By restricting the inversion
for \mbox{$z>0$} to a family of densities of four parameters
\mbox{$\vec{\Theta}=(\bar{z},\sigma_{\rm z},s,k)$},
\begin{multline}
  \label{eq:gramcharlier}
  p(z|\lambda,\vec{\Theta})=\\
  \lambda\,
  \e^{-\frac{\epsilon}{z}}\,\e^{-\frac{x^2}{2}}\,
  \left(
    1+\frac{s}{6}He_3(x)+\frac{k}{24}He_4(x)
  \right)
  ~;~
  x=\frac{z-\bar{z}}{\sigma_{\rm z}}\;,
\end{multline}
however, the problem is tractable for our lens samples; we set
\mbox{$p(z|\lambda,\vec{\Theta})=0$} for \mbox{$z\le0$}.  The
normalisation $\lambda$ is defined through
$\int_0^\infty\d z\,p_{\rm d}(z|\lambda,\vec{\Theta})=1$, and the
functions \mbox{$He_3(x)=x^3-3x$} and \mbox{$He_4(x)=x^4-6x^2+3$} are
Hermite polynomials. This family of models \Ref{eq:gramcharlier} is a
Gram-Charlier series that approximates the narrow redshift
distributions of samples inside photometric redshift bins by a normal
distribution and corrections for a skewness or kurtosis. In addition,
the exponential function with \mbox{$\epsilon=10^{-2}$} asserts
\mbox{$p_{\rm d}(z|\lambda,\vec{\Theta})\to0$} for \mbox{$z\to 0$}.
This correction influences the distribution function only very close
to \mbox{$z=0$}. The aim of our technique is to determine the
parameters $\vec{\Theta}$ for given $p_i$.

\begin{figure}
  \begin{center}
    \includegraphics[width=85mm]{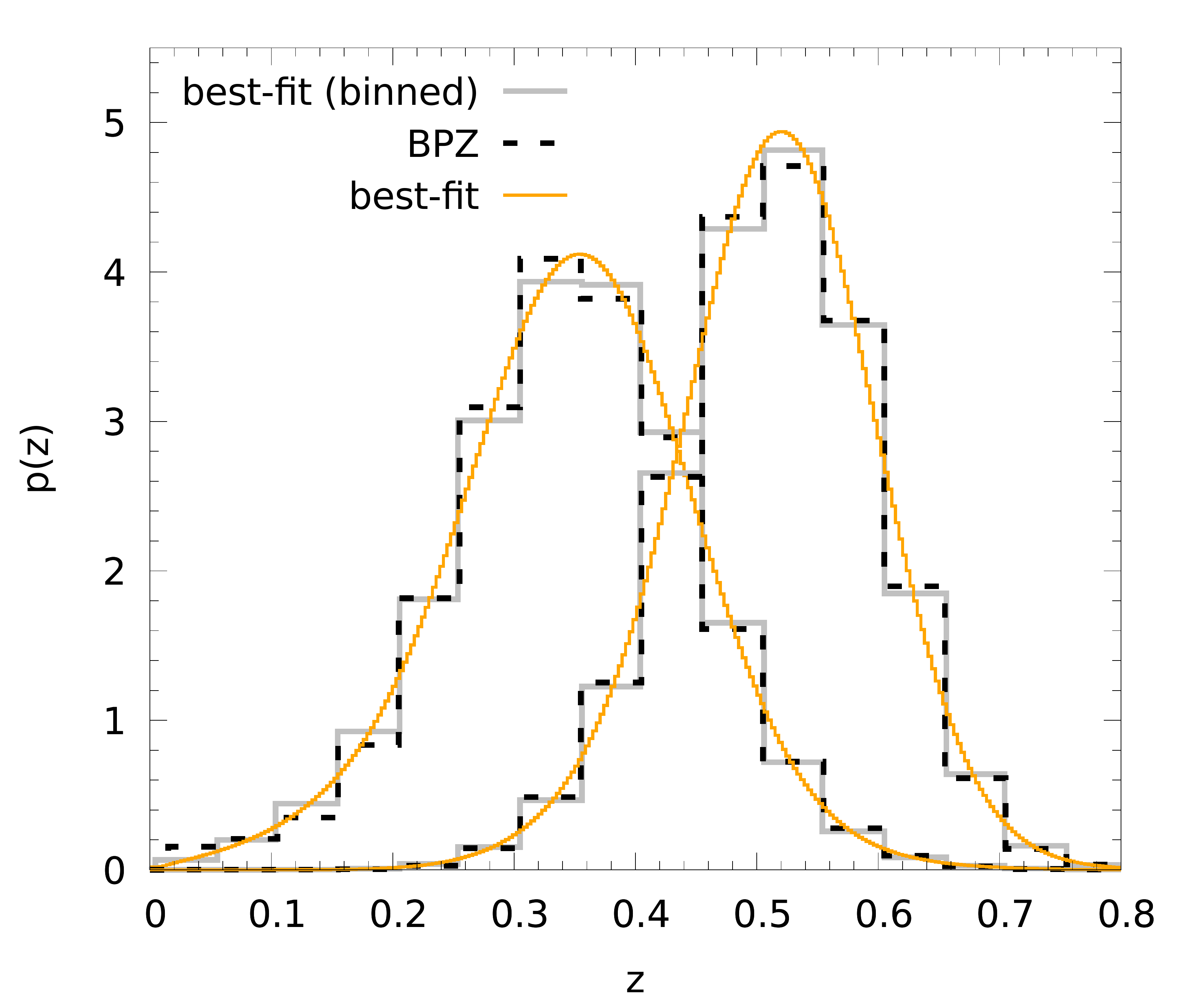}
  \end{center}
  \caption{\label{fig:pofzfit} Two examples of maximum-likelihood
    fits of models of $p(z)$ to the combined binned posterior of the
    \texttt{BPZ} redshift. The left-hand peak is for the low-$z$ bin
    of lenses SM1 to SM6 combined; the right-hand peak corresponds to
    the high-$z$ bin. The dashed histograms are the average
    \texttt{BPZ} posteriors of lenses; the solid histograms are the
    binned best fits. The smooth curves are the unbinned best fits as
    used in the analysis.}
\end{figure}

To determine the best-fitting parameters in \Ref{eq:gramcharlier} for
a given \texttt{BPZ} posterior, we find the binned distribution
\begin{equation}
  \hat{p}_i(\lambda,\vec{\Theta}):=
  \int\limits_{i\,\Delta z}^{(i+1)\,\Delta z}\d z\;p(z|\lambda,\vec{\Theta})
\end{equation}
that has the minimal Kullback-Leibler divergence with respect to
$p_i$, which means finding the $\vec{\Theta}_{\rm ml}$ that minimise
\begin{equation}
  d_{\rm KL}(\vec{\Theta})=
  -\sum_{i=0}^{N_{\rm bin}-1}p_i\ln{\left(
      \frac{\hat{p}_i(\lambda,\vec{\Theta})}{p_i}\right)}
  ={\rm const}-\sum_{i=0}^{N_{\rm bin}-1}p_i
  \ln{\hat{p}_i(\lambda,\vec{\Theta})}\;.
\end{equation}
These values are the maximum-likelihood parameters in the family of
model distributions \citep{knight1999mathematical}.

Figure \ref{fig:pofzfit} shows two examples of this fitting procedure
for a galaxy sample SM1-SM6 in the two photo-$z$ bins low-$z$ and
high-$z$. In this figure the dashed histograms are the binned
\texttt{BPZ} posteriors $p_i$, the solid histograms are the
best-fitting binned models $\hat{p}_i$, and the solid orange lines are
the best-fitting $p(z|\lambda,\Vec{\Theta})$   used in the analysis.

\section{$V_{\rm max}$-estimator for photo-$z$ selected galaxies}
\label{sect:vmaxdetails}

For our modification of the $V_{\rm max}$-estimator, we consider
galaxies that are selected in magnitude
\mbox{$m_{\rm min}\le m<m_{\rm max}$} and from a photo-$z$ interval
\mbox{$z_{\rm min}\le z_{\rm ph}<z_{\rm max}$}. Due to the latter
there is a non-negligible probability that   galaxies below
$z\le z_{\rm min}$ or above $z\ge z_{\rm max}$ are also selected;  the
selection thus does not clearly define a survey volume. To implement
the volume uncertainty in an estimator of the galaxy number density,
we assume that we know of every galaxy $i$ the probability density
$p_i(z|z_{\rm ph})$ of its true redshift $z$ given its photometric
redshift, and that we know the probability density $p_i(z_{\rm ph}|z)$
of photometric redshifts given the true galaxy redshift. We  first 
study how galaxies are selected.

To define the selection of galaxies, let
$p_i(z,\vec{\theta})\,\d V=C_i(z)\,S(\vec{\theta})\,\d V$ be the
probability that the $i$th galaxy can be observed inside a comoving
volume element $\d V$ in the line-of-sight direction $\vec{\theta}$ at
redshift $z$. The volume element is in a cosmological context
\begin{equation}
  \d V=
  f^2_K(z)\,
  \frac{\d\chi(z)}{\d z}\,
  \d\Omega\,\d z\;,
\end{equation}
where $\d\Omega$ is an infinitesimal solid angle on the celestial
sphere, $f_K(z)$ is the comoving angular diameter distance at $z$, and
$\chi(z)$ is the comoving distance of a galaxy at $z$. The radial
selection $C_i(z)$ depends on the survey selection function and the
intrinsic properties of the galaxy, and the angular selection
$S(\vec{\theta})$ shall be identical for all galaxies.

Due to the assumed independence of the angular selection on galaxy
properties and redshift, all incompleteness effects varying with the
position on the sky (e.g. gaps) are accounted for by the
effective survey area
\begin{equation}
  \label{eq:omegaeff}
  \Omega_{\rm eff}
  =\int\d\Omega\;S(\vec{\theta})
  =\int\limits_{-\pi/2}^{\pi/2}\int\limits_0^{2\pi}
  \d\delta\,\d\alpha\;\cos{\delta}\,
  S\!(\vec{\theta}(\alpha,\delta))\;;
\end{equation}
the right-hand side uses spherical $(\alpha,\delta)$ coordinates on
the sky. The effective area of our data, accounting only for gaps
($S=0$) or no gaps ($S=1$), is summarised in Table \ref{tab:omeff}.

For an estimator of the average number density, we assume a
homogeneous distribution of selected galaxies throughout the observed
volume and, in a preliminary step, imagine that we have only galaxies
identical to the $i$th galaxy; the comoving number density of these
galaxies shall be $\bar{n}_i$. Then the total detected number $N_i$ of
galaxies would be
\begin{equation}
  N_i=
  \int\d V\;\bar{n}_i\,C_i(z)\,S(\vec{\theta})=
  \bar{n}_i\,V_i
\end{equation}
with the effective volume
\begin{equation}
  \label{eq:Vi}
  V_i:=
  \Omega_{\rm eff}\,\int_0^\infty\d z\;C_i(z)\,
  f^2_K(z)\,\frac{\d\chi(z)}{\d z}\;.
\end{equation}
In other words, $N_i/V_i$ is, apart from sampling noise, an estimator
of $\bar{n}_i$. Since we have only $N_i=1$ galaxy $i$ in reality,
$1/V_i$ estimates its contribution to the total number density of all
$N_{\rm g}$ galaxies, which is
\begin{equation}
  \label{eq:nest}
  \bar{n}^{\rm est}_{\rm g}=
  \sum_{i=1}^{N_{\rm g}}\frac{1}{V_i}\;.
\end{equation}
The individual (radial) selection of the various galaxies is encoded
in the $V_i$. The original $V_{\rm max}$-method considers for each
$V_i$ the redshift interval over which a galaxy, similar to galaxy
$i$, would be selected.

This interval is not clearly defined for galaxies with
uncertain redshift, however , and we have to modify the original $V_{\rm max}$
approach by the following two steps. Firstly, for galaxies selected by
($i^\prime$-band) magnitude $m$ and photometric redshift $z_{\rm ph}$,
we calculate the uncertainty in the inferred absolute magnitude $M$ of
a galaxy.  Let $(m_i,z_{{\rm ph},i})$ be the apparent magnitude and
photometric redshift of galaxy $i$. The luminosity distance
$D_{\rm L}(z)=(1+z)\,f_K(z)$ and therefore the absolute magnitude $M$
of galaxy $i$ is uncertain for its given $z_{{\rm ph},i}$, reflected
by a probability $p(M|m_i,z_{{\rm ph},i})$. This $M$ probability is
given by marginalising the relation
\begin{equation}
  \label{eq:Mmz}
  M(m,z)=
  m-5\,\log_{10}{\left(\frac{D_{\rm L}(z)}{h^{-1}\,\rm Mpc}\right)}-25-K(z)-A(z)
\end{equation}
over the $z$-uncertainty $p_i(z|z_{{\rm ph},i})$,
\begin{equation}
  \label{eq:PM}
  p(M|m_i,z_{{\rm ph},i})=
  \int\limits_0^\infty\d z\;p_i(z|z_{{\rm ph},i})\,
  \delta_{\rm D}\Big(M-M(m_i,z)\Big)\;,
\end{equation}
where $\delta_{\rm D}(x)$ is the Dirac delta function. The
$K$-correction and dust extinction are assumed to be irrelevant here,
which means we set \mbox{$K(z)=A(z)=0$}.

Secondly, to account for the $M$ uncertainty in $V_i$ we assume an
ensemble of galaxies with an intrinsic $M$-distribution \Ref{eq:PM}
instead of one particular value $M$ as in the original
$V_{\rm max}$-estimator, and we consider their selection probability
when located at different $z$.  Placing this ensemble at $z$, we find
a joint distribution
\mbox{$p_i(M,z_{\rm ph}|z)=p(M|m_i,z_{{\rm ph},i})\times p_i(z_{\rm
    ph}|z)$} for $M$ and $z_{\rm ph}$.\footnote{This is not to be
  confused with the posterior distribution of $(M,z)$ of the observed
  galaxy $i$. Rather, we consider similar galaxies with intrinsic $M$
  variations at a $z$ different to $z_i$.} From this ensemble, we
select galaxies in intervals of $m$ and $z_{\rm ph}$ or, more
generally, with selection function \mbox{$0\le s(m,z_{\rm ph})\le
  1$}. Therefore, the selection probability of an $i$th-like galaxy at
$z$ is
\begin{equation}
  \label{eq:Ci}
  C_i(z)=
  \int\limits_0^\infty\d z_{\rm ph}\;\d M\,
  p_i(z_{\rm ph}|z)\,p(M|m_i,z_{{\rm ph},i})\,
  s(m(M,z),z_{\rm ph})\;,
\end{equation}
where $m(M,z)$ is $M(m,z)$ inverted with respect to $m$.  This allows
us to compute $V_i$ (Eq. \ref{eq:Vi}) and, if necessary, the density
$\bar{n}_{\rm g}$ (Eq. \ref{eq:nest}) of all galaxies.  

To calculate the integrals in an efficient way we suggest the
following numerical method. Using Eq. \Ref{eq:Ci}, the expression
\Ref{eq:Vi} reads
\begin{multline}
  V_i=
  \Omega_{\rm eff}\,
  \int\limits_{-\infty}^{+\infty}\d M\;p(M|m_i,z_{{\rm ph},i})\;
  \int\limits_0^\infty\d z\;
  \int\limits_0^\infty\,\d z_{\rm ph}\\
  \times f^2_K(z)\,\frac{\d\chi(z)}{\d z}
  \,p_i(z_{\rm ph}|z)\,
  s(m(M,z),z_{\rm ph})\;.
\end{multline}
Applying our specific hard selection $s(m,z_{\rm ph})\in\{0,1\}$ with
$[z_{\rm min},z_{\rm max}]\times[m_{\rm min},m_{\rm max}]$ to the
integral limits, we obtain
\begin{multline}
  V_i=
  \Omega_{\rm eff}\, \int\limits_{-\infty}^{+\infty}\d
  M\;p(M|m_i,z_{{\rm ph},i})\\
  \times\int\limits_{z(M,m_{\rm
      min})}^{z(M,m_{\rm max})}\!\!\!\!\d z\;
  f^2_K(z)\,\frac{\d\chi(z)}{\d z}
  \int\limits_{z_{\rm
      min}}^{z_{\rm max}}\!\d z_{\rm ph}
  \;p_i(z_{\rm ph}|z)\;,
\end{multline}
where the redshift $z(M,m)$ is implicitly defined by
Eq. \Ref{eq:Mmz}. We approximate the integration over $M$ by a
Monte Carlo integral that draws a set of \mbox{$j=1\ldots n_{\rm mc}$}
values \mbox{$M_{ij}\sim p(M|m_i,z_{{\rm ph},i})$} (Eq. \ref{eq:PM})
through \mbox{$z_{ij}\sim p(z|z_{{\rm ph},i})$} and
$M_{ij}=M(m_i,z_{ij})$ to obtain
\begin{equation}
  \label{eq:Viapprox}
  V_i\approx
  \frac{\Omega_{\rm eff}}{n_{\rm mc}}\sum_{j=1}^{n_{\rm mc}}
  \int\limits_{z(M_{ij},m_{\rm min})}^{z(M_{ij},m_{\rm max})}\!\!\!\d z\;
  f_K^2(z)\,\frac{\d\chi(z)}{\d z}\,
  F_i[z_{\rm min},z_{\rm max}|z]\;,
\end{equation}
where $F_i[z_{\rm min},z_{\rm max}|z]$ denotes the probability that
galaxy $i$ at redshift $z$ has
$z_{\rm ph}\in[z_{\rm min},z_{\rm max}]$:
\begin{equation}
  \label{eq:Fi}
  F_i[z_{\rm min},z_{\rm max}|z]:=
  \int_{z_{\rm min}}^{z_{\rm max}}\d z_{\rm ph}\; 
  p_i(z_{\rm ph}|z)\;.
\end{equation}
The approximation in Eq. \Ref{eq:Viapprox} reduces for
\mbox{$n_{\rm mc}\to\infty$} and \mbox{$z\equiv z_{\rm ph}$} (no
photo-$z$ errors) to the observable volume of a galaxy in the original
$V_{\rm max}$-estimator.

To speed up the computation of \Ref{eq:Viapprox}, we numerically
compute, for a broad range of $z$ values, the volume
\begin{equation}
  \label{eq:Viz}
  V_i[z]:=
  \Omega_{\rm eff}\,
  \int_0^z\!\!\!\d z^\prime\;
  f_K^2(z^\prime)\,\frac{\d\chi(z^\prime)}{\d z^\prime}\,
  F_i[z_{\rm min},z_{\rm max}|z^\prime]\;,
\end{equation}
which we then interpolate to obtain
\begin{equation}
  V_i\approx\frac{1}{n_{\rm mc}}\,
  \sum_{j=1}^{n_{\rm mc}}
  \left(V_i[z(M_{ij},m_{\rm max})]-V_i[z(M_{ij},m_{\rm min})]\right)\;.
\end{equation}
In the interpolation,  for all galaxies for the photo-$z$
errors, we adopt the (truncated) Gaussian model
\begin{equation}
  p_i(z_{\rm ph}|z)=
  \frac{\sqrt{2}}{\sqrt{\pi}\sigma_{\rm z}}
  \left[1+\erf{\frac{z}{\sigma_{\rm z}\sqrt{2}}}\right]^{-1}
  \e^{-\frac{(z_{\rm ph}-z)^2}{2\sigma_{\rm z}^2}}
\end{equation}
with \mbox{$\sigma_{\rm z}=(1+z)\,\delta_{\rm z}$} and
$\delta_{\rm z}=0.04$ \citep{2012MNRAS.421.2355H} for
\mbox{$z_{\rm ph}\ge0$}, and \mbox{$p_i(z_{\rm ph}|z)=0$}
otherwise. The factor in the square brackets with the error function
$\erf{x}$ is the normalisation due to the truncation of the normal
distribution below \mbox{$z_{\rm ph}=0$}. With this model of photo-$z$
errors, we can employ the same
\begin{multline}
  F_i[z_{\rm min},z_{\rm, max}|z]=\\
  \left[1+\erf{\frac{z}{\sigma_{\rm z}\sqrt{2}}}\right]^{-1}
  \times\left[\erf{\frac{z_{\rm max}-z}{\sigma_{\rm z}\sqrt{2}}}-
    \erf{\frac{z_{\rm min}-z}{\sigma_{\rm z}\sqrt{2}}}\right]
\end{multline}
and the same $V_i[z]$ for all galaxies in the sample. Moreover, as
estimated for $p_i(z|z_{{\rm ph},i})$ and needed for $M_{ij}$, we use the
\texttt{BPZ} posterior density of $z$ for the $i$th galaxy in
\cfhtlens.

By means of this analysis we find for our lens low-$z$ (high-$z$)
samples values of $\bar{n}_{\rm g}$ that are on average
\mbox{$\sim3\,\%$} (\mbox{$\sim1\,\%$}) larger than the
traditional $V_{\rm max}$-estimator (assuming $z\equiv z_{\rm
  ph}$). All results are shown in Fig. \ref{fig:ng} as black data
points.

\section{Construction of galaxy bias templates}
\label{app:templateconstruction}

We use smooth templates of $b(k)$ and $r(k)$ to stabilise the
deprojection of the projected galaxy bias. They represent the average
scale-dependent galaxy bias of a specifically selected galaxy
population with mean redshift $\bar{z}_{\rm d}$. Details of the
derivation can be found in SH18. We give a brief summary here only.

Our templates separate galaxy bias in the two-halo regime from galaxy
bias in the one-halo regime. In the two-halo regime, the bias is a
stochastic linear bias with two free parameters
$b_{\rm ls},r_{\rm ls}$: $b_{\rm ls}$ is the large-scale bias factor,
and $r_{\rm ls}$ is the correlation factor.

In the one-halo regime, galaxies populate halos of virial mass
$m$. The mean comoving number density of halos at redshift
$\bar{z}_{\rm d}$ is $n(m)\,\d m$ for the mass interval
$m\in[m,m+\d m]$ according to \cite{1999MNRAS.308..119S}.  The density
profile of the dark matter for a halo with mass $m$,
\begin{equation}
  \rho_{\rm m}(r,m)\propto
  r_{\rm vir}^{-1}\,
  \,\left[1+\frac{r\,c(m,\bar{z}_{\rm d})}{r_{\rm vir}}\right]^{-2}\;,
\end{equation}
follows a Navarro-Frenk-White (NFW) density profile, but is
truncated at the virial radius $r_{\rm vir}$
\citep{2002PhR...372....1C,1996ApJ...462..563N}; for the concentration
parameter $c(m,z)$ at redshift $z=\bar{z}_{\rm d}$ we adopt the
relation in \cite{2002MNRAS.337..875T}. The formalism uses the density
profile as a Fourier transform and normalises it by its viral mass:

\begin{equation}
  \tilde{u}_{\rm m}(k,m)=
  \frac{\int_0^{r_{\rm vir}}\d r\;r\,k^{-1}\,\rho_{\rm m}(r,m)\,\sin{(k\,r)}}
  {\int_0^{r_{\rm vir}}\d r\;r^2\,\rho_{\rm m}(r,m)}\;.
\end{equation}
For the radial distribution of satellite galaxies inside a halo, we
adopt the normalised profile
\begin{equation}
  \tilde{u}_{\rm g}(k,m)=
  \left[\tilde{u}_{\rm m}(k,m)\right]^\zeta\;,
\end{equation}
where only for $\zeta=1$ satellites have the same radial profile as
dark matter. Then we write the biasing functions in the one-halo
regime as
\begin{eqnarray}
  \label{eq:b1hpq}
  [b^{\rm 1h}_{pq}(k)]^2&=&
  \frac{\int_0^\infty\d m\;n(m)\,m^2\,\tilde{u}_{\rm g}^{2p}(k,m)\,b(m)^2\,\left(1+\frac{V(m)}{\ave{N|m}}\right)}
  {\int_0^\infty\d m\;n(m)\,m^2\,\tilde{u}^2_{\rm m}(k,m)}\;;
  \\\nonumber
  r^{\rm 1h}_{pq}(k)&=&
  \frac{\int_0^\infty\d m\;n(m)\,m^2\,\tilde{u}_{\rm m}(k,m)\,\tilde{u}_{\rm g}^q(k,m)\,b(m)}
  {\int_0^\infty\d m\;n(m)\,m^2\,\tilde{u}^2_{\rm m}(k,m)}\,\frac{1}{b^{\rm 1h}_{pq}(k)\,}\;,                 
\end{eqnarray}
for the mean number of galaxies inside halos of mass $m$,
\begin{equation}
  \ave{N|m}=\frac{m\,b(m)\,\bar{n}_{\rm g}}{\bar{\rho}_{\rm m}}=:\frac{m\,b(m)}{m_{\rm piv}\,b(m_{\rm piv})}\;,
\end{equation}
expressed by the mean biasing function $b(m)$
\citep{2012MNRAS.426..566C} and a normalisation with the pivotal mass
$m_{\rm piv}$.  At $m_{\rm piv}$ the mean number of galaxies inside a
halo is \mbox{$\ave{N|m_{\rm piv}}=1$}. The mean number of galaxy
pairs is given by the normalised excess variance $V(m)$ in
\mbox{$\ave{N(N-1)|m}=\ave{N|m}\,[\ave{N|m}+V(m)]$}. If there are no
central galaxies, we set \mbox{$p=q=1$} in the biasing functions,
whereas if there are central galaxies in the selected galaxy
population for every halo with \mbox{$N\ge1$}, we use
\begin{eqnarray}
  \label{eq:pq}
  p&=&\left\{
    \begin{array}{ll}
      1 & ,\,{\rm for~} \ave{N(N-1)|m}>1\;\\
      1/2& ,\,{\rm otherwise}
    \end{array}
  \right.\;;
  \\
  \nonumber
  q&=&\left\{
    \begin{array}{ll}
      1 & ,\,{\rm for~} \ave{N|m}>1\;\\
      0 & ,\,{\rm otherwise}
    \end{array}
  \right.\;.
\end{eqnarray}

The more realistic case is that a galaxy population has central
galaxies, but not for all halos that they populate. We achieve this by
another parameter \mbox{$f_{\rm cen}\in[0,1]$} that mixes halos with
central galaxies ($p,q$ as in Eq. \ref{eq:pq}) and halos with only
satellites ($p=q=1$) by the mixing ratio $f_{\rm cen}$. This mixes the
biasing functions in the one-halo regime according to
\begin{eqnarray}
  [b^{\rm 1h}(k)]^2&=&
   f_{\rm cen}\,[b_{pq}^{\rm 1h}(k)]^2+(1-f_{\rm cen})\,[b_{11}^{\rm 1h}(k)]^2\;,
  \\
  r^{\rm 1h}(k)&=&
    \frac{b^{\rm 1h}_{pq}(k)}{b^{\rm 1h}_{11}(k)}\,\frac{f_{\rm cen}}{\sqrt{{\cal S}_{pq}(k)}}\,r^{\rm 1h}_{pq}(k)+
    \frac{1-f_{\rm cen}}{\sqrt{{\cal S}_{pq}(k)}}\,r^{\rm 1h}_{11}(k)\;,
\end{eqnarray}
where we used the shorthand
\begin{equation}
  {\cal S}_{pq}(k):=
   1-f_{\rm cen}+f_{\rm cen}\left(\frac{b^{\rm 1h}_{pq}(k)}{b^{\rm 1h}_{11}(k)}\right)^2\;.
\end{equation}
The galaxy bias in the two-halo regime is unaffected by the mixing
parameter $f_{\rm cen}$. Unbiased galaxies have \mbox{$p=q=\zeta=1$},
\mbox{$b(m)=1$}, and \mbox{$V(m)=0$} for all $m$.

For greater flexibility and no strong constraints on the mass
dependence of the HOD, both $b(m)$ and $V(m)$ are interpolated between
$10^4$ and $10^{16}\,h^{-1}\,\msol$ based on $i=1\ldots22$ free
interpolation points $b_i$ or $V_i$ at $m_i$,
\begin{equation}
  b(m)=
  b_i+\frac{b_{i+1}-b_i}{\ln{m_{i+1}}-\ln{m_i}}\,
  \left(\ln{m}-\ln{m_i}\right)
\end{equation}
and
\begin{equation}
  V(m)=
  V_i+\frac{V_{i+1}-V_i}{\ln{m_{i+1}}-\ln{m_i}}\,
  \left(\ln{m}-\ln{m_i}\right)
\end{equation}
for \mbox{$m_i\le m\le m_{i+1}$}. The function $b(m)$ has to obey
\begin{equation}
  \label{eq:bmnorm}
  \int_0^\infty\d m\;n(m)\,m\,b(m)=\int_0^\infty\d m\;n(m)\,m\;,
\end{equation}
which we use to normalise every set of parameters $\{b_i\}$. In total,
we have $49$ free template parameters
\mbox{$\vec{\Theta}=(\ldots,b_i,\ldots,V_i,\ldots,\zeta,m_{\rm
    piv},b_{\rm ls},r_{\rm ls},f_{\rm cen})$}.

Finally, the biasing functions in the one- and two-halo regime are
consistently combined into full templates by the relations
\begin{equation}
  [b(k;\vec{\Theta})]^2=
  \left(1-W_{\rm m}(k)\right)[b^{\rm 1h}(k)]^2+W_{\rm m}(k)\,b_{\rm ls}^2\;
\end{equation}
and
\begin{multline}
  r(k;\vec{\Theta})=\\
  \sqrt{(1-W_{\rm m}(k))(1-W_{\rm g}(k))}\,r^{\rm 1h}(k)
  +\sqrt{W_{\rm m}(k)W_{\rm g}(k)}\,r_{\rm ls}\;,
\end{multline}
using the relative weight of the two-halo term in the matter power
spectrum $P_{\rm m}(k;\chi)$ at $\bar{z}_{\rm d}$,
\begin{equation}
  W_{\rm m}(k)=
  \frac{P_{\rm m}^{\rm 2h}(k,\chi)}{P_{\rm m}(k,\chi)}=
  \left(1+\frac{P_{\rm m}^{\rm 1h}(k,\chi)}{P_{\rm m}^{\rm 2h}(k,\chi)}\right)^{-1}\;,    
\end{equation}
and the expression
\begin{equation}
  W_{\rm g}(k):=\left(\frac{b_{\rm ls}}{b(k;\vec{\Theta})}\right)^2\,W_{\rm m}(k)\;.
\end{equation}
We estimate $W_{\rm m}(k)$ by applying the model of the matter power
spectrum in \cite{2000MNRAS.318..203S},
\begin{equation}
  \frac{P_{\rm m}^{\rm 1h}(k,\chi)}{P_{\rm m}^{\rm 2h}(k,\chi)}=
  \frac{\int_0^\infty\d m\;n(m)\,m^2\,\tilde{u}_{\rm m}^2(k,m)}
  {P_{\rm lin}(k,\bar{z}_{\rm d})\,
    \left(\int_0^\infty\d m\;n(m)\,m\,
      \tilde{u}_{\rm m}(k,m)\,b_{\rm h}(m,\bar{z}_{\rm d})\right)^2}\;,
\end{equation}
where $P_{\rm lin}(k,z)$ denotes the linear matter power spectrum at
redshift $z$ in our fiducial cosmology, and $b_{\rm h}(m.z)$ is the
bias factor of halos with mass $m$ at redshift $z$, for which we
utilise here the model in \cite{2005ApJ...631...41T}.

\section{Template prior}
\label{app:templateprior}

\begin{figure}
  \begin{center}
    \includegraphics[width=95mm]{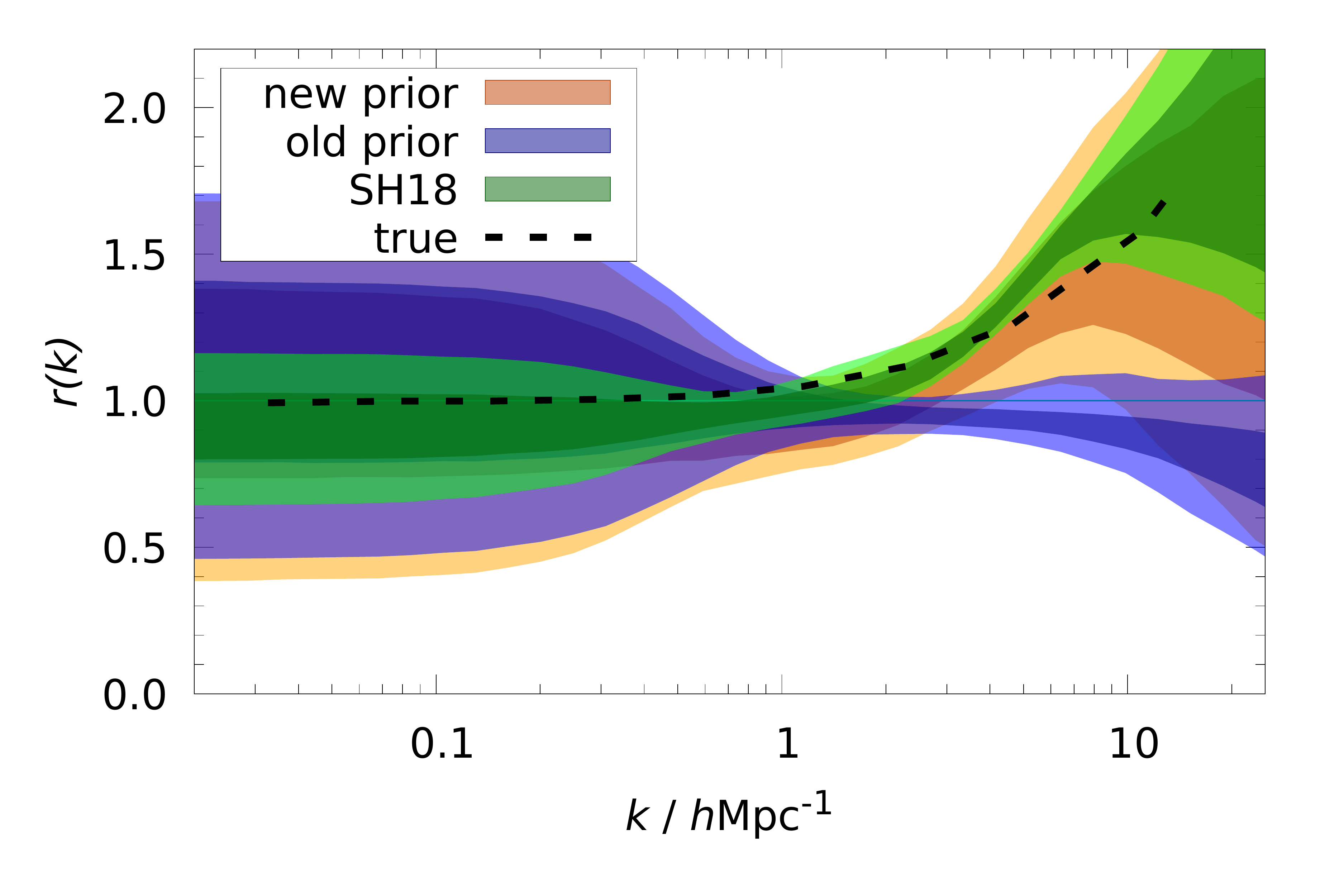}
  \end{center}
  \caption{\label{fig:priortest} Effect of different $b(m)$ prior
    distributions on the reconstructed $r(k)$ for \cfhtlens-like noisy
    data (the orange `new prior' and the blue `old prior') in
    comparison to the green SH18-like data with high
    signal-to-noise ratio. The posterior constraints indicate 68\% and 95\%
    credibility regions. The dashed line is the true $r(k)$ of the H15
    semi-analytic galaxies used in this test (SM4 high-$z$).}
\end{figure}

We modify the SH18 prior for the mean biasing function $b(m)$ due to
the following reason.  The fiducial \mbox{$\sim1000\,\rm deg^2$}
survey in SH18 effectively assumes data with measurement noise roughly
a factor of \mbox{$\sim\sqrt{10}$} smaller than in our
\cfhtlens~analysis. To make sure that the original setup in SH18 also
works for noisier data, we repeated the tests in SH18 with the
original synthetic data but using a noise covariance $\mat{C}$ inflated by a
factor of $10$. While the reconstructed $b(k)$ is still consistent
with the true galaxy bias in the synthetic data, we found problems for
$r(k)$ on scales smaller than \mbox{$k\gtrsim3\,h\,\rm
  Mpc^{-1}$}. This is demonstrated for one example in
Fig. \ref{fig:priortest}. The green contours SH18 show the constraints
for $r(k)$ and data with a high signal-to-noise ratio, as in SH18, which is
consistent with the true $r(k)$. The blue contours, labelled `old
prior', show the reconstruction with the noisier data and the original
$b(m)$ prior in SH18. Clearly, for the old prior and noisier data, the
$r(k)$ constraints are too low compared to the true correlation factor
(dashed line) on small spatial scales. After a series of tests of our
code to possibly identify bugs in the  Markov chain Monte Carlo (MCMC)
sampler, we found that the original uniform prior within
$b(m)\in[0,300]$ is too informative so that it dominates and biases
the reconstruction on small scales where our noisier data now makes
the likelihood subdominant.

We fixed this problem by using equal probability for different orders of
magnitude in the $b(m)$ prior. As discussed in Section 2.9 of
\cite{gelman2003bayesian}, a uniform prior on a logarithmic scale,
which means a prior distribution
\mbox{$p_{\rm b}(b(m_i))\propto b(m_i)^{-1}$}, often provides a less
informative alternative, which we adopt for each interpolation point
$m_i$ of $b(m)$ and the orange contours for the `new prior'. In addition, to
restrict the dynamic range of $b(m)$, we reject MCMC proposals for
$b(m)$ for which the ratio \mbox{$\max{\{b(m)\}}/\min{\{b(m)\}}$}
exceeds $10^4$. As a result, the new prior increases the uncertainty
on small spatial scales and gives a reconstruction that is now
consistent with the correct correlation factor (dashed
line). Furthermore, the prior is subdominant for the higher
signal-to-noise data in SH18 and both the old and our new prior give
 similar constraints for $r(k)$ (not shown).

Our new prior for $b(m)$ is easily implemented into the MCMC code by
changing from $b(m)$ to $\ln{b(m)}$ in the parametrisation of the
galaxy-bias templates. Its practical effect is to prefer small values
of $b(m)$ for halo-mass scales $m$ where the likelihood is
subdominant (i.e. it suppresses galaxy numbers in these halos). For
comparison, the original prior in SH18 prefers a uniform linear-scale
scatter around \mbox{$b(m)=1$} in these cases.

\section{Distribution of systematic errors}
\label{sect:normerror}

\begin{figure*}
  \begin{center}
    \includegraphics[width=80mm]{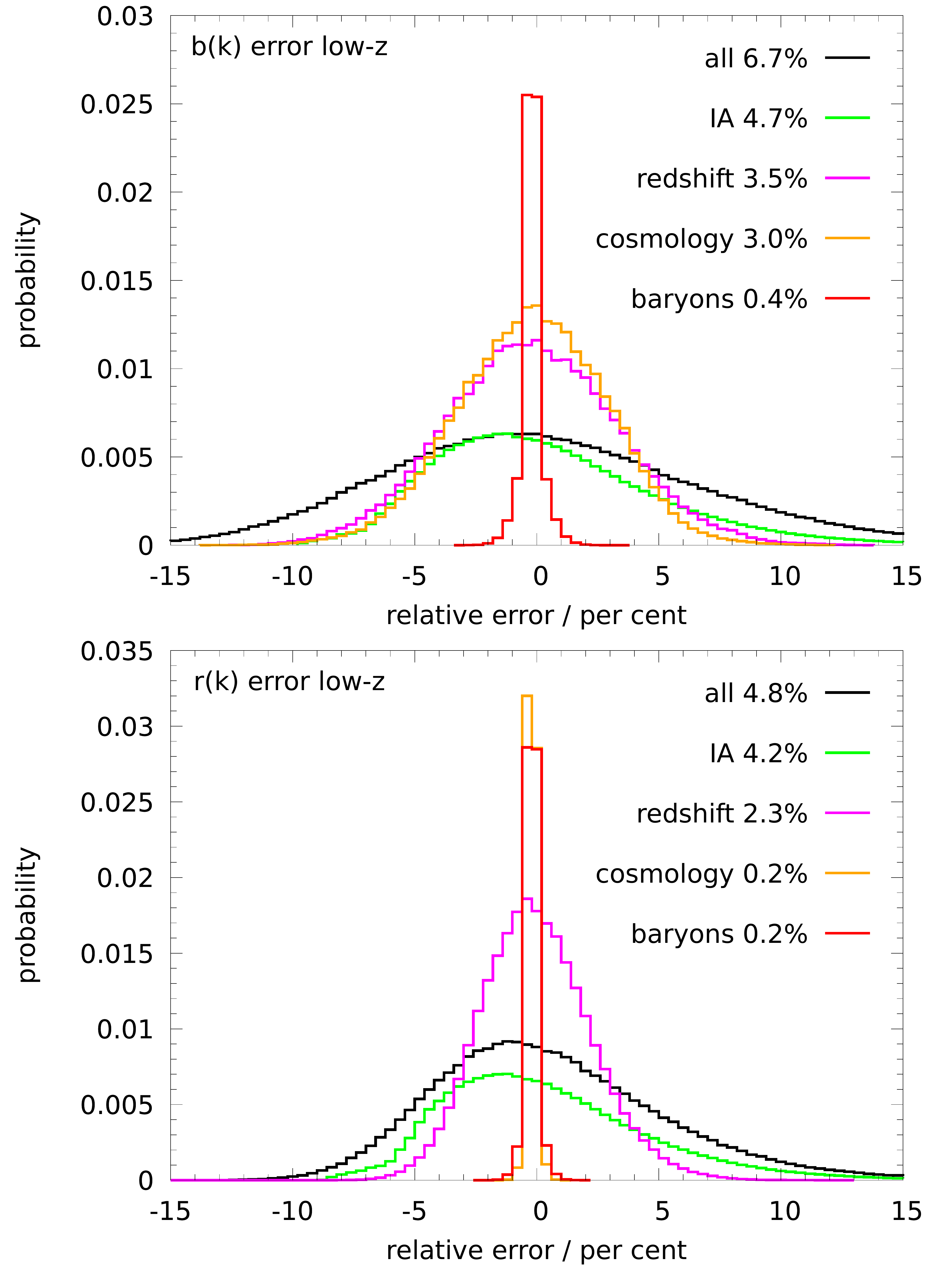}\vspace{0.1cm}\includegraphics[width=80mm]{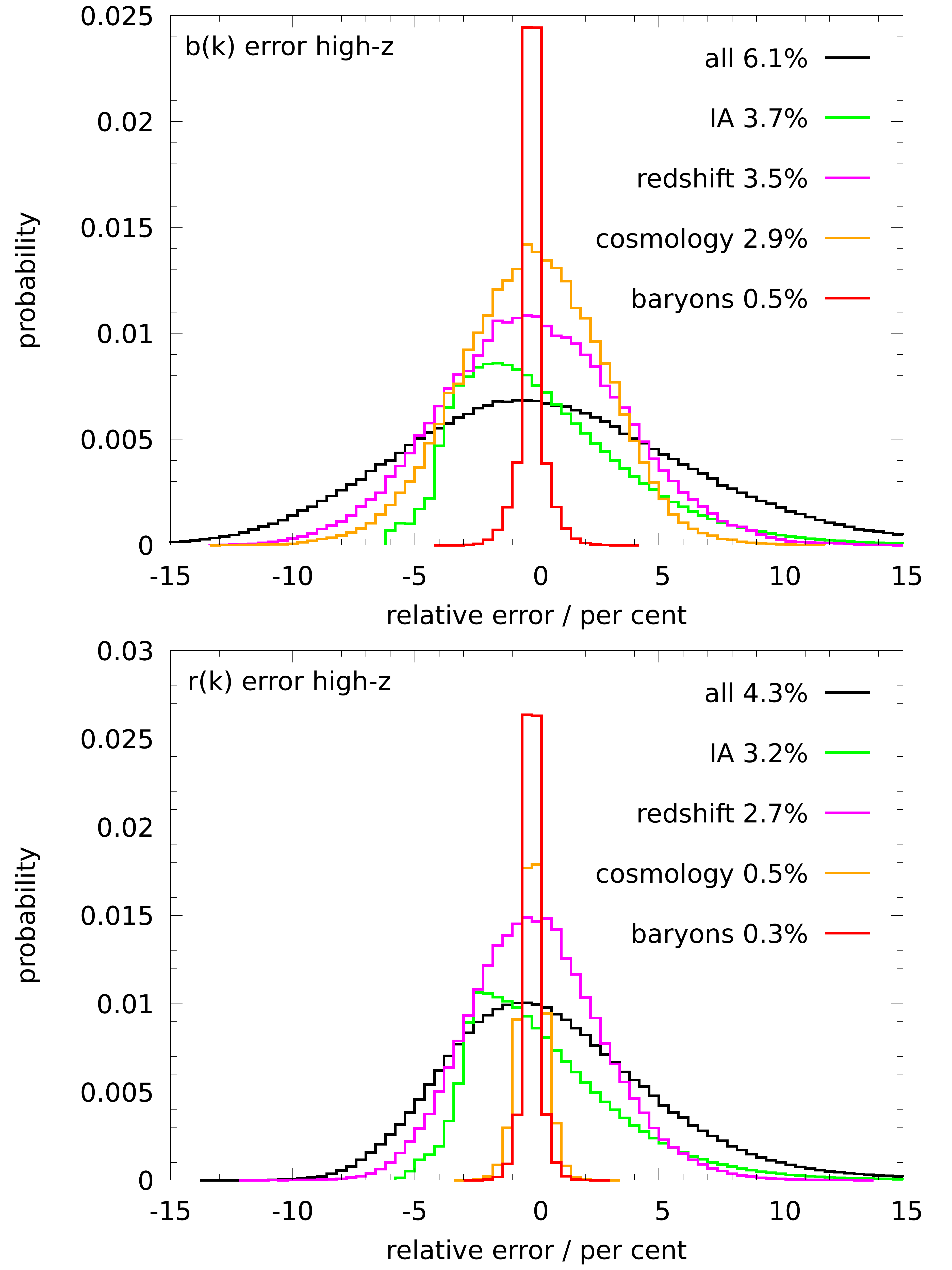}
  \end{center}
  \caption{\label{fig:normerror} Distribution of fractional errors in
    the reconstructed $b(k)$ (\emph{top}) and $r(k)$ (\emph{bottom})
    for unbiased galaxies. The baseline for systematic errors of
    parameters is summarised in Table \ref{tab:errorbase}.  The
    left-hand panels show results for the low-$z$ samples; the
    right-hand panels are for the high-$z$ samples. The lines labelled
    `all' (see legend for colours) include errors from all parameters,
    whereas the other lines show the errors from varying one parameter
    only. The numbers in each panel's legend give the RMS error. A
    positive error means that the inferred galaxy bias is
    systematically too high on all scales.}
\end{figure*}

The reconstructed biasing functions rely on parameters for the
fiducial cosmology, redshift distributions of lenses and sources, and
a model for the intrinsic source alignment (IA).  For
Fig. \ref{fig:normerror}, we vary these fiducial parameters for
unbiased galaxies in the low-$z$ and high-$z$ redshift bin to simulate
the distribution of amplitude errors in $b(k)$ and $r(k)$. As baseline
for the error model, we assume the RMS values in Table
\ref{tab:errorbase} for (i) the mean and width of the redshift
distributions, (ii) the set of fiducial cosmological parameters, (iii)
the baryon physics in the non-linear matter power spectrum, and (iv) 
the amplitude of the IA.

The error distribution for all parameter uncertainties combined is
given by the histograms labelled `all', while the other histograms show the
propagated errors with just one parameter variation switched on at a
time. The  `all' histograms adopt uncorrelated errors between (i) to
(iv) so that the RMS error $\sigma$ of `all', shown in the figure
panels, is the quadratic sum \mbox{$\sigma^2=\sum_i\sigma_i^2$} of the
individual RMS values $\sigma_i$, which are also indicated in the
panels. Realistically, however, we expect the errors to be correlated
to some degree. In the worst case, the maximum combined RMS error we
can achieve is given by the linear sum \mbox{$\sigma=\sum_i\sigma_i$},
which provides a conservative upper limit.

In total we find a systematic RMS error for the $b(k)$ amplitude
between $6.1\%$ (uncorrelated) and $11.6\%$ (conservative), and for
$r(k)$ an error between $4.3\%$ and $6.9\%$. The errors in the low-$z$
bin are about $0.5\%$ larger than those for high-$z$ due to
IA. For $b(k)$, the main contributors to the systematic error are IA,
redshift errors, and uncertainties in the fiducial cosmology.  For
$r(k)$, the cosmology uncertainties are less relevant, while IA and
redshift errors are still important.

\section{Supplemental data on template parameters}
\label{app:template_prms}

\begin{figure*}
  \centering  \includegraphics[scale=0.8,clip=false,angle=0]{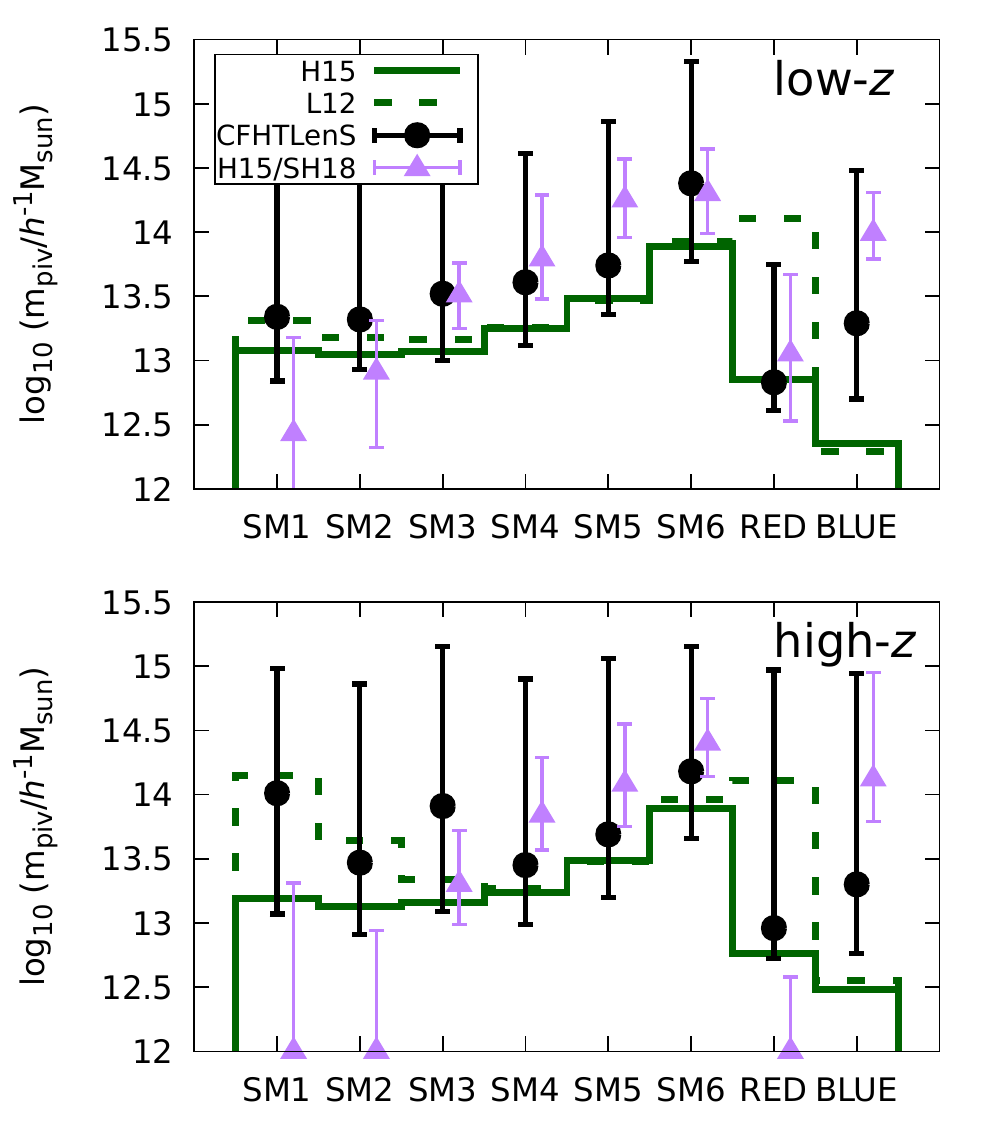}
   \includegraphics[scale=0.8,clip=false,angle=0]{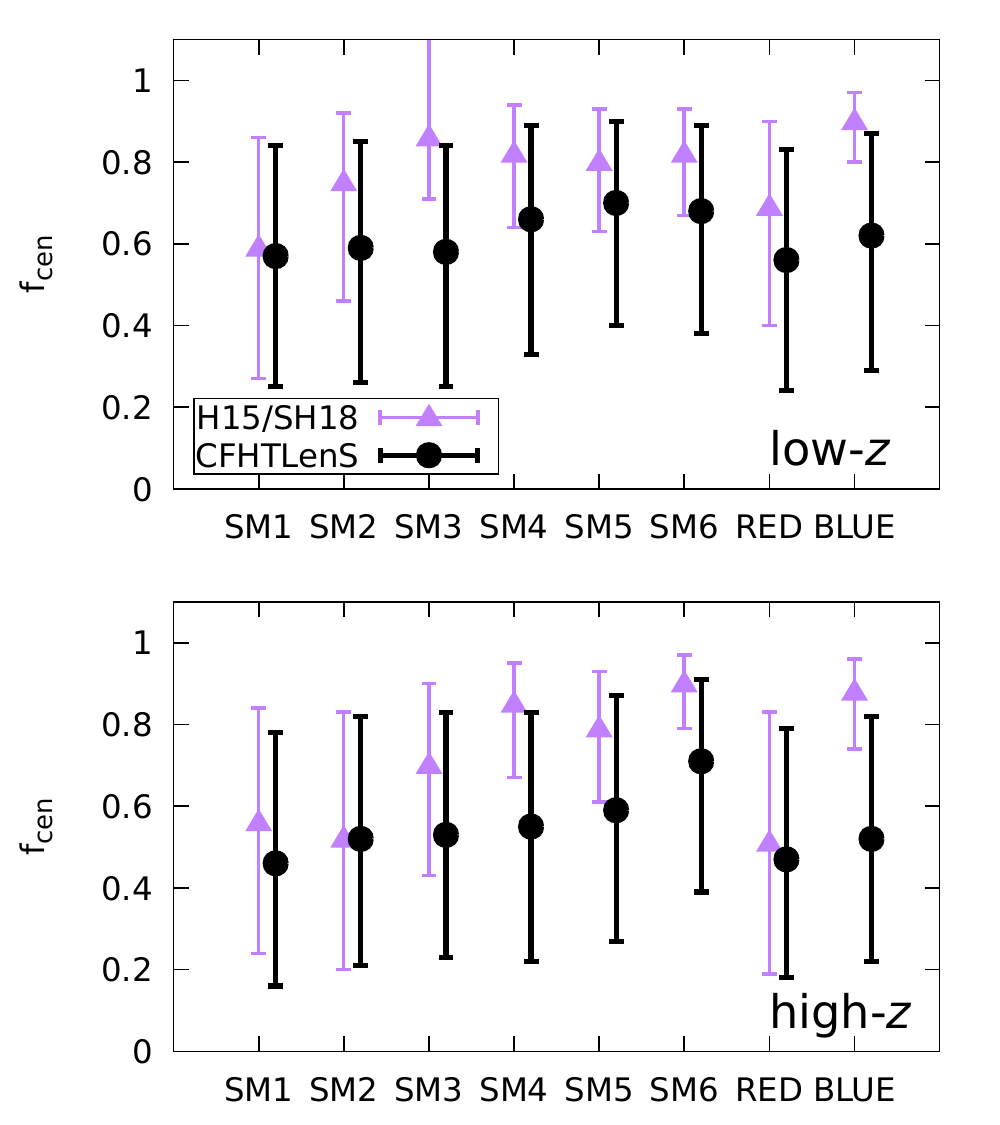}
   \caption{\label{fig:fcenmpiv} Posterior constraints of
     $m_{\rm piv}$ (left panels) and $f_{\rm cen}$ (right panels) with
     $68\%$ CIs as median and error bars for all galaxy samples
     indicated on the $x$-axes; top panels are for low-$z$ samples and
     bottom panels for high-$z$ samples. The circles are
     \cfhtlens~measurements, and the triangles are measurements from
     SH18 in a $10^3\,\rm deg^2$ mock survey with H15 galaxies and
     galaxy selections as in \cfhtlens. The green lines show
     $m_{\rm piv}$ in H15 (solid) and L12 (dotted) directly from the
     simulation snapshots.}
\end{figure*}

\begin{figure*}
  \centering  \includegraphics[scale=0.33,clip=false,angle=0]{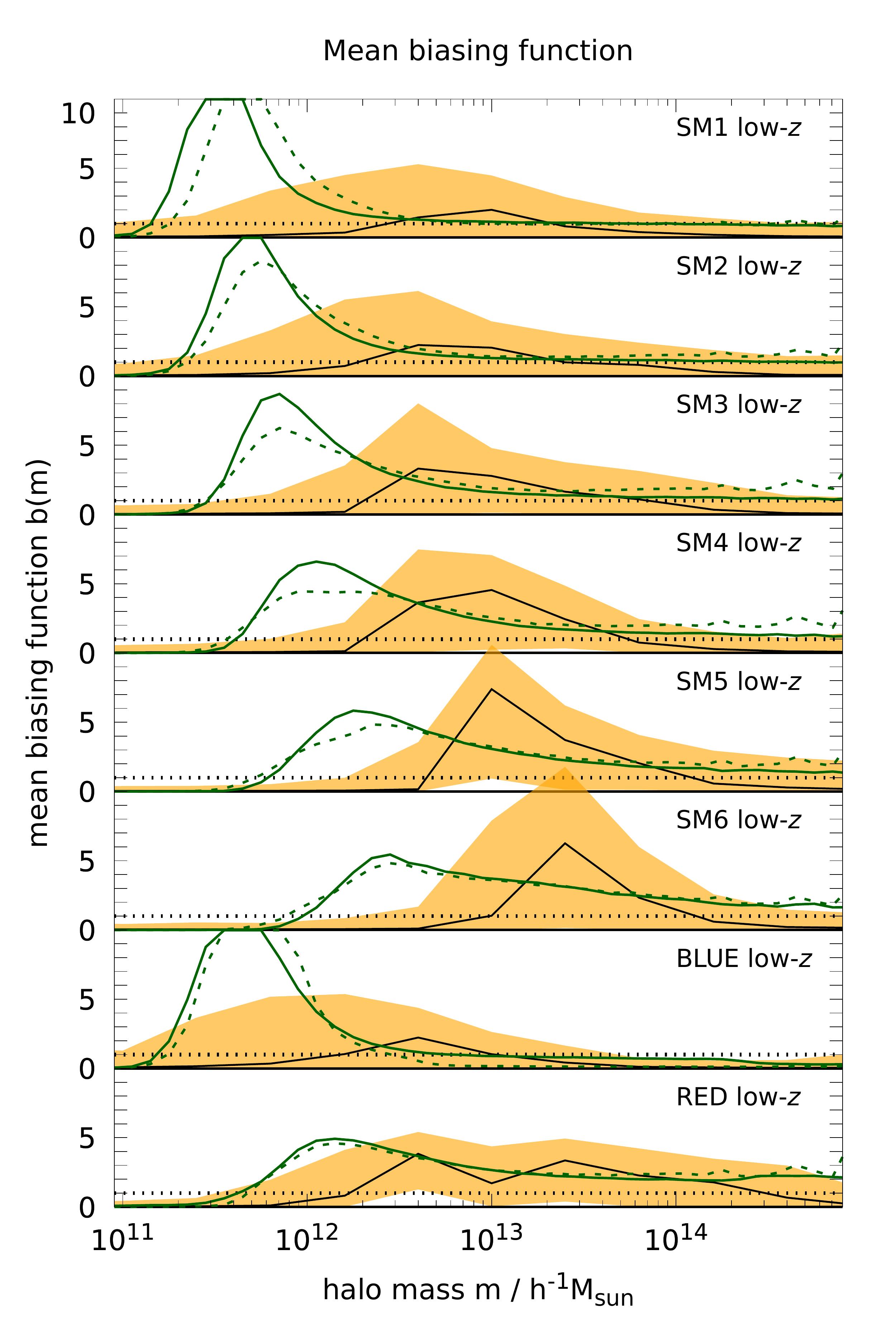}\includegraphics[scale=0.33,clip=false,angle=0]{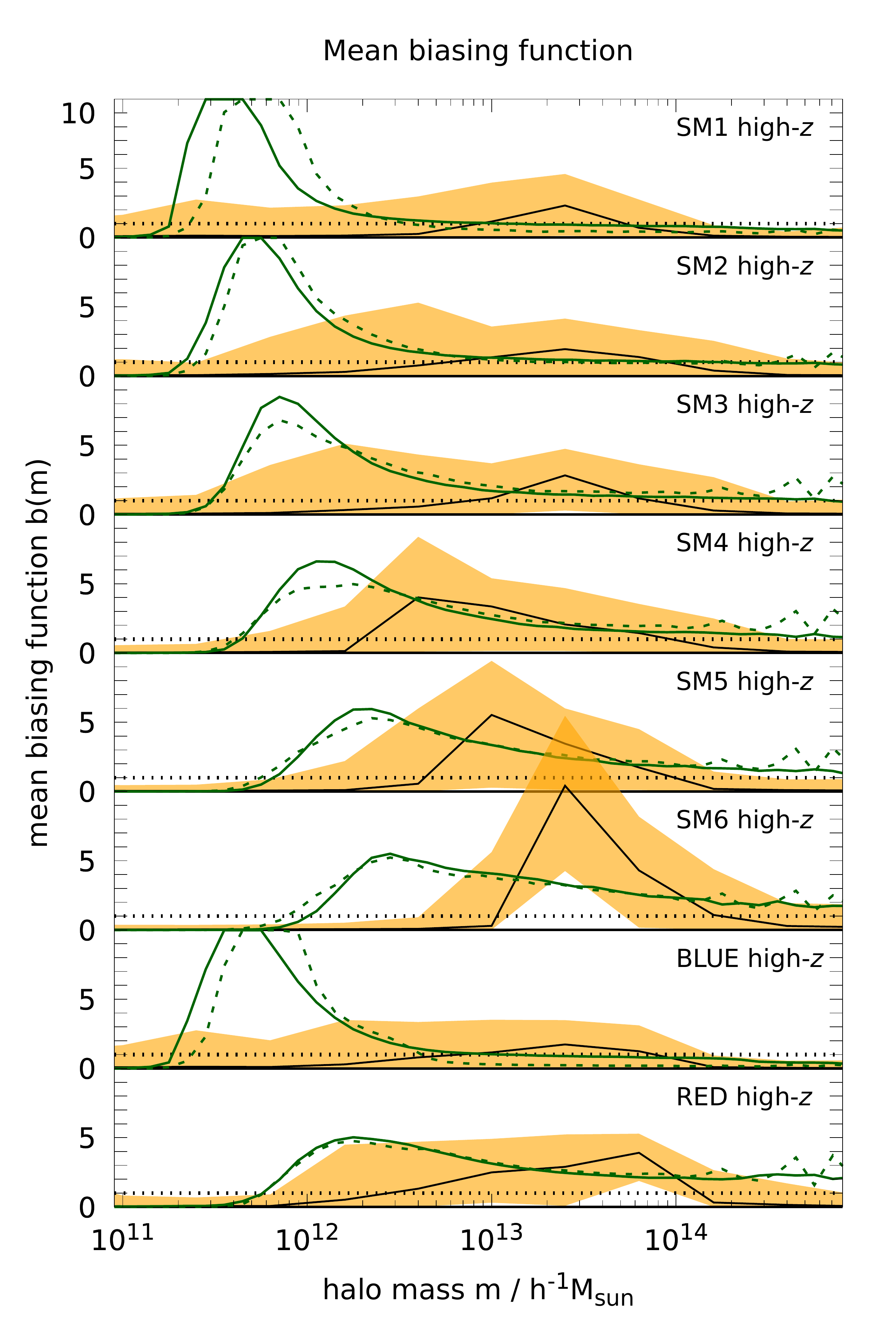}
  \caption{\label{fig:summary_bm} Posterior density of the mean
    biasing function
    \mbox{$b(m)=\ave{N|m}\bar{\rho}_{\rm m}(\bar{n}_{\rm g}m)^{-1}$}
    as function of halo mass $m$. The low-$z$ samples are in the left
    panel, the high-$z$ samples in the right panel. The dotted lines
    indicate \mbox{$b(m)=1$} for each sample. The shaded regions are
    the $68\%$ CIs around the median (solid lines). The green lines
    show for comparison the $b(m)$ in the SAMs (solid line: H15;
    dotted lines: L12) directly from the $\ave{N|m}$ in the simulation
    snapshots.}
\end{figure*}

\begin{figure*}
  \centering
  \includegraphics[scale=0.22,clip=false,angle=0]{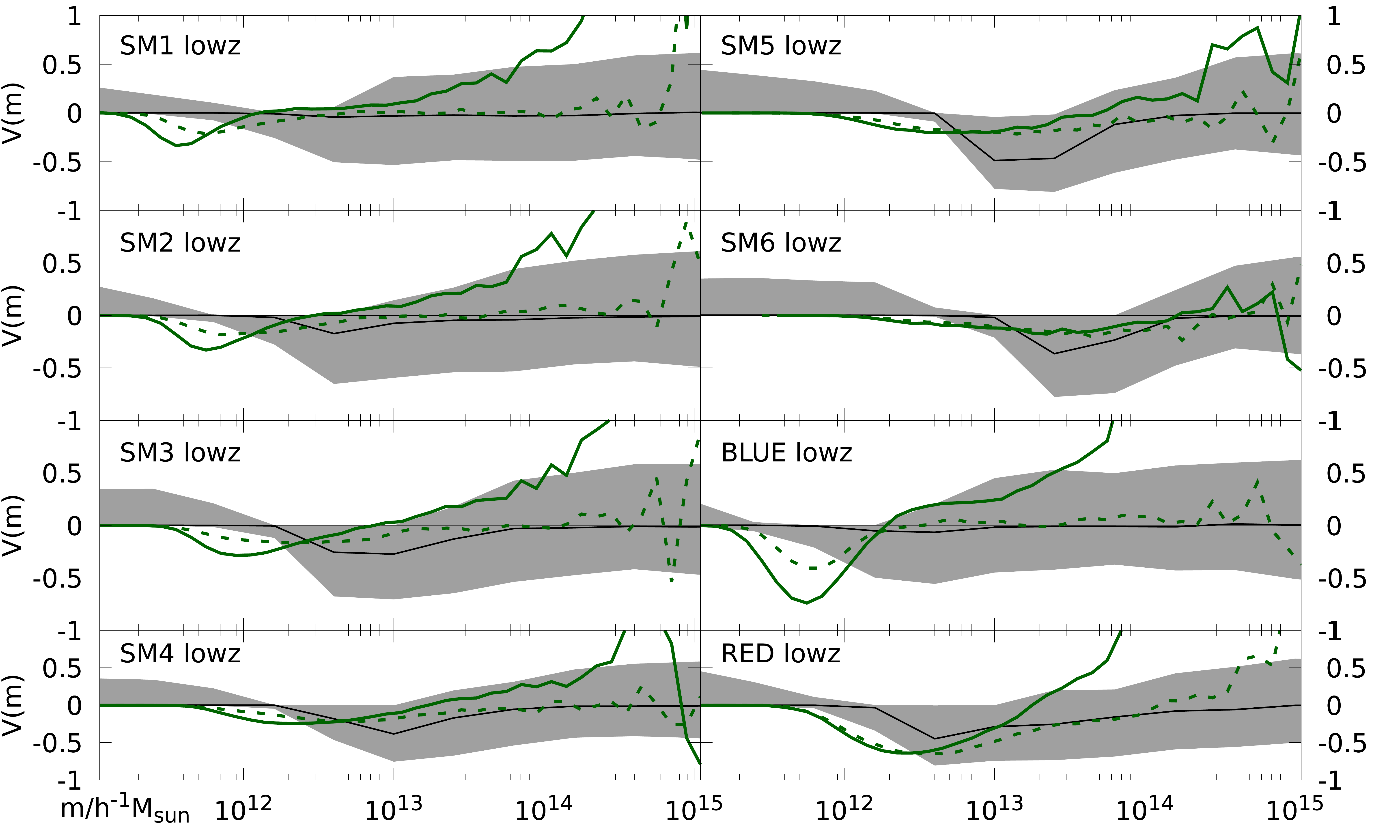}
  \\
   \includegraphics[scale=0.22,clip=false,angle=0]{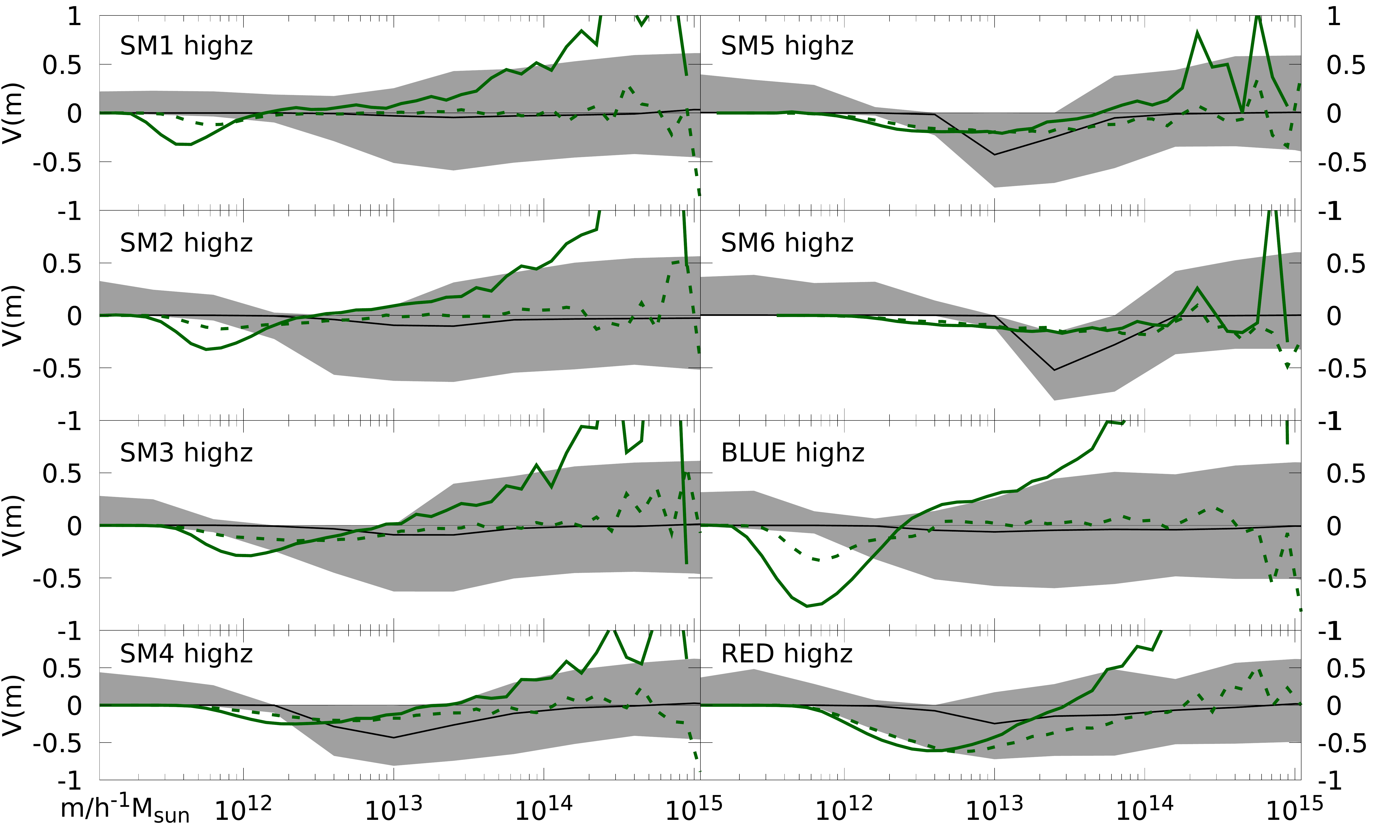}
   \caption{\label{fig:summary_vm} Posterior density of the normalised
     excess variance
     \mbox{$V(m)=[\ave{N(N-1)|m}-\ave{N|m}^2]\,\ave{N|m}^{-1}$}
     ($y$-axis) as a function of halo mass $m/h^{-1}\,\msol$ ($x$-axis)
     for the low-$z$ galaxy samples (top) and high-$z$ samples
     (bottom).  The shaded regions are the $68\%$ CIs around the
     median (black solid lines). The green lines show $V(m)$ for H15
     (solid) and L12 (dashed) directly from the $\ave{N(N-1)|m}$ in
     the simulation snapshots.}
 \end{figure*}

 Figures \ref{fig:fcenmpiv} to \ref{fig:summary_vm} summarise our
 posterior constraints on the template parameters and compare them,
 where data are available, to the corresponding values in the
 SAMs. The \cfhtlens~error-bars or shaded regions denote $68\%$ CIs
 around the posterior median. We note here again that the purpose of
 the templates is not an accurate modelling of the HOD of galaxies,
 but the stabilisation of the biasing functions in the deprojection of
 the ratio statistics.  With this in mind, we find the following:
\begin{itemize}
\item The CIs of pivotal mass, $m_{\rm piv}$, shown with filled
  circles in the left panels of Fig. \ref{fig:fcenmpiv}, span more
  than one order of magnitude.  The median values are almost all above
  the SAM values (green lines), directly determined from their
  $\ave{N|m}$ in the simulation snapshots. This could indicate that
  the mean halo mass of \cfhtlens~galaxies is higher than in the
  SAMs. However, these values are probably biased high to some
  degree. This can be seen by the triangle data points, taken from a
  mock galaxy bias reconstruction in SH18 using H15 galaxies in a
  $10^3\,\rm deg^2$ \cfhtlens-like mock survey. These data are
  also somewhat higher than the true values for H15 (solid green lines) in
  the mock data.
\item The fraction $f_{\rm cen}$ of {halos} hosting central
  galaxies in the right panels of Fig. \ref{fig:fcenmpiv} falls
  between $20\%$ to $80\%$, basically reflecting the Bayesian prior,
  with a weak increase in the median from SM1 to SM6 in both redshift
  bins (filled circles).  Again, the triangle data points are, for
  comparison, from the mock analysis in SH18 for H15 data. They are
  consistent with \cfhtlens~but better confined, roughly zooming in to
  a posterior interval from $70\%$ to $90\%$.
\item Although the exact shape of $b(m)$ in Fig. \ref{fig:summary_bm}
  is poorly confined (orange regions), high values of $b(m)$ are
  typically permitted only within some range of halo mass. This range
  shifts for low-$z$ (less clear for high-$z$) from a few $10^{12}$ to
  a few $10^{13}\,h^{-1}\,\msol$ between SM1 and SM6, and the range
  narrows for larger stellar masses.  The BLUE samples suppress $b(m)$
  for high halo masses \mbox{$m\gtrsim10^{14}\,h^{-1}\,\msol$}, and
  the RED samples have a strongly varying $b(m)$ over a broad range of
  halo masses except below a few $10^{11}\,h^{-1}\,\msol$.

  This is qualitatively also seen in the SAMs. The green lines (solid
  lines: H15; dashed lines: L12) are $b(m)$ for the SAMs, constructed
  from the $\ave{N|m}$ at the simulation snapshots by using the
  definition
  \mbox{$b(m)=\Omega_{\rm m}\bar{\rho}_{\rm
      crit}\,\ave{N|m}\,(\bar{n}_{\rm g}\,m)^{-1}$}. For SM1-SM6, the
  SAM $b(m)$ basically behaves as \cfhtlens: $b(m)$; it is long-tailed and
  peaked around a characteristic halo mass with relatively high galaxy
  numbers per halo mass, and the peak moves to higher masses for
  higher stellar mass. Crucially, the SAMs are offset with respect to
  the \cfhtlens~peaks; in other words  \cfhtlens~galaxies of the same stellar mass
  seemingly prefer to populate halos of higher mass. However, this
  tendency towards halos of higher mass is at least partly an artefact
  of the crude modelling in the templates because the peak position in
  $b(m)$ for SM5-SM6 low-$z$ and high-$z$ in the analysis by SH18 with
  H15-based mock data (their Figure D.1) is similarly offset with
  respect to the true $b(m)$ of H15.

  The offset might also be present for the BLUE and RED samples, but
  this is far less clear here.  As is true for \cfhtlens, the SAM BLUE
  galaxies prefer to populate low-mass halos (they are preferentially
  field galaxies), while RED galaxies prefer halos of higher mass
  (many red galaxies are satellites).
\item The normalised excess variance $V(m)$, grey shaded regions in
  Fig. \ref{fig:summary_vm}, is consistent with a Poisson variance of
  galaxy numbers inside halos for all galaxy samples ($V(m)=0$) or
  possibly a sub-Poisson variance around $10^{13}$ to
  $10^{14}\,h^{-1}\,\msol$; SM6 high-$z$ is a prominent example.  We note
  that for $m\lesssim10^{12}\,h^{-1}\,\msol$ we have $\ave{N|m}\ll1$,
  and thus a lower limit of $V(m)\ge-\ave{N|m}\approx0$ for all
  samples, which is visible in the plots. The SH18 constraints ($68\%$
  CIs) with the H15-based mock data are similar to the
  \cfhtlens~results (green dashed lines).

  The green lines (H15: solid; L12: dashed) are the corresponding
  trends for $V(m)$ in the SAM galaxy samples. They also indicate a
  sub-Poisson variance around $10^{12}\,h^{-1}\,\msol$, but with an
  offset relative to \cfhtlens~that is similar to the offset seen for
  $b(m)$. At higher halo mass of $m\gtrsim10^{14}\,h^{-1}\,\msol$ and
  for the small stellar masses SM1-SM4, H15 shows a super-Poisson
  variance ($V(m)>0$) that is not visible in L12. This might be
  related to the different SAM physics of satellites inside galaxy
  clusters. The SAM statistics notably becomes noisy for halo masses
  above a few $10^{14}\,h^{-1}\,\msol$ where the simulation box
  quickly runs out of halos.
\item There is no evidence that supports a distribution of satellites
  different to that of matter inside halos for all stellar mass
  samples (\mbox{$\zeta\approx1$}), although the median $\zeta$ tends
  to decrease for increasing stellar mass. Typical values are
  $\zeta=1.28^{+0.44}_{-0.45}$ ($1.28^{+0.39}_{-0.39}$) for SM1 and
  $\zeta=0.98^{+0.56}_{-0.48}$ ($1.16^{+0.50}_{-0.49}$) for SM6 in
  the low-$z$ (high-$z$) samples.
\end{itemize}

\end{document}